# *De novo* inference of diversity genes and analysis of non-canonical V(DD)J recombination in immunoglobulins

Yana Safonova[1,*] and Pavel A. Pevzner[2]

[1] Center for Information Theory and Applications, University of California in San Diego, San Diego, USA

[2] Computer Science and Engineering Department, University of California in San Diego, San Diego, USA

* Corresponding author: isafonova@eng.ucsd.edu

## Abstract

The V(D)J recombination forms the immunoglobulin genes by joining the variable (V), diversity (D), and joining (J) germline genes. Since variations in germline genes have been linked to various diseases, personalized immunogenomics aims at finding alleles of germline genes across various patients. Although recent studies described algorithms for *de novo* inference of V and J genes from immunosequencing data, they stopped short of solving a more difficult problem of reconstructing D genes that form the highly divergent CDR3 regions and provide the most important contribution to the antigen binding. We present the IgScout algorithm for *de novo* D gene reconstruction and apply it to reveal new alleles of human D genes and previously unknown D genes in camel, an important model organism in immunology. We further analyze non-canonical V(DD)J recombination that results in unusually long tandem CDR3s and thus expands the diversity of the antibody repertoires. We demonstrate that tandem CDR3s represent a consistent and functional feature of all analyzed immunosequencing datasets, reveal ultra-long tandem CDR3s, and shed light on the mechanism responsible for their formation.

## INTRODUCTION

Antibodies provide specific binding to an enormous range of antigens and represent a key component of the adaptive immune system. The *antibody repertoire* is generated by *somatic recombination* of the V (*variable*), D (*diversity*), and J (*joining*) germline gene segments. Immunosequencing has emerged as a method of choice for generating millions of reads that sample antibody repertoires and provide insights into monitoring immune response to disease and vaccination (Turchaninova et al., 2016).

Information about all germline genes in an individual is a pre-requisite for analyzing immunogenomics data. However, nearly all immunogenomics studies rely on the population-level germline genes rather than germline genes in a specific individual that the immunosequencing data originated from. This approach is deficient since the set of known germline genes is incomplete (particularly for non-Europeans) and contains alleles that resulted from sequencing and annotation errors (Wang et al., 2008, Ralph and Matsen, 2017). Moreover, it is non-trivial to figure out which known allele(s) is present in a specific individual since the widespread practice of aligning each read to its closest germline gene results in high error rates (Ralph and Matsen, 2017). These errors hide the identity of the individual germline genes, make it difficult to analyze *somatic hypermutations* (*SHM*) and complicate studies of antibody evolution (Yaari et al., 2012, McCoyet al., 2015, Cui et al., 2016).

*Personalized immunogenomics* (i.e., identifying individual germline genes) is important since variations in germline genes have been linked to various diseases (Watson and Breden, 2012), differential response to infection, vaccination, and drugs (Parameswaran et al., 2013, Chang et al., 2017), aging (Boyd et al., 2013), and disease susceptibility (Kidd et al., 2012, Watson and Breden, 2012, Avnir et al., 2016). However, since the International ImMunoGeneTics (IMGT) database is incomplete even in the case of well-studied human germline genes (Collins et al., 2015), there exist still unknown human allelic variants that are difficult to differentiate from SHMs. In the case of immunologically important but less studied model organisms, such as camels or sharks, the germline genes remain largely unknown. Unfortunately, since assembling the highly repetitive immunoglobulin locus from whole genome sequencing data faces challenges (Luo et al., 2017), the efforts like the 1000 Genomes Project have resulted only in limited progress towards inferring the population-wide census of germline genes (Luo et al., 2017, Yu et al., 2017, Watson et al., 2017).

In addition to personalized immunogenomics, the incompleteness of the IMGT database negatively affects analysis of monoclonal antibodies. Existing tools for antibody sequencing from tandem mass spectra (Bandeira et al., 2008, Castellana et al., 2010) rely on a comprehensive database of V, D, and J genes to assemble tandem mass spectra into an intact antibody. Lack of such databases for many species limits applications of Valens (Digital Proteomics), SuperNova (Protein Metrics), and other software tools for antibody sequencing.

Although the personalized immunogenomics approach was first proposed by Boyd et al., 2010, the manual analysis in this study did not result in a software tool for inferring germline genes. Gadala-Maria et al., 2015 developed the TIgGER algorithm for inferring germline genes and used it to discover eleven novel allelic V segments. However, Gadala-Maria et al., 2015 stopped short of *de novo* reconstruction of the germline genes and acknowledged that it is important to develop algorithms for finding diverged alleles that TIgGER is not able to find. In the case of V and J genes, this challenge was addressed by Corcoran et al., 2016, Zhang et al., 2016, and Ralph and Matsen, 2017. However, as Ralph and Matsen, 2017 commented, the more challenging task of *de novo* reconstruction of D genes remains elusive. This is unfortunate since D genes contribute to the *complementarity determining region 3* (*CDR3*) that covers the junctions between V, D, and J genes and represents the highly divergent part of antibodies. We describe the IgScout algorithm for *de novo* inference of D genes and apply it to diverse immunosequencing datasets.

Although many studies analyzed patterns of V-D-J pairing (Kidd at el., 2017; Kirik et al., 2017), there is still a shortage of studies of unusual recombination events such as *V(DD)J recombination* incorporating two D genes into a single unusually long *tandem CDR3*. Meek et al., 1989 were the first to reveal a few tandem CDR3s, thus confirming the V(DD)J recombination conjecture put forward by Kurosawa and Tonegawa, 1982. However, since tandem CDR3s are rare, they remained elusive for the next two decades and Corbett et al., 1997 and Watson et al., 2006 even argued that tandem CDR3s found in Meek et al., 1989 represent artifacts. However, Briney et al., 2012 and Larimore et al., 2012 demonstrated that tandem CDR3s do exist (at frequency 1 per 800 B-cells) by analyzing high-throughput immunosequencing datasets.

As emphasized in Briney et al., 2012, detecting V(DD)J recombination has to be done with caution since it is often confused with standard V(D)J recombination. Although they came up with a heuristic for detecting tandem CDR3s, there is still no software for detecting tandem CDR3s and it remains unclear how many tandem CDR3s found in Briney et al., 2012 represent false positives. We thus extended the functionality of the IgScout algorithm to finding tandem CDR3s and revealed that V(DD)J recombination is a functional (rather than aberrant) feature with frequency varying from 1 per 200 to 1 per 2500 B-cells across various datasets. Finally, we revealed *ultra-long tandem CDR3s* and shed light on the mechanism responsible for their formation.

## RESULTS

**Immunosequencing datasets.** We analyzed the following datasets described in the Supplemental Note "Immunosequencing datasets":

- **HEALTHY**: 14 datasets from 14 healthy human donors,
- **ALLERGY**: 24 datasets from six allergy patients (Landais et al., 2017),
- **HIV**: 13 datasets from two HIV-infected patients (Levin et al., 2017),
- **PROJECT10**: 600 datasets from various humans resulting from ten NCBI projects
- **CAMEL**: 6 datasets from three healthy camels (Li et al., 2016).

**Constructing CDR3 datasets.** We illustrate the work of IgScout using one of the HEALTHY datasets (Set 1) containing heavy chain repertoires extracted from *peripheral blood mononuclear cells* (*PBMC*). The IgReC tool (Shlemov et al., 2017) extracted 228,619 distinct CDR3s from this dataset. To minimize impact of sequencing and amplification errors, we clustered similar CDR3s (differing by at most 3 mismatches) and constructed consensus for each cluster resulting in 98,576 *consensus CDR3* of average length 46 nucleotides.

Each CDR3s typically starts from a short suffix of a V gene and ends with a short prefix of a J gene. Since these suffixes and prefixes negatively affect reconstruction of D genes, IgScout trims them as described in the Supplemental Note "Preprocessing CDR3 datasets." This procedure reduces the average length of CDR3s from 46 to 30 nucleotides and results in the set of strings $CDR3_{trimmed}$. We refer to the number of strings in $CDR3_{trimmed}$ as $|CDR3_{trimmed}|$.

**Overview of human D genes.** The human immunoglobulin (IGH) locus contains 27 D genes that vary in length from 11 to 37 nucleotides. Since two pairs of human D genes are identical, there exist only 25 distinct D genes. Since the IMGT database refers to D genes using rather long names and since these names do not reveal the ordering of D genes in the IGH loci (that is important for analyzing tandem CDR3s), it is difficult to visualize the IgScout results across all D genes and across multiple immunosequencing datasets. We thus renamed distinct human D genes from D1 to D25 in the increasing order of their positions in the IGH locus. The IMGT database also contains seven alleles of D genes denoted D2*, D2**, D3*, D8*, D10*, D15*, and D19*. See Table 1 and Supplemental Note "Information about human D genes" for details.

| Name | IMGT name | Position (bp) | Length (nt) | Name | IMGT name | Position (bp) | Length (nt) |
|---|---|---|---|---|---|---|---|
| D1 | IGHD1-1 | 1,124,217 | 17 | D14 | IGHD2-15 | 1,145,762 | 31 |
| D2 | IGHD2-2 | 1,126,893 | 31 | D15 | IGHD3-16 | 1,148,085 | 37 |
| D3 | IGHD3-3 | 1,129,360 | 31 | D16 | IGHD4-17 | 1,149,211 | 16 |
| D4 | IGHD4-4 | 1,130,497 | 16 | D5 | IGHD5-18 | 1,150,177 | 20 |
| D5 | IGHD5-5 | 1,131,462 | 20 | D17 | IGHD6-19 | 1,152,020 | 21 |
| D6 | IGHD6-6 | 1,133,309 | 18 | D18 | IGHD1-20 | 1,152,528 | 17 |
| D7 | IGHD1-7 | 1,133,812 | 17 | D19 | IGHD2-21 | 1,155,168 | 28 |
| D8 | IGHD2-8 | 1,136,508 | 31 | D20 | IGHD3-22 | 1,157,688 | 31 |
| D9 | IGHD3-9 | 1,139,038 | 31 | D21 | IGHD4-23 | 1,158,849 | 19 |
| D10 | IGHD3-10 | 1,139,222 | 31 | D22 | IGHD5-24 | 1,159,816 | 20 |
| D4 | IGHD4-11 | 1,140,103 | 16 | D23 | IGHD6-25 | 1,162,180 | 18 |
| D11 | IGHD5-12 | 1,141,070 | 23 | D24 | IGHD1-26 | 1,162,685 | 20 |
| D12 | IGHD6-13 | 1,142,577 | 21 | D25 | IGHD7-27 | 1,178,168 | 11 |
| D13 | IGHD1-14 | 1,143,081 | 17 | | | | |

**Table 1. Positions and lengths of human D genes.** Since the IGH locus starts at the end of the 14th chromosome, positions are given with respect to its reverse complementary sequence. Green and orange cells correspond to two duplicated and identical D genes IGHD4-4*01 – IGHD4-11*01 (D4) and IGHD5-5*01 – IGHD5-18*01 (D5).

**Frequent *k*-mers in D genes.** The problem of inferring germline genes can be formulated as the Trace Reconstruction Problem (Mitzenmacher, 2009) in information theory described in the Methods section. IgScout is a heuristic for solving this problem that is inspired by the RepeatScout algorithm for *de novo* repeat finding (Price et al., 2005) and that is based on analyzing frequent *k*-mers in CDR3s. We illustrate the work of IgScout using *k*-mers of size 15 (all human D genes are longer than 15 nucleotides except for 11 nucleotide long gene D25).

The human D genes contain 305 15-mers. We classify a *k*-mer as *known* if it occurs in a human D gene (from D1 to D25), *mutated* if it differs from a known *k*-mer by a single substitution, and *trimmed* if it contains a known (*k*-2)-mer. All other *k*-mers are called *foreign*. 27% of strings in the $CDR3_{trimmed}$ dataset contain a known 15-mer and 35% contain either a known, or a mutated, or a trimmed 15-mer.

We classify a *k*-mer as *common* if its abundance exceeds *fraction* * $|CDR3_{trimmed}|$ (the default value *fraction*=0.001). Figure 1 and the Supplemental Note "Common *k*-mers" present distributions of frequencies of all common 15-mers in various datasets. Although the vast majority of common *k*-mers are known, mutated, or trimmed, some of them are foreign. These foreign common *k*-mers have to be treated with caution since they may trigger false positive inferences of D genes.

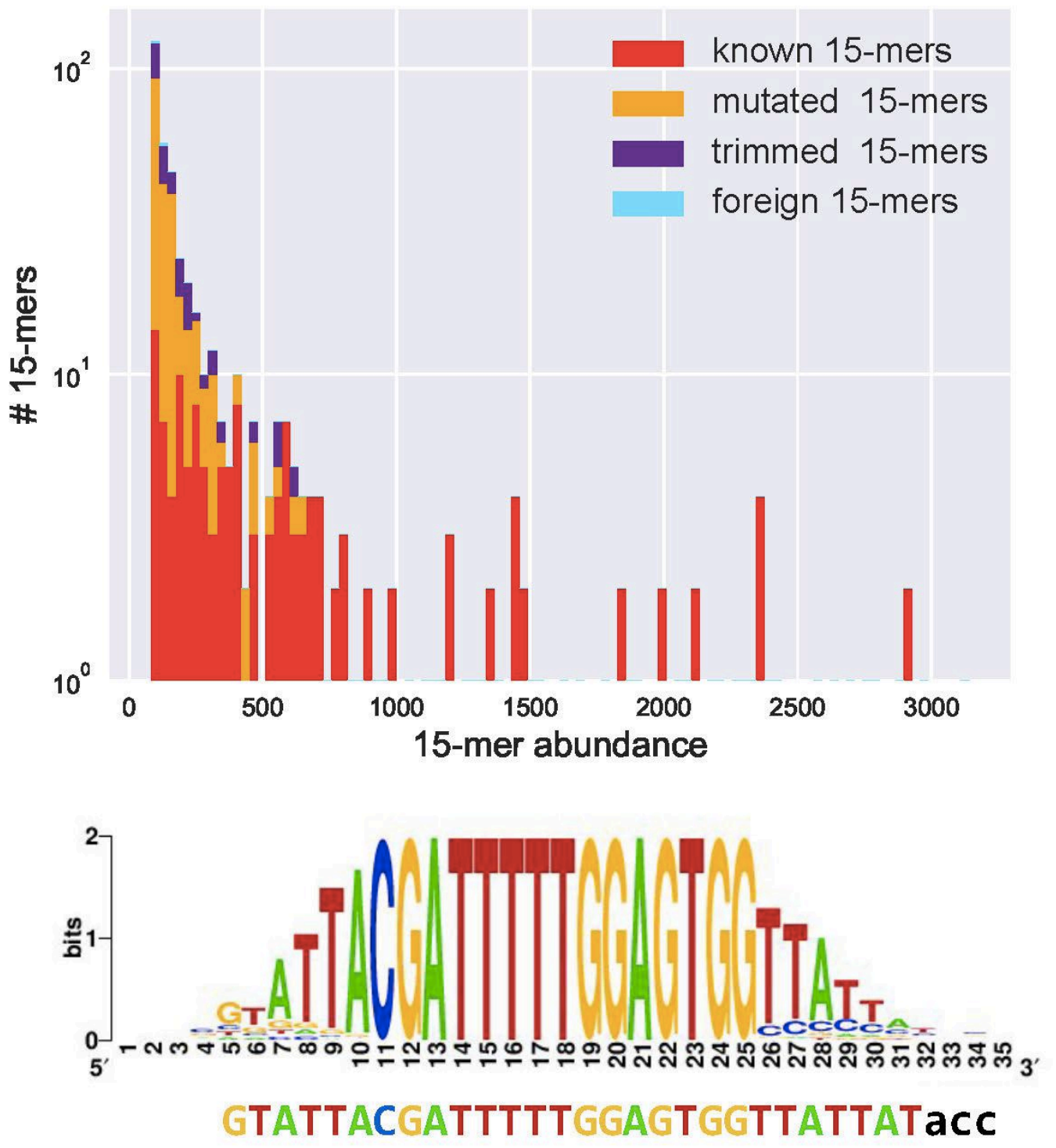

**Figure 1. Abundances of all 443 common 15-mers (top) and the motif logo constructed for the most abundant 15-mer CGATTTTTGGAGTGG in the *CDR3$_{trimmed}$* dataset (bottom).** (Top) The *CDR3$_{trimmed}$* dataset contains 91% of all 15-mers appearing in human D genes (all 15-mers in human D genes are unique, i.e., appear in a single D gene). 443 common 15-mers in the *CDR3$_{trimmed}$* set have abundances varying from 83 to 3141. The *y*–axis represents the number of common 15-mers with given abundance (in logarithmic scale). Red, yellow, violet, and blue bars represent the number of common 15-mers with given abundance among known, mutated, trimmed, and foreign 15-mers, respectively There exist 175 known, 195 mutated, 70 trimmed, and 3 foreign common 15-mers. The histogram represents 100 bins of width 30 each. (Bottom) The ATTACGATTTTTGGAGTGGTTAT is the initial 28-nucleotide long sequence formed by positions in the motif logo with high information content (Crooks et al., 2004). After extending this 28-mer, IgScout reconstructed the 30-mer GTATTACGATTTTTGGAGTGGTTATTAT that is a substring of the 33-nucleotide long D3 (IGHD3-3) gene GTATTACGATTTTTGGAGTGGTTATTATacc shown below the logo.

**From frequent *k*-mers to D gene reconstruction.** IgScout selects a most abundant *k*-mer in the *CDR3$_{trimmed}$* dataset, aligns all CDR3 that contain this *k*-mer (using this *k*-mer as the alignment seed), and constructs the *motif logo* of the resulting alignment (Figure 1). It further trims all positions of the motif logo with the *information content* below *IC* (the default value *IC*=0.5) and computes the consensus string. Afterwards, it extends the consensus strings to the right and to the left to construct a putative D gene as described in the Methods section. Finally, the algorithm removes the sequences that contain *k*-mers from the identified putative D gene from the set *CDR3$_{trimmed}$,* finds a most abundant *k*-mer in the resulting dataset, and iterates. IgScout stops when a most abundant *k*-mer is not a common *k*-mer (see Supplemental Notes "IgScout pseudocode," "IgScout parameters," and "Benchmarking IgScout on simulated immunosequencing datasets"). Figure 2 demonstrates that IgScout reconstructs many known human D genes.

Similarly to the existing tools for reconstructing V and J genes (that typically trim a few nucleotides in the beginning/end of the reconstructed genes), IgScout also trims a few nucleotides in the beginning/end of the reconstructed D genes. Although lowering the *IC* threshold would reduce the number of trimmed nucleotides, we decided not to do it since lowering this parameter may result in erroneous reconstructions and since the trimmed nucleotides hardly affect the downstream applications of IgScout. See Supplemental Note: "How trimmed (rather than complete) D genes affect the downstream analysis of immunosequencing datasets".

Indeed, the personalized immunogenomics applications (such as the discovery of "deficient" germline variants that lead to poor responses to vaccination (Avnir et al., 2016)) are hardly affected by the fact that all existing tools for inferring the V, D, and J genes trim a few nucleotides from the ends. Reconstruction of monoclonal antibodies from tandem mass spectra and various proteogenomics applications are also hardly affected by this trimming. Moreover, in the case of human germline genes (and other genomes with well-characterized germline genes) the trimmed nucleotides can be tentatively reconstructed based on similarity with known germline genes (as has been done in previous studies of V and J genes). The antibody analysis and engineering in model organisms can also be done with partial D genes.

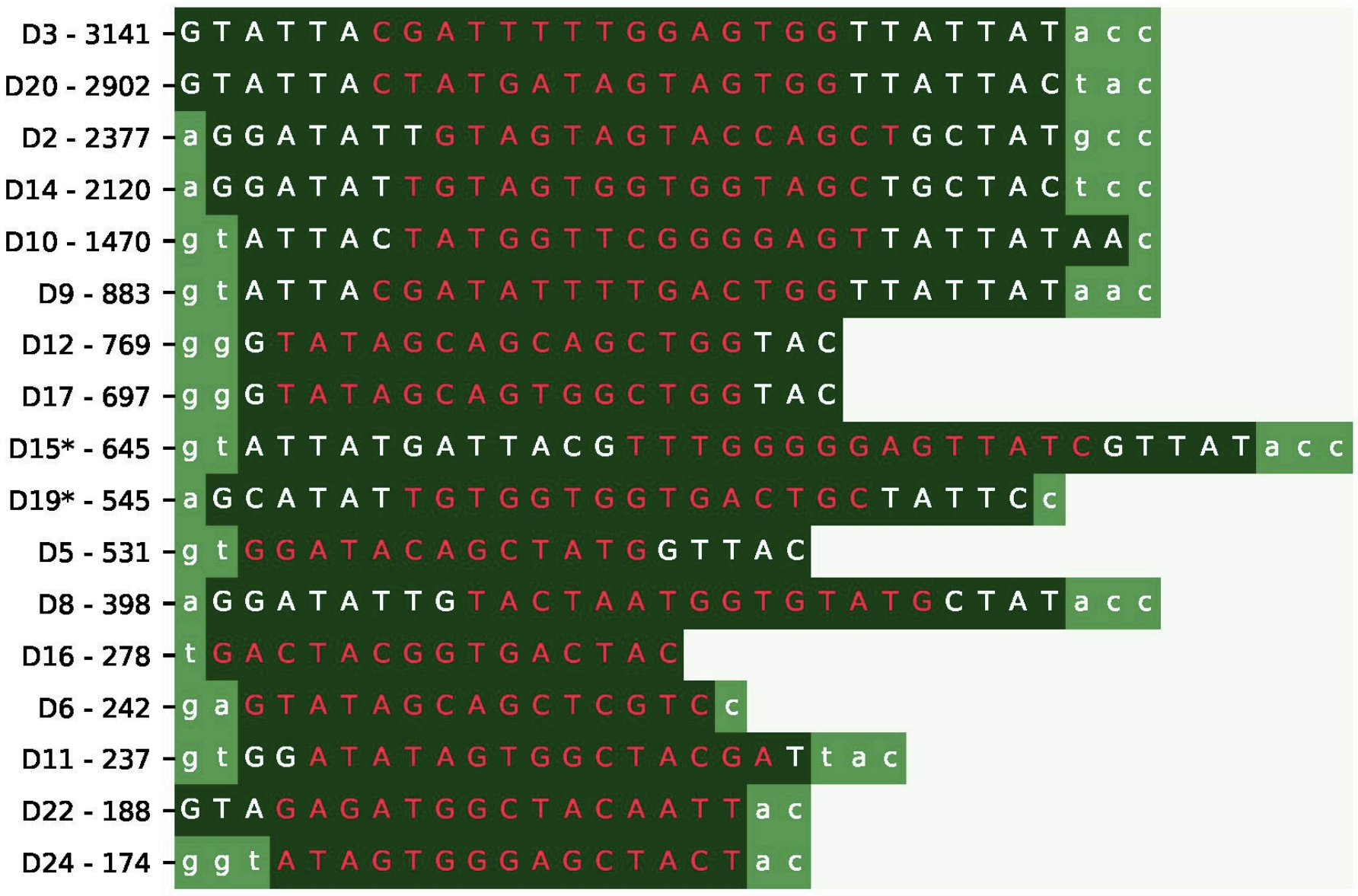

**Figure 2. IgScout results on the *CDR3$_{trimmed}$* dataset.** Each row shows a reconstructed string (strings are inferred in the order from the top to the bottom). Dark green segments correspond to reconstructed substrings of human D genes (flanking non-reconstructed nucleotides are shown in standard green). The most frequent 15-mers that were used for reconstructing the corresponding D genes are shown in red (their abundances are shown on the left). The reconstructed substring of the D2 gene (IGHD2-2) also occurs in D2* and D2** genes. 17 strings reconstructed by IgScout represent substrings of 17 human D genes. IgScout misses short prefixes and suffixes of D genes: 1.4 nucleotides on the left and 1.7 nucleotides on the right, on average for the Set 1 dataset (0.9 nucleotides on the left and 1.5 nucleotides on the right, on average after combining reconstructions over all HEALTHY datasets). IgScout did not reconstruct eight human D genes: D1 (IGHD1-1), D4 (IGHD4-4), D7 (IGHD1-7), D13 (IGHD1-13), D18 (IGHD1-20), D21 (IGHD4-23), D23 (IGHD6-25), and D25 (IGHD7-27) that contributed to few CDR3 in the Set 1. These genes have the following abundances of their most frequent 15-mers: 43 for D1, 59 for D4, 83 for D7, 0 for D13, 33 for D18, 75 for D21, 0 for D23, and 0 for D25.

**Reconstruction of human D genes.** Figure 3 illustrates that IgScout reconstructed 18 out of 25 human D genes across all HEALTHY datasets, Supplemental Note "Summary of IgScout results across diverse immunosequencing datasets" describes inference of 20 human D genes across multiple immunosequencing datasets. Supplemental Note "Reconstructing variants of human D genes" describes inference of 5 allelic variants of the D7, D10, D15, D16, and D21 genes, However, since variations in D7, D16, and D21 genes affect the first or last nucleotides of the corresponding D genes, they likely represent computational artifacts caused by abundant nucleotides at the flanking positions of the D genes within CDR3s. In contrast, variations of the D10 and D15 genes (referred to as $D10^{+}$ and $D15^{+}$, respectively) have mutations in the middle of D genes (Figure 3). Thus, since they were inferred from multiple datasets (Set 5 and Set 7 for $D10^{+}$, and Set 5, Set 7, Set 9, and Set 13 for $D15^{+}$), they likely represent previously unknown allelic variants. Supplemental Note "Reconstructing variants of human D genes" illustrates that 50 (46) samples among 600 samples in the PROJECTS10 dataset support $D10^{+}$ ($D15^{+}$) variants and presents two more variants $D10^{++}$ and $D15^{++}$.

To demonstrate that D10+ and D15+ indeed represent new variants of D10 and D15 genes, we analyzed 40 whole genome sequencing datasets from the population-wide study of esophageal cancer (PRJNA427604 project) and searched for exact occurrences of D10+ and D15+ in reads. Both variations were detected in 5 out of 40 datasets (SRR6435661, SRR6435676, SRR6435686, SRR6435691, and SRR6435692) with the number of reads supporting D10+ (D15+) varying from 8

to 14 (30 to 58) across these five datasets. See Supplemental Note "How IgScout results are affected by the number of consensus CDR3s and cell types".

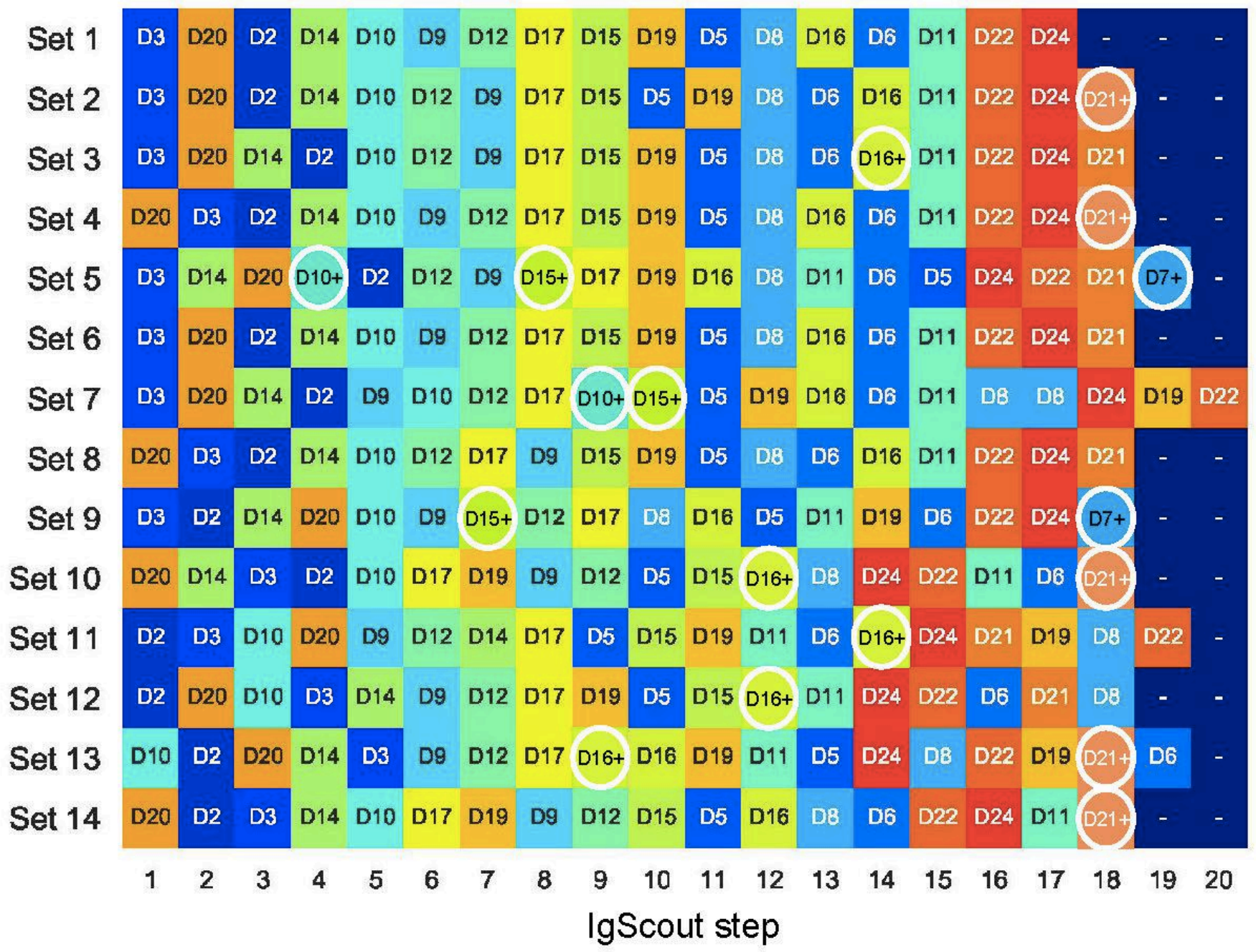

```
         D10+  GTATTACTATGGTTCAGGGAGTTATTATAAC
IGHD3-10 D10   GTATTACTATGGTTCGGGGAGTTATTATAAC
         D10*  GTATTACTATGTT-CGGGGAGTTATTATAAC

         D15+  GTATTATGATTACATTTGGGGGAGTTATCGTTATACC
IGHD3-16 D15   GTATTATGATTACGTTTGGGGGAGTTATGCTTATACC
         D15*  GTATTATGATTACGTTTGGGGGAGTTATCGTTATACC
```

**Figure 3. *De novo* reconstructions of D genes across all HEALTHY datasets (top) and allelic variants D10⁺ and D15⁺ inferred by IgScout (bottom).** (Top) Genes reconstructed at consecutive steps of IgScout for all HEALTHY datasets. Rows correspond to the datasets and columns correspond to the IgScout steps. Each cell is marked by a reconstructed D gene (each D gene is assigned a unique color). Cells marked with the "+" sign refer to strings that differ from known D genes by and correspond to putative novel variants (shown within a circle). (Bottom) Allelic variants $D10^{+}$ and $D15^{+}$ inferred by IgScout. Differences from human D genes and their allelic variants listed in the IMGT database are shown in red.

**Reconstruction of camel D genes.** Although camel V genes were inferred in Conrath et al., 2003, camel D genes remain unknown. We analyzed six CAMEL datasets from three camels (VH and VHH libraries for each camel) labeled as Camel 1VH, 1VHH, 2VH, 2VHH, 3VH, and 3VHH (Li et al., 2016). While the VH libraries contain the heavy chain of the conventional (both heavy and light chain) camel antibodies, the VHH libraries contain the heavy chains of the single-chain *nano antibodies* (Leslie, 2018).

We extracted camel CDR3s by aligning camel antibody repertoires against the known camel V and J genes using the IgReC tool (Shlemov et al., 2017). For the Camel 1VH dataset, IgScout constructed 60,066 consensus CDR3 sequences of average length 48 nucleotides. The $CDR3_{trimmed}$ dataset for Camel 1VH has total length 1,400,360 nucleotides (the average length 23 nt).

IgScout reconstructed four D genes in the case of the Camel 1VH dataset that we refer to as D1, D2, D3, and D4 (see Supplemental Note "Reconstructing camel D genes"). It reconstructed four putative D genes in datasets Camel 1VHH, and Camel 2VH, and three putative D genes in the remaining three camel datasets (17 strings in total) that are largely consistent with genes D1, D2, D3, and D4 derived from the Camel 1VH dataset (previous studies assumed that the camel genome has a single germline D gene (Conrath et al., 2003). Supplemental Note "Reconstructing camel D genes" illustrates that all camel D genes are shared between the VH and VHH datasets. Supplemental Note "Usage of camel D genes" demonstrates that the camel D genes have strikingly different usage in the VH and VHH antibodies.

**D gene usage.** 25 human D genes form a set of strings that we refer to as *D-Genes*. Given an arbitrary string *Target*, a string *D* from *D-Genes,* and a parameter *k*, we say that a string *Target* is *formed* by *D* if it contains a *k*-mer from *D* but does not contain *k*-mers from other strings in *D-Genes* (the default value *k*=11). We classify a CDR3 as *traceable* if it is formed by a D gene and *non-traceable*, otherwise. The percentage of traceable CDR3s is rather conservative across all HEALTHY datasets: ≈60% of CDR3s in the HEALTHY datasets are traceable (Supplemental Note "Traceable CDR3s").

Given a set of strings *Strings* and a string *D* from *D-Genes*, we define *usage*(*Strings*, *D-Genes*, *D*) as the fraction of traceable strings in *Strings* formed by the string *D*. We are interested in *usage*($CDR3_{trimmed}$, *D – Genes*, *D*) for each human D gene. Supplemental Note "Traceable CDR3s" analyzes the usage of all human D genes across all HEALTHY datasets. Supplemental Note "D gene classification by IgScout and IgBlast" compares IgScout and IgBlast classification of D genes forming CDR3s.

We analyzed the usage of known and novel allelic variants ($D10^+$ and $D15^+$) across all HEALTHY datasets. Figure 4 reveals that usage of allelic variants of D2 and D3 is consistent across all datasets with D2* and D3 as dominant variants. However, the Set 5 has different dominant variants as compared to other datasets: D8* (compared to D8 in all other datasets); $D10^+$ (compared to D10 in all other datasets); and D19 (compared to D19* in all other datasets). The variant $D15^+$ is dominant in Sets 5, 7, 9, and 13, while the D15 gene is dominant in the remaining eight datasets.

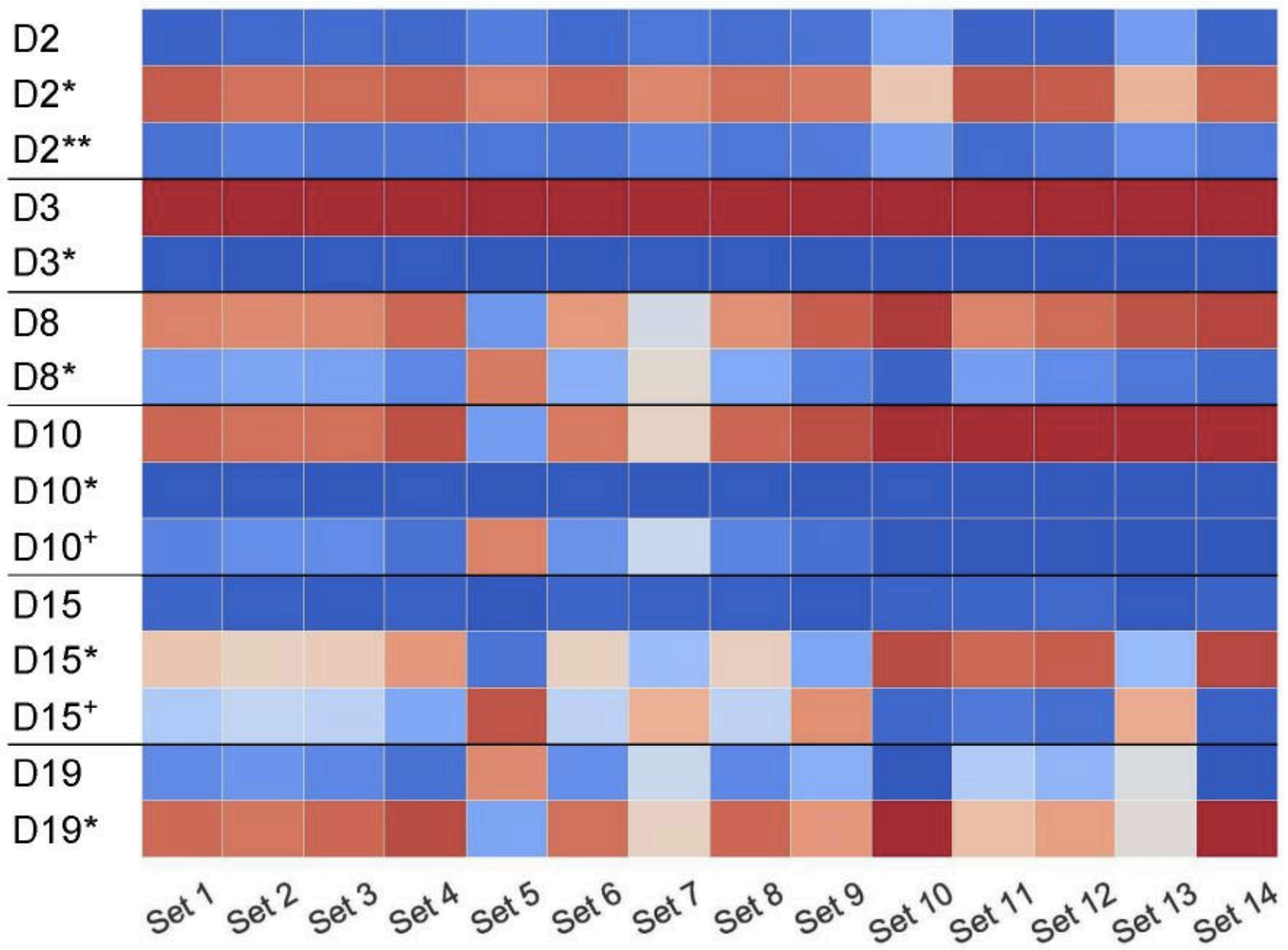

**Figure 4. Usage of D genes with known and novel allelic variants across all HEALTHY datasets.** Horizontal black lines sub-partition the matrix into six sub-matrices corresponding to allelic variants of D2 (IGHD2-2), D3 (IGHD3-3), D8 (IGHD2-8), D10 (IGHD3-10), D15 (IGHD3-16), and D18 (IGHD1-20). For each D gene and each dataset, we computed the percentage of usage of each variant. Values in cells vary from 0 (blue) to 100 (red). White cells correspond to values ≈50% and likely represent cases when a single individual carries different variants of a given D genes on two different chromosomes.

**Tandem CDR3s.** Given strings *D* and *D',* and a parameter *k*, we say that a string *Target* is *formed* by *D* and *D'* if it contains *k*-mers from both *D* and *D'* and a *k*-mers from *D'* starts after a *k*-mer from *D* ends. Since tandem CDR3s represent a small fraction of all CDR3s, we set the default value *k*=11 (rather than *k*=15 for all CDR3s) to increase the number of identified tandem CDR3s. Although a smaller value of *k* may lead to identification of *pseudo-tandem* CDR3s, the Methods section describes how to filter out such pseudo-random CDR3s.

There exists 187 *tandem CDR3s* formed by two D genes in the $CDR3_{trimmed}$ dataset (Figure 5). We denote the longest substring between a tandem CDR3 *Target* and *D* (*Target* and *D'*) as $D_{match}$ ($D'_{match}$) and represent a tandem CDR3 *Target* as a concatenate of five strings *prefix* $*$ $D_{match}$ $*$ *middle* $*$ $D'_{match}$ $*$ *suffix*. We define the *span* of a tandem CDR3 formed by $D$ and $D'$ as the substring $D_{match}$ $*$ *middle* $*$ $D'_{match}$ and *inter-D insertion* as the substring *middle* (Figure 5).

Briney et al., 2012 emphasized that detecting tandem CDR3s has to be done with caution since they are often confused with *pseudo-tandem* CDR3s formed by the standard V(D)J recombination (Figure 5). The Methods section describes how IgScout detects pseudo-tandem CDR3s.

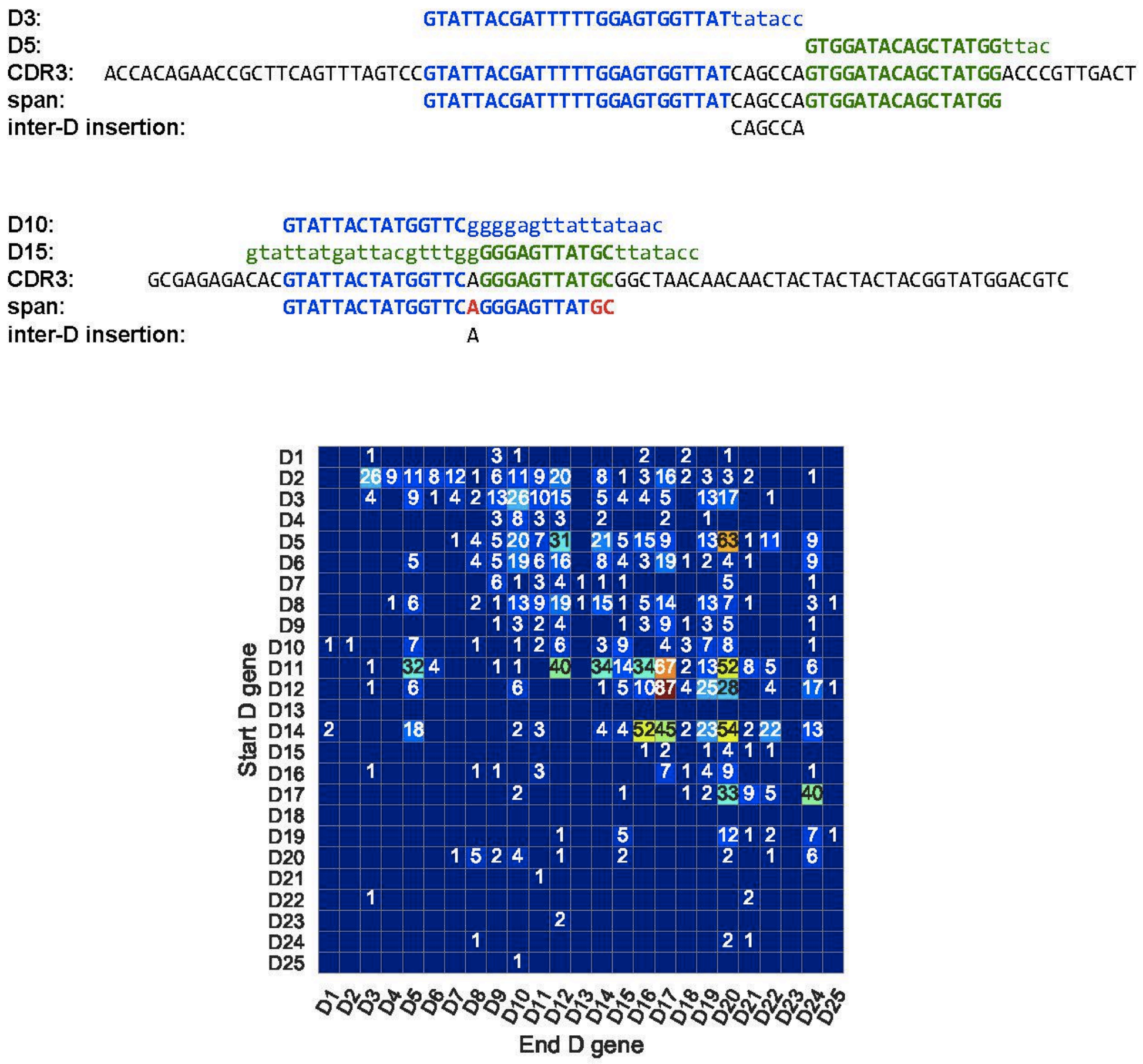

**Figure 5. A tandem CDR3 formed by genes D3 (IGHD3-3) and D5 (IGHD5-5) (top), a pseudo-tandem CDR3 formed by genes D10 and D15 (middle), and the tandem matrix for all tandem CDR3s across all HEALTHY datasets (bottom).** (Top) A tandem CDR3 with $D_{match}$=GTATTAGGATTTTTGGAGTGGTTAT, *middle*=CAGCCA, and $D'_{match}$=GTGGATACAGCTATGG. (Middle) The pseudo-tandem CDR3, formed by genes D10 (IGHD3-10) and D15 (IGHD3-16). This CDR3 was formed by a single gene D10 (IGHD3-10) with three mutations (shown in red). IgScout filters out most pseudo-tandem CDR3s. (Bottom) The number in a cell $(i,j)$ shows the total number of tandem CDR3s formed by genes $D_i$ and $D_j$ across all HEALTHY datasets. Empty cells correspond to pairs of D genes that do not form tandem CDR3s. Genes D4 and D5 appear in two copies in the IGH loci. The second copy of D4 (IGHD4-11) appears between D10 (IGHD3-10) and D11 (IGHD5-12). The second copy of D5 (IGHD5-18) appears between D16 (IGHD4-17) and D17 (IGHD6-19). The vast majority of tandem CDR3 correspond to cells in the upper half of the matrix. The only populated column in the lower part of the tandem matrix corresponds to the D5 gene and likely results from tandem CDR3s formed by the second copy of D5 in the IGH locus.

**Tandem bias**. There exists 114 tandem CDR3s in the Set 1 dataset and 1900 tandem CDR3s across all HEALTHY datasets. Figure 5 represents all tandem CDR3s as a *tandem matrix* and reveals that the vast majority of them correspond to cells in the upper half of this matrix. If tandem CDR3s were computational artifacts, we would expect similar numbers of CDR3s in the upper and lower parts of the tandem matrix. We define the *tandem bias* as $N_{upper}$ / ($N_{upper}$ + $N_{lower}$), where $N_{upper}$ and $N_{lower}$ is the sum of entries in the upper and lower parts of the tandem matrix, respectively (we assume that the main diagonal belongs to the lower part of the matrix). The tandem bias varies from 0.03 % to 0.21% across various datasets (see Supplemental Note: "Analysis of tandem CDR3s).

Since most pairs of D genes in tandem CDR3s contribute to the upper part of the tandem matrix (and thus follow the order of D genes in the IGH locus), some entries in the lower part of the tandem matrix may reveal possible duplications of D genes, e.g., the D20 row in the lower part of the tandem matrix in Figure 5 reveals many tandem CDR3s. Analysis of the hepatitis patient 1776 in the PROJECTS10 dataset (Galson et al., 2016) revealed particularly many entries in the D20 column in the lower part of the tandem matrix, suggesting a duplication of the D20 gene in this patient (see Supplemental Note "Analysis of tandem CDR3s"). Kidd et al., 2016 analyzed biases in the D-J pairing and also suggested that D20 may be duplicated in some individuals.

**Ultra-long tandem CDR3s reveal unusual recombination events.** 1900 tandem CDR3s across all HEALTHY datasets contain 1081 distinct inter-D insertions, varying in length from 0 to 153 nucleotides. The two longest inter-D insertions (denoted $I_1$ and $I_2$) appear in the Set 1 and have length 153 nucleotides. They are formed by genes D9 and D10, differ by a single nucleotide, and appear in CDR3s differing by 6 nucleotides. Surprisingly, the inter-D insertion $I_2$ coincides with the sequence of the IGH locus between the D9 and D10 genes. Germline D genes are flanked by 12-nucleotide long *recombination signal sequences* (*RSSs*) and the inter-D insertion $I_2$ starts with the right RSS of D9 and ends with the left RSS of D10 (Supplemental Note "Ultra-long tandem CDR3s").

Thus, ultra-long tandem CDR3s reveal unusual *RSS skipping* events during somatic recombination: skipping the right RSS of D9 and left RSS of D10 led to a tandem CDR3 representing a concatenate D9 + $I_2$ + D10. The existing immunosequencing protocols are likely to miss ultra-long immunoglobulins since they are not designed to capture the abnormally long variable regions (exceeding ~400 nt). We captured reads containing ultra-long tandem CDR3s because the 300-nucleotide long paired reads (overlapping by only 50 nucleotides) in the Set 1 are longer than reads used in most other immunosequencing datasets. Thus, even if ultra-long tandem CDR3s were common, they would likely remain below the radar of most immunosequencing studies.

**Tandem CDR3s contribute to adaptive immune response.** We investigated whether tandem CDR3s contribute to the adaptive immune response by analyzing their *isotypes*. Since IgG, IgA, and IgE isotypes occur in plasma and memory B cells subjected to the antibody-antigen interactions, these isotypes they indicate (in difference from IgM isotypes common in memory and naïve B cells) that the corresponding antibodies participate in the adaptive immune response.

We inferred isotypes in the ALLERGY and HIV datasets using markers described in Levin et al., 2017 (Figure 6), the vast majority of tandem CDR3s from the ALLERGY dataset correspond to the IgM isotype and thus are produced by memory and naïve B cells. In contrast, ~60% of tandem CDR3s in the HIV dataset correspond to the IgG type. This observation suggests that tandem CDR3s in the HIV-infected patients arise from immunoglobulins that are produced by plasma cells and thus might contribute to the immune response against HIV antigens.

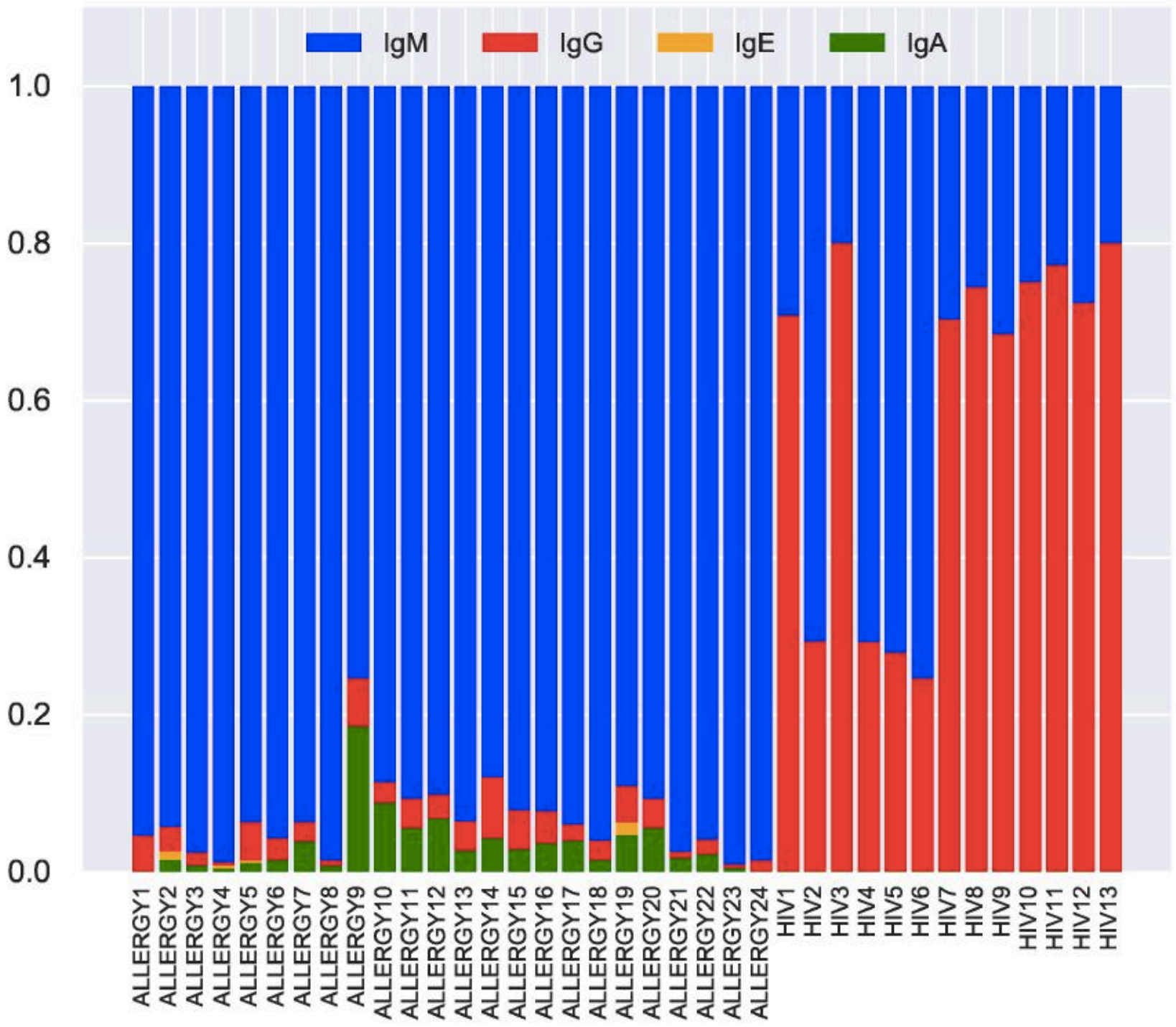

**Figure 6. Fractions of IgM, IgG, IgE, and IgA isotypes representing tandem CDR3s in the repertoires from the ALLERGY and HIV datasets.**

## DISCUSSION

Since many human germline alleles remain unknown (particularly for non-European subjects), missing alleles may mislead clinical decisions (Xochelli et al., 2015) and lead to erroneous derivation of clonal lineages due to misinterpretations of SHMs. Thus, finding new germline alleles and building personalized sets of germline genes for each individual is important for downstream analysis of immunosequencing datasets.

Although there exists a number of tools for inferring V and J genes (Corcoran et al., 2016, Zhang et al., 2016, Ralph and Matsen, 2017), a more difficult problem of reconstructing D genes remains open. IgScout aims to reconstruct all D genes explaining a large percentage of the VDJ recombination in an antibody repertoire rather than to reconstruct all D genes. The IMGT database reflects the *genomic diversity* of D genes but not their *recombinant diversity* (information about rearrangements, transcription, and translation of D genes). Since assemblies of the highly repetitive IGH loci are fragmented and error-prone (Matsuda et al., 1998; Watson et al., 2012; Watson et al., 2013. Luo et al., 2017), reconstruction of all germline genes from the whole-genome sequencing data is a difficult problem. Although the IGH locus is extremely diverse (Watson et al., 2017), it remains largely unknown how it varies across the human population Moreover, even in the case when the IGH locus is correctly assembled, prediction of the functional germline genes is a non-trivial problem (Wang et al., 2008; Collins et al., 2015).

Immunosequencing datasets reflect the recombinant diversity of antibody repertoires and thus complement the genomic datasets. If some D genes do not contribute to the VDJ recombination (e.g., our analysis suggests that genes D1, D13, D18, D23, and D25 do no significantly contribute to VDJ recombination in any of the analyzed datasets), they have limited contribution to immune response. In this paper, we focused on reconstructing D genes shaping the recombinant diversity rather than all D genes.

IgScout reconstructed 20 out of 25 human D genes across multiple datasets and missed genes D1, D13, D18, D23, D25 that form a small number of CDR3s (less than 0.1% each) across all analyzed datasets. It remains unclear whether some of these genes ever contribute to any CDR3s, for example genes D13 and D23 do not form any CDR3s in most datasets (few CDR3s formed by these D genes in some datasets may represent computational artifacts).

IgScout revealed four new allelic variants ($D10^{+}$, $D10^{++}$, $D15^{+}$, and $D15^{++}$), thus increasing the number of known variants of human D genes from seven to eleven. These new variants are unlikely to be computational artifacts since they were found in dozens immunosequencing datasets from distinct individuals and many whole genome sequencing datasets. The frequency of the already known Single Nucleotide Polymorphisms (SNPs) in D genes exceeds the frequency of SNPs in the entire human genome by two orders of magnitude (12 SNPs for all D genes of total length only 288 nucleotides). One possible explanation is the selection pressure for variants of D genes forming CDR3s that are particularly effective against emerging pathogens.

Although IgScout revealed four novel variants of human D genes and inferred camel D genes, these genes will not be included in the IMGT database since they haven't been experimentally confirmed yet. Similarly to Gadala-Maria et al., 2015, we argue that, like in other areas of genomics, the time has come to add such prediction to the IMGT database. For example, the lion's share of genes in genomic databases represent computational predictions that haven never been experimentally confirmed. We argue that IMGT should classify alleles with varying levels of supporting evidence, not unlike classification systems used in other biological databases. Since the number of computationally predicted but not yet experimental confirmed germline genes has been steadily increasing, lack of depository for such predictions makes it difficult to analyze diverse immunosequencing datasets.

Although IgScout is not specifically designed for reconstructing V and J genes, it turned out that its applications are not limited to reconstructing D genes (see Supplemental Note "De novo reconstruction of human J genes"). In addition to de novo reconstruction of D genes, it also detects tandem CDR3s. Briney et al., 2012 postulated that tandem CDR3s mostly appear in naïve B cells and thus do not contribute to adaptive immune response. In contrast, our analysis revealed that tandem CDR3s contribute to adaptive immune response since ~80% of them in the HIV dataset have been contributing to antigen binding.

## METHODS

**Inferring germline genes as the Trace Reconstruction Problem.** In information theory, a string *S* yields a collection of *traces*, where each trace is independently obtained from *S* by substituting each symbol in *S* by another symbol from a fixed alphabet with a given probability $\delta$. Given the traces and the value $\delta$, the *Trace Reconstruction Problem* (Mitzenmacher, 2009) is to reconstruct the original string *S*. *De novo* reconstruction of D genes results in a more complex version of the Trace Reconstruction Problem where traces are generated by multiple strings and each trace is obtained from one of these strings by (i) randomly trimming it from both sides, (ii) adding a randomly generated prefix in the front of the string, and (iii) adding a randomly generated suffix in the end of the string. Given a set of such traces (modeled by a set of trimmed CDR3s extracted from an immunosequencing dataset), the goal is to reconstruct the original set of strings.

**Extending the consensus string.** IgScout trims all positions of the motif logo with the information content below *IC* and computes the consensus string. Afterwards, it extracts the first *k*-mer of the consensus string and finds all CDR3s that contain this *k*-mer. If the position preceding the first *k*-mer in these reads has information content exceeding a threshold, IgScout adds the most frequent nucleotide at this position to the consensus and iterates. Afterwards, it applies a similar procedure to the last *k*-mer of the consensus string. The resulting extended consensus is reported as a putative D gene (Figure 1).

**Detecting pseudo-tandem CDR3s.** Given strings *Span* and *S,* we define $distance_t(Span, Target)$ as the minimum Hamming distance between *t*-mers in *Span* and *S*. Given a parameter $\Delta$ (the default value $\Delta$=5) we define the $\Delta$-*distance* between strings *Span* and *Target* as $distance_t(S, Target)$ for $t=|Span|-\Delta$, where $|Span|$ stands for the length of the string *Span*. Finally, we define the *Δ-distance* between a string *Span* and a set of strings *Strings* as the minimum $\Delta$-distance between *Span* and all strings in *Strings*.

We computed the $\Delta$-distance between the spans of all 187 identified tandem CDR3s in $CDR3_{trimmed}$ and all string in *D-Genes*. 73 out of these 187 CDR3s can be explained as CDR3s originating from a single D gene (for the $\Delta$-distance threshold 3). However, the remaining 114 CDR3s have $\Delta$-distance at least 9. We thus classify a CDR3 sequence *Target* formed by genes *D* and *D'* as pseudo-tandem if the Δ-distance between the span of this pseudo-tandem CDR3 and *D-Genes* does not exceed a predefined threshold (the default value is 3), and (truly) tandem, otherwise. See Supplementary Note "List of tandem CDR3s."

**Availability of data and materials.** IgScout is available at https://github.com/Immunotools.

**Competing interests.** The authors declare no competing financial interests.

**Funding.** Y.S. was supported by the Data Science Fellowships at UCSD. The work of P.A.P. was supported by the NIH 2-P41-GM103484PP grant.

**Author contributions.** Y.S. implemented the IgScout algorithm and performed benchmarking. Y.S. and P.A.P. conceived the study, developed the IgScout algorithm, designed the computational experiments, and wrote the manuscript.

**Acknowledgements.** Authors are grateful to Dmitry Chudakov for providing us with the datasets Set 1 – Set 9.

## REFERENCES

1. Avnir Y, Watson CT, Glanville J, Peterson EC, Tallarico AS, Bennett AS, Qin K, Fu Y, Huang CY, Beigel JH, Breden F, Zhu Q, Marasco WA. IGHV1-69 polymorphism modulates anti-influenza antibody repertoires, correlates with IGHV utilization shifts and varies by ethnicity. Scientific Rep. 2016; 6: 20842.
2. Bandeira N, Pham V, Pevzner P, Arnott D, Lill JR. Automated de novo protein sequencing of monoclonal antibodies. Nat Biotech. 2008; 26(12): 1336-8.
3. Bolotin DA, Poslavsky S, Mitrophanov I, Shugay M, Mamedov IZ, Putintseva EV, Chudakov DM. MiXCR: software for comprehensive adaptive immunity profiling. Nat Methods. 2015; 12(5): 380-381.
4. Boyd SD, Gaëta BA, Jackson KJ, Fire AZ, Marshall EL, Merker JD, Maniar JM, Zhang LN, Sahaf B, Jones CD, Simen BB, Hanczaruk B, Nguyen KD, Nadeau KC, Egholm M, Miklos DB, Zehnder JL, Collins AM. Individual variation in the germline Ig gene repertoire inferred from variable region gene rearrangements. J Immunol. 2010; 184(12): 6986-92.
5. Boyd SD, Liu Y, Wang C, Martin V, Dunn-Walters DK. 2013. Human lymphocyte repertoires in ageing. Curr Opin Immunol 25(4): 511–515.
6. Briney BS, Willis JR, Hicar MD, Thomas JW, Crowe JE. Frequency and genetic characterization of V(DD)J recombinants in the human peripheral blood antibody repertoire. Immunology. 2012; 137(1): 56–64.
7. Brochet X, Lefranc MP, Giudicelli V. IMGT/V-QUEST: the highly customized and integrated system for IG and TR standardized V-J and V-D-J sequence analysis. Nucleic Acids Res. 2008; 36(Web Server issue): W503-8.
8. Castellana NE, Pham V, Arnott D, Lill JR, Bafna V. Template Proteogenomics: Sequencing Whole Proteins Using an Imperfect Database. Mol Cell Proteomics. 2010; 9: 1260-70.
9. Chang CJ, Chen CH, Chen BM, Su YC, Chen YT, Hershfield M, Lee MTM, Cheng TL, Chen YT, Roffler SR, Wu JY. A genome-wide association study identifies a novel susceptibility locus for the immunogenicity of polyethylene glycol. Nat Comm. 2012; 8: 522.
10. Collins AM, Wang Y, Roskin KM, Marquis CP, Jackson KJ. The mouse antibody heavy chain repertoire is germline-focused and highly variable between inbred strains. Philos Trans R Soc Lond Ser B Biol Sci. 2010; 370: 20140236.
11. Conrath K, Wernery U, Muyldermans S, Nguyen V. Emergence and evolution of functional heavy-chain antibodies in Camelidae. Dev Comp Immunol. 2003; 27(2): 87–103.
12. Corbett SJ, Tomlinson IM, Sonnhammer EL, Buck D, Winter G. Sequence of the human immunoglobulin diversity (D) segment locus: a systematic analysis provides no evidence for the use of DIR segments, inverted D segments, "minor" D segments or D-D recombination. J Mol Biol. 1997; 270(4): 587-97.
13. Corcoran MM, Phad GE, Vázquez Bernat N, Stahl-Hennig C, Sumida N, Persson MA, Martin M, Karlsson Hedestam GB. Production of individualized V gene databases reveals high levels of immunoglobulin genetic diversity. Nat Commun. 2016;7:13642.
14. Crooks GE, Hon G, Chandonia J-M, Brenner SE. WebLogo: A Sequence Logo Generator. Genome Research. 2004;14(6):1188-90.
15. Cui A, Di Niro R, Vander Heiden JA, Briggs AW, Adams K, Gilbert T, O'Connor KC, Vigneault F,

Shlomchik MJ, Kleinstein SH. A Model of Somatic Hypermutation Targeting in Mice Based on High-Throughput Ig Sequencing Data. J Immunol. 2016;197(9):3566–74.
16. Elhanati Y, Sethna Z, Marcou Q, Callan CG Jr, Mora T, Walczak AM. Inferring processes underlying B-cell repertoire diversity. Philos Trans R Soc Lond B Biol Sci. 2015;370:1676.
17. Gadala-Maria D, Yaari G, Uduman M, Kleinstein SH. Automated analysis of high-throughput B-cell sequencing data reveals a high frequency of novel immunoglobulin V gene segment alleles. PNAS USA. 2015; 112(8):E862-70.
18. Galson JD, Trück J, Clutterbuck EA, Fowler A, Cerundolo V, Pollard AJ, Lunter G, Kelly DF. B-cell repertoire dynamics after sequential hepatitis B vaccination and evidence for cross-reactive B-cell activation. Genome Medicine. 2016;8:68.
19. Ellebedy AH, Jackson KJ, Kissick HT, Nakaya HI, Davis CW, Roskin KM, McElroy AK, Oshansky CM, Elbein R, Thomas S, Lyon GM, Spiropoulou CF, Mehta AK, Thomas PG, Boyd SD, Ahmed R. Defining antigen-specific plasmablast and memory B cell subsets in blood following viral infection and vaccination of humans. Nature Immunol. 2016;17(10):1226-34.
20. Kidd MJ, Chen Z, Wang Y, Jackson KJ, Zhang L, Boyd SD, Fire AZ, Tanaka MM, Gaëta BA, Collins AM. The inference of phased haplotypes for the immunoglobulin H chain V region gene loci by analysis of VDJ gene rearrangements. J Immunol. 2012;188(3):1333-40.
21. Kidd MJ, Jackson KJ, Boyd SD, Collins AM. DJ Pairing during VDJ Recombination Shows Positional Biases That Vary among Individuals with Differing IGHD Locus Immunogenotypes. J Immunol. 2016;196(3):1158-64.
22. Kirik U, Greiff L, Levander F, Ohlin M. Data on haplotype-supported immunoglobulin germline gene inference. Data Brief. 2017;13:620-40.
23. Kurosawa Y, Tonegawa S. Organization, structure, and assembly of immunoglobulin heavy chain diversity DNA segments. J Exp Med. 1982;155(1):201-8.
24. Landais E, Murrell B, Briney B, Murrell S, Rantalainen K, Berndsen ZT, Ramos A, Wickramasinghe L, Smith ML, Eren K, de Val N, Wu M, Cappelletti A, Umotoy J, Lie Y, Wrin T, Algate P, Chan-Hui PY, Karita E; IAVI Protocol C Investigators; IAVI African HIV Research Network, Ward AB, Wilson IA, Burton DR, Smith D, Pond SLK, Poignard P. HIV Envelope Glycoform Heterogeneity and Localized Diversity Govern the Initiation and Maturation of a V2 Apex Broadly Neutralizing Antibody Lineage. Immunity. 2017; 47(5):990-1003.e9.
25. Larimore K, McCormick MW, Robins HS, Greenberg PD. Shaping of human germline IgH repertoires revealed by deep sequencing. J Immunol. 2012;189(6):3221-30.
26. Leslie M., Mini-antibodies discovered in sharks and camels could lead to drugs for cancer and other diseases. Science. 2018;360.
27. Levin M, Levander F, Palmason R, Greiff L, Ohlin M. Antibody-encoding repertoires of bone marrow and peripheral blood-a focus on IgE. J Allergy Clin Immunol. 2017;139(3):1026-30.
28. Li X, Duan X, Yang K, Zhang W, Zhang C, Fu L, Ren Z, Wang C, Wu J, Lu R, Ye Y, He M, Nie C, Yang N, Wang J, Yang H, Liu X, Tan W. Comparative Analysis of Immune Repertoires between Bactrian Camel's Conventional and Heavy-Chain Antibodies. PLoS One. 2016;11(9):e0161801.
29. Luo S, Yu JA, Li H, Song YS. Worldwide genetic variation of the IGHV and TRBV immune receptor gene families in humans. bioRxiv. 2017.
30. Matsuda F, Ishii K, Bourvagnet P, Kuma Ki, Hayashida H, Miyata T, Honjo T. The complete nucleotide sequence of the human immunoglobulin heavy chain variable region locus. J Exp Med. 1998;188(11): 2151-62.
31. McCoy CO, Bedford T, Minin VN, Bradley P, Robins H, Matsen IV FA. Quantifying evolutionary constraints on B-cell affinity maturation. Philos Trans R Soc Lond B Biol Sci. 2015;370(1676): 20140244.
32. Mitzenmacher M. A survey of results for deletion channels and related synchronization channels. Probability Surveys. 2009;6:1–33.
33. Meek KD, Hasemann CA, and Carpa DJ. Novel rearrangements at the immunoglobulin D locus. Inversions and fusions add to IgH somatic diversity. J Exp Med. 1989;170(1):39-57.
34. Murugan A, Mora T, Walczak AM, Callan CG Jr. Statistical inference of the generation probability of T-cell receptors from sequence repertoires. Proc Natl Acad Sci USA. 2012;109(40):16161-6.
35. Parameswaran P, Liu Y, Roskin KM, Jackson KK, Dixit VP, Lee JY, Artiles KL, Zompi S, Vargas MJ, Simen BB, Hanczaruk B, McGowan KR, Tariq MA, Pourmand N, Koller D, Balmaseda A, Boyd SD, Harris E, Fire AZ. Convergent antibody signatures in human dengue. Cell Host Microbe. 2013;13(6):691-700.
36. Price AL, Jones NC, Pevzner PA. De novo identification of repeat families in large genomes. Bioinformatics. 2005;1:i351-8.
37. Ralph DK, Matsen FA 4th. Consistency of VDJ Rearrangement and Substitution Parameters Enables Accurate B Cell Receptor Sequence Annotation. PLoS Comput Biol. 2016;12(1):e1004409.
38. Rubelt F, Bolen CR, McGuire HM, Vander Heiden JA, Gadala-Maria D, Levin M, Euskirchen GM, Mamedov MR, Swan GE, Dekker CL, Cowell LG, Kleinstein SH, Davis MM. Individual heritable differences result in unique cell lymphocyte receptor repertoires of naïve and antigen-experienced cells. Nat Commun 2016;7:11112.
39. Shlemov A, Bankevich S, Bzikadze A, Turchaninova MA, Safonova Y, Pevzner PA. Reconstructing Antibody Repertoires from Error-Prone Immunosequencing Reads. J Immunol. 2017;199(9):3369-80.
40. Souto-Carneiro MM, Sims GP, Girschik H, Lee J, Lipsky PE. Developmental changes in the human heavy chain CDR3. J Immunol. 2005;175(11):7425-36.
41. Turchaninova MA, Davydov A, Britanova OV, Shugay M, Bikos V, Egorov ES, Kirgizova VI, Merzlyak EM, Staroverov DB, Bolotin DA, Mamedov IZ, Izraelson M, Logacheva MD, Kladova O, Plevova K, Pospisilova S, Chudakov DM. High-quality full-length immunoglobulin profiling with unique molecular barcoding. Nat Protocols. 2016;11(9):1599-616.
42. Wang Y, Jackson KJL, Sewell WA, Collins AM. Many human immunoglobulin heavy-chain IGHV gene polymorphisms have been reported in error. Immunol Cell Biol. 2008;86(2):111–5.
43. Watson LC, Moffatt-Blue CS, McDonald RZ, Kompfner E, Ait-Azzouzene D, Nemazee D, Theofilopoulos AN, Kono DH, Feeney AJ. Paucity of V-D-D-J rearrangements and VH replacement events in lupus prone and nonautoimmune TdT-/- and TdT+/+ mice. J Immunol. 2006;177:1120–8.
44. Wang Y, Jackson KJ, Sewell WA, Collins AM. Many human immunoglobulin heavy-chain IGHV gene polymorphisms have been reported in error. Immunol Cell Biol. 2008;86(2):111-5.
45. Watson CT and Breden F. The immunoglobulin heavy chain locus: Genetic variation, missing data, and implications for human disease. Genes Immun. 2012;13(5):363–73.
46. Watson CT, Matsen FA 4th, Jackson KJL, Bashir A, Smith ML, Glanville J, Breden F, Kleinstein SH, Collins AM, Busse CE. Comment on "A Database of Human Immune Receptor Alleles Recovered from Population Sequencing Data". J Immunol. 2017;198(9):3371–3.
47. Xochelli A, Agathangelidis A, Kavakiotis I, Minga E, Sutton LA, Baliakas P, Chouvarda I, Giudicelli V, Vlahavas I, Maglaveras N, Bonello L, Trentin L, Tedeschi A, Panagiotidis P, Geisler C, Langerak AW, Pospisilova S, Jelinek DF, Oscier D, Chiorazzi N, Darzentas N, Davi F, Ghia P, Rosenquist R, Hadzidimitriou A, Belessi C, Lefranc MP, Stamatopoulos K. Immunoglobulin heavy variable (IGHV) genes and alleles: new entities, new names and implications for research and prognostication in chronic lymphocytic leukemia. Immunogenetics. 2015;67(1):61–6.
48. Yaari G, Uduman M, Kleinstein SH. Quantifying selection in high-throughput Immunoglobulin sequencing data sets. Nucleic Acids Res. 2012;40(17):e134.
49. Ye J, Ma N, Madden TL, Ostell JM. IgBLAST: an immunoglobulin variable domain sequence analysis tool. Nucleic Acids Res. 2013;41:W34-40.
50. Yu Y, Ceredig R, Seoighe C. A Database of Human Immune Receptor Alleles Recovered from Population Sequencing Data. J Immunol. 2017;198(9):3758.
51. Zhang W, Wang IM, Wang C, Lin L, Chai X, Wu J, Bett AJ, Dhanasekaran G, Casimiro DR, Liu X. IMPre: An Accurate and Efficient Software for Prediction of T- and B-Cell Receptor Germline Genes and Alleles from Rearranged Repertoire Data. Front Immunol. 2016;7:457.

# Supplemental Notes

## Supplemental Note: Immunosequencing datasets

We analyzed the following immunosequencing datasets:

- **HEALTHY:** 14 datasets from healthy individuals labeled as Set 1 – Set 14 (Table A1).
- **ALLERGY:** 24 datasets from allergy patients available from the NCBI project PRJEB18926 and labeled as ALLERGY 1 – ALLERGY 24 (Table A2).
- **HIV:** 13 datasets from HIV-infected patients available from the NCBI project PRJNA396773 and labeled as HIV 1 – HIV 13 (Table A3).
- **PROJECTS10**: 600 datasets from various human subjects corresponding to ten NCBI projects (Table A4).
- **CAMEL**: 6 camel datasets labeled as Camel 1VH, Camel 1VHH, Camel 2VH, Camel 2VHH, Camel 3VH, and Camel 3VHH (Table A5).

Datasets Set 1 – Set 9 were generated in Dr. Chudakov's lab at Moscow Institute for Bioorganic Chemistry to study aging of the adaptive immune system. Datasets Set 10 – Set 14 were generated to study immunological response to vaccines *(39)*. These datasets contain heavy chain repertoires extracted from peripheral blood mononuclear cells (PBMC) of fourteen healthy individuals. Although B cells from peripheral blood contain SHMs, we do not expect to see large clonal lineages in healthy donors.

| dataset | accession number | # reads | # distinct CDR3s | # consensus CDR3s | # trimmed CDR3s |
|---|---|---|---|---|---|
| Set 1 | TBA | 1,611,497 | 228,619 | 98,576 | 82,653 |
| Set 2 | TBA | 1,497,830 | 226,206 | 93,561 | 75,472 |
| Set 3 | SRR5851422 | 783,971 | 80,741 | 39,930 | 33,123 |
| Set 4 | TBA | 1,231,238 | 176,250 | 111,752 | 95,278 |
| Set 5 | TBA | 1,213,516 | 218,157 | 141,518 | 118,862 |
| Set 6 | TBA | 2,062,940 | 209,257 | 90,465 | 75,978 |
| Set 7 | TBA | 2,263,605 | 277,715 | 152,999 | 124,837 |
| Set 8 | TBA | 1,748,496 | 163,215 | 80,212 | 67,382 |
| Set 9 | TBA | 1,392,370 | 256,232 | 153,251 | 132,595 |
| Set 10 | SRR3620050 | 1,309,906 | 379,695 | 129,162 | 102,768 |
| Set 11 | SRR3620092 | 613,907 | 181,511 | 102,186 | 84,430 |
| Set 12 | SRR3620100 | 599,674 | 184,143 | 112,820 | 115,005 |
| Set 13 | SRR3620109 | 602,833 | 213,507 | 158,332 | 130,560 |
| Set 14 | SRR3620118 | 497,441 | 212,070 | 144,299 | 119,731 |

**Table A1. Information about the HEALTHY immunosequencing datasets.** The "# distinct CDR3s" column refers to the number of distinct CDR3s extracted from reads. The "# consensus CDR3s" column refers to the number of distinct consensus CDR3s. The "# trimmed CDR3s" column shows the number of trimmed CDR3s that are longer than $k$ (the default value $k$ = 15). Some of the listed datasets are in the process of uploading to SRA.

| dataset | accession number | # reads | # distinct CDR3s | # consensus CDR3s | # trimmed CDR3s |
|---|---|---|---|---|---|
| **Donor 1** | | | | | |
| ALLERGY1 | ERR1812282 | 1,249,203 | 213,573 | 104,981 | 89,215 |
| ALLERGY2 | ERR1812283 | 1,566,025 | 292,102 | 160,637 | 137,836 |
| ALLERGY3 | ERR1812288 | 1,782,715 | 291,796 | 173,419 | 150,577 |
| ALLERGY4 | ERR1812289 | 1,372,999 | 263145 | 172,202 | 149,626 |
| **Donor 2** | | | | | |
| ALLERGY5 | ERR1812284 | 1,313,874 | 353,957 | 189,172 | 163,373 |
| ALLERGY6 | ERR1812285 | 1,578,854 | 411,139 | 227,437 | 196,932 |
| ALLERGY7 | ERR1812290 | 644,711 | 185,269 | 133,524 | 113,985 |
| ALLERGY8 | ERR1812291 | 1,208,581 | 259,113 | 173,590 | 148,412 |
| **Donor 3** | | | | | |
| ALLERGY9 | ERR1812286 | 1,260,585 | 174,620 | 72,466 | 62,208 |
| ALLERGY10 | ERR1812287 | 2,366,528 | 270,805 | 95,473 | 81,552 |
| ALLERGY11 | ERR1812292 | 2,116,149 | 350,726 | 184,033 | 157,660 |
| ALLERGY12 | ERR1812293 | 1,842,407 | 308,770 | 167,897 | 143,617 |
| **Donor 4** | | | | | |
| ALLERGY13 | ERR1812294 | 1,935,709 | 225,119 | 98,917 | 83,730 |
| ALLERGY14 | ERR1812295 | 1,526,356 | 207,544 | 101,642 | 85,928 |
| ALLERGY15 | ERR1812300 | 783,249 | 174,985 | 129,828 | 108,518 |
| ALLERGY16 | ERR1812301 | 1,107,910 | 228,960 | 169,861 | 142,102 |
| **Donor 5** | | | | | |
| ALLERGY17 | ERR1812296 | 1426885 | 269077 | 125,283 | 108,452 |
| ALLERGY18 | ERR1812297 | 2140711 | 390376 | 190,216 | 166,223 |
| ALLERGY19 | ERR1812302 | 942524 | 110675 | 57,414 | 47,864 |
| ALLERGY20 | ERR1812303 | 1383359 | 118322 | 52,666 | 42,870 |
| **Donor 6** | | | | | |
| ALLERGY21 | ERR1812298 | 2,349,277 | 391,293 | 203,739 | 175,067 |
| ALLERGY22 | ERR1812299 | 2,137,156 | 357,480 | 187,634 | 160,635 |
| ALLERGY23 | ERR1812304 | 1,018,489 | 183,855 | 114,532 | 97,957 |
| ALLERGY24 | ERR1812305 | 818,062 | 136,659 | 81,853 | 69,686 |

**Table A2. Information about the ALLERGY immunosequencing datasets.** The first two datasets within each group represent the bone marrow samples (BM) and the second two datasets represent the peripheral blood

samples (PBMC). For example, for the datasets ALLERGY 1 and ALLERGY 2 represent BM samples and the datasets ALLERGY 3 and ALLERGY 4 represent PBMC samples.

| dataset | accession number | # reads | # distinct CDR3s | # consensus CDR3s | # trimmed CDR3s |
|---|---|---|---|---|---|
| HIV1 | SRR5888724 | 775,005 | 128,433 | 26,887 | 21,696 |
| HIV2 | SRR5888725 | 1,961,141 | 246,330 | 55,271 | 42,302 |
| HIV3 | SRR5888726 | 893,865 | 115,323 | 25,235 | 19,981 |
| HIV4 | SRR5888727 | 1,914,113 | 241,801 | 54,492 | 41,689 |
| HIV5 | SRR5888728 | 1,666,263 | 200,016 | 45,763 | 35,284 |
| HIV6 | SRR5888729 | 1,896,887 | 215,481 | 47,593 | 36,450 |
| HIV7 | SRR5888730 | 812,033 | 124,228 | 24,200 | 18,677 |
| HIV8 | SRR5888731 | 1,446,869 | 155,623 | 30,938 | 25,815 |
| HIV9 | SRR5888732 | 1,856,458 | 198,851 | 39,164 | 32,349 |
| HIV10 | SRR5888733 | 1,138,382 | 115,075 | 28,911 | 24,168 |
| HIV 11 | SRR5888734 | 1,371,172 | 145,683 | 32,407 | 26,809 |
| HIV 12 | SRR5888735 | 1,460,715 | 128,252 | 19,334 | 16,503 |
| HIV 13 | SRR5888736 | 1,485,469 | 108,508 | 25,021 | 20,506 |

**Table A3. Information about the HIV immunosequencing datasets.**

| NCBI project | Reference | # datasets |
|---|---|---|
| PRJEB18926 | *(21)* | 24 |
| PRJNA396773 | *(28)* | 13 |
| PRJNA308641 | *(37)* | 107 |
| PRJNA324093 | *(39)* | 95 |
| PRJNA248475 | *(40)* | 32 |
| PRJNA308566 | *(41)* | 142 |
| PRJNA355402 | *(42)* | 93 |
| PRJNA393446 | *(43)* | 42 |
| PRJNA349143 | *(45)* | 24 |
| PRJNA430091 | *(46)* | 28 |

**Table A4. Information about the PROJECTS10 immunosequencing datasets.** The "# datasets" column shows the number of datasets in each project.

| dataset | accession number | # reads | # distinct CDR3s | # consensus CDR3s | # trimmed CDR3s |
|---|---|---|---|---|---|
| Camel 1VH | SRR3544217 | 369,502 | 183,973 | 60,006 | 46,326 |
| Camel 1VHH | SRR3544218 | 339,758 | 157,938 | 43,880 | 39,701 |
| Camel 2VH | SRR3544219 | 288,099 | 170,899 | 74,368 | 58,118 |
| Camel 2VHH | SRR3544220 | 281,403 | 164,087 | 74,846 | 68,808 |
| Camel 3VH | SRR3544221 | 347,291 | 176,854 | 79,382 | 61,918 |
| Camel 4VHH | SRR3544222 | 343,485 | 150,724 | 59,322 | 53,799 |

**Table A5. Information about the CAMEL immunosequencing datasets.**

## Supplemental Note: Preprocessing CDR3 datasets

Although IgScout performs a more aggressive error correction (clustering CDR3s that differ by at most 3 mismatches) than the error correction implemented in IgReC *(31)* repertoire construction tool, the resulting consensus CDR3s may still contain amplification errors. However, these errors do not corrupt our analysis since they typically result in the low abundance *k*-mers that are not considered by IgScout.

To exclude suffixes (prefixes) of V (J) genes from the constructed set of CDR3s, we trimmed prefixes (suffixes) of CDR3s if they represent suffixes of V genes (prefixes of J genes). If fragments of known V and J genes were not found in a CDR3, we nevertheless cropped it by 10 nucleotides from the start (the end) to remove suffixes (prefixes) of mutated or still unknown V and J genes.

## Supplemental Note: Information about human D genes

All human D genes are located in a 30 kp long region in the human IGH locus. Figure A1 shows allelic variants of human D genes listed in the IMGT database.

```
D2     AGGATATTGTAGTAGTACCAGCTGCTATGCC     D10    GTATTACTATGGTTCGGGGAGTTATTATAAC
D2*    AGGATATTGTAGTAGTACCAGCTGCTATACC     D10*   GTATTACTATG-TTCGGGGAGTTATTATAAC
D2**   TGGATATTGTAGTAGTACCAGCTGCTATGCC

D3     GTATTACGATTTTTGGAGTGGTTATTATACC     D15    GTATTATGATTACGTTTGGGGGAGTTATGCTTATACC
D3*    GTATTAGCATTTTTGGAGTGGTTATTATACC     D15*   GTATTATGATTACGTTTGGGGGAGTTATCGTTATACC

D8     AGGATATTGTACTAATGGTGTATGCTATACC     D19    AGCATATTGTGGTGGTGATTGCTATTCC
D8*    AGGATATTGTACTGGTGGTGTATGCTATACC     D19*   AGCATATTGTGGTGGTGACTGCTATTCC
```

**Figure A1. Allelic variants of human D genes listed in the IMGT database.** Differences between various variants are highlighted in red. Alleles of human D genes differ from the main variants in a single mutation (D2*, D2**, D10*, D15*, and D19*) or two mutations at adjacent positions (D3* and D8*).

We say that two $k$-mers are $\delta$-*similar* if the Hamming distance between them does not exceed the parameter $\delta$. To evaluate similarities between D genes, we constructed their *similarity graph* $SimGraph_{k,\delta}$ in which each D gene corresponds to a vertex and two vertices are connected by an edge if the corresponding D genes contains $\delta$-similar $k$-mers. The weight of the edge connecting two D genes in the similarity graph is defined as the number of $\delta$-similar $k$-mers between these genes.

Figure A2 shows non-trivial connected components of the similarity graphs $SimGraph_{15,1}$ and $SimGraph_{15,2}$ for human D genes and illustrates that connected components contain D genes from the same family of D genes.

Since the similarity graphs $SimGraph_{15,1}$ and $SimGraph_{15,2}$ for camel D genes are empty, we reduced the parameter $k$ from 15 to 8 and constructed the graph $SimGraph_{8,2}$ for camel D genes. As Figure A (left) illustrates, all four camel D genes are similar to each other with respect to shared 8-mers.

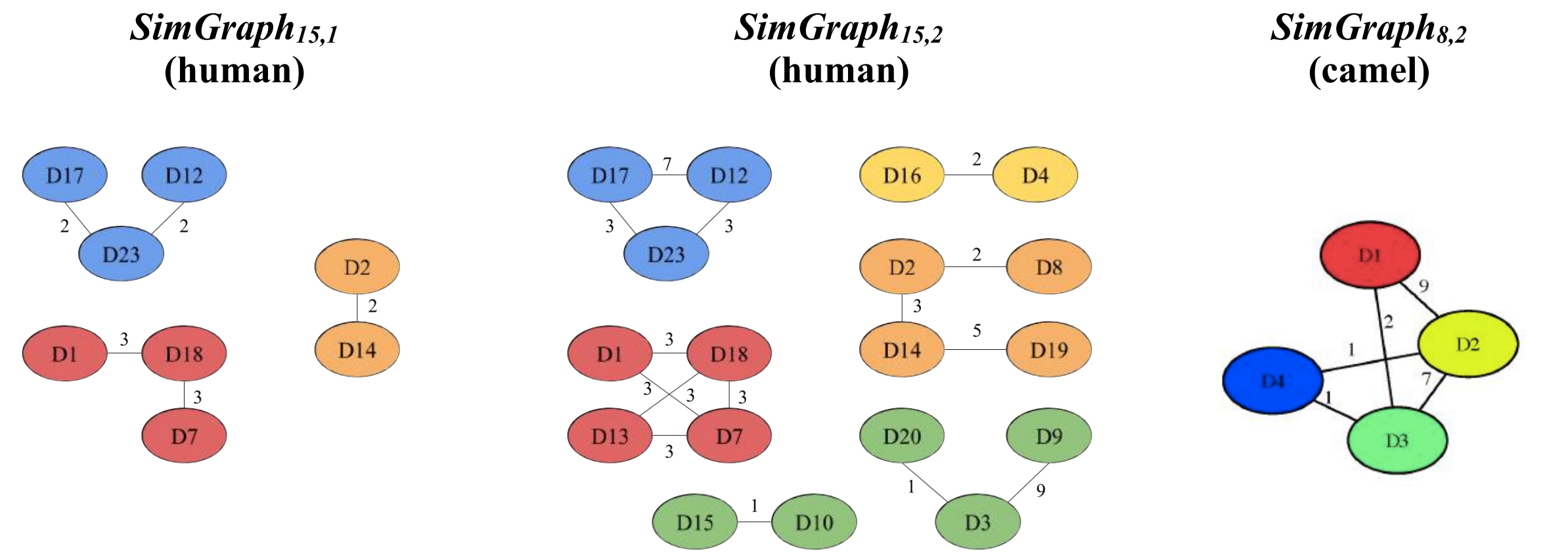

**Figure A2. Non-trivial connected components in the similarity graphs $SimGraph_{15,1}$ for human (left), $SimGraph_{15,2}$ for human (middle), and $SimGraph_{8,2}$ for camel (right) D genes.** We assigned an individual color to each family of human D genes and to each camel D gene. The graph $SimGraph_{15,4}$ consists of a single connected component containing all human D genes. Human D genes form seven gene families: F1 shown in red: D1 (D1-1), D7 (D1-7), D13 (D1-14), D18 (D1-20), D24 (D1-26); F2 shown in orange: D2 (D2-2), D8 (D2-8), D14 (D2-15), D19 (D2-21); F3 shown in green: D3 (D3-3), D9 (D3-9), D10 (D3-10), D15 (D3-16), D20 (D3-22); F4 shown in yellow: D4 (D4-4), D16 (D4-17), D21 (D4-23); F5: D5 (D5-5), D11 (D5-12), D22 (D5-24); F6 shown in blue: D6 (D6-6), D12 (D6-13), D17 (D6-19), D23 (D6-25); F7: D25 (D7-27).

## Supplementary Note: Common *k*-mers

Table A6 provides information about common 15-mers in all HEALTHY datasets. Figure A3 illustrates that most 15-mers from human D genes have high abundances and low ranks.

| dataset | known 15-mers | | mutated 15-mers | | trimmed 15-mers | | foreign 15-mers | |
|---|---|---|---|---|---|---|---|---|
| | # | min / max | # | min / max | # | min / max | # | min / max |
| Set 1 | 175 | 83 / 3141 | 195 | 83 / 645 | 70 | 83 / 604 | 3 | 83 / 134 |
| Set 2 | 174 | 77 / 2850 | 185 | 76 / 587 | 68 | 76 / 556 | 3 | 94 / 104 |
| Set 3 | 174 | 34 / 1070 | 165 | 34 / 222 | 63 | 34 / 199 | 2 | 35 / 41 |
| Set 4 | 177 | 99 / 3921 | 193 | 96 / 739 | 83 | 96 / 728 | 7 | 96 / 131 |
| Set 5 | 169 | 120 / 4699 | 159 | 119 / 2252 | 82 | 119 / 1204 | 22 | 121 / 252 |
| Set 6 | 176 | 76 / 2483 | 173 | 76 / 547 | 59 | 77 / 520 | 3 | 91 / 114 |
| Set 7 | 174 | 128 / 4313 | 143 | 126 / 1001 | 64 | 126 / 877 | 2 | 129 / 130 |
| Set 8 | 168 | 72 / 2371 | 168 | 68 / 523 | 65 | 68 / 491 | 3 | 70 / 98 |
| Set 9 | 180 | 134 / 6505 | 234 | 133 / 1877 | 106 | 136 / 1728 | 8 | 135 / 278 |
| Set 10 | 180 | 104 / 4627 | 193 | 103 / 1319 | 89 | 103 / 763 | 6 | 112 / 185 |
| Set 11 | 176 | 86 / 4007 | 163 | 85 / 890 | 80 | 86 / 513 | 4 | 85 / 131 |
| Set 12 | 176 | 121 / 5241 | 162 | 116 / 1094 | 75 | 116 / 675 | 3 | 122 / 217 |
| Set 13 | 177 | 134 / 4663 | 176 | 131 / 1143 | 80 | 131 / 1066 | 5 | 135 / 175 |
| Set 14 | 175 | 123 / 5650 | 182 | 120 / 1309 | 93 | 120 / 1084 | 8 | 122 / 183 |

**Table A6. Information about known, mutated, trimmed, and foreign *k*-mers among common 15-mers across all HEALTHY datasets.** The "#" columns show the number of 15-mers of each type. The "min / max" columns refer to the minimal / maximal abundance of 15-mers of each type.

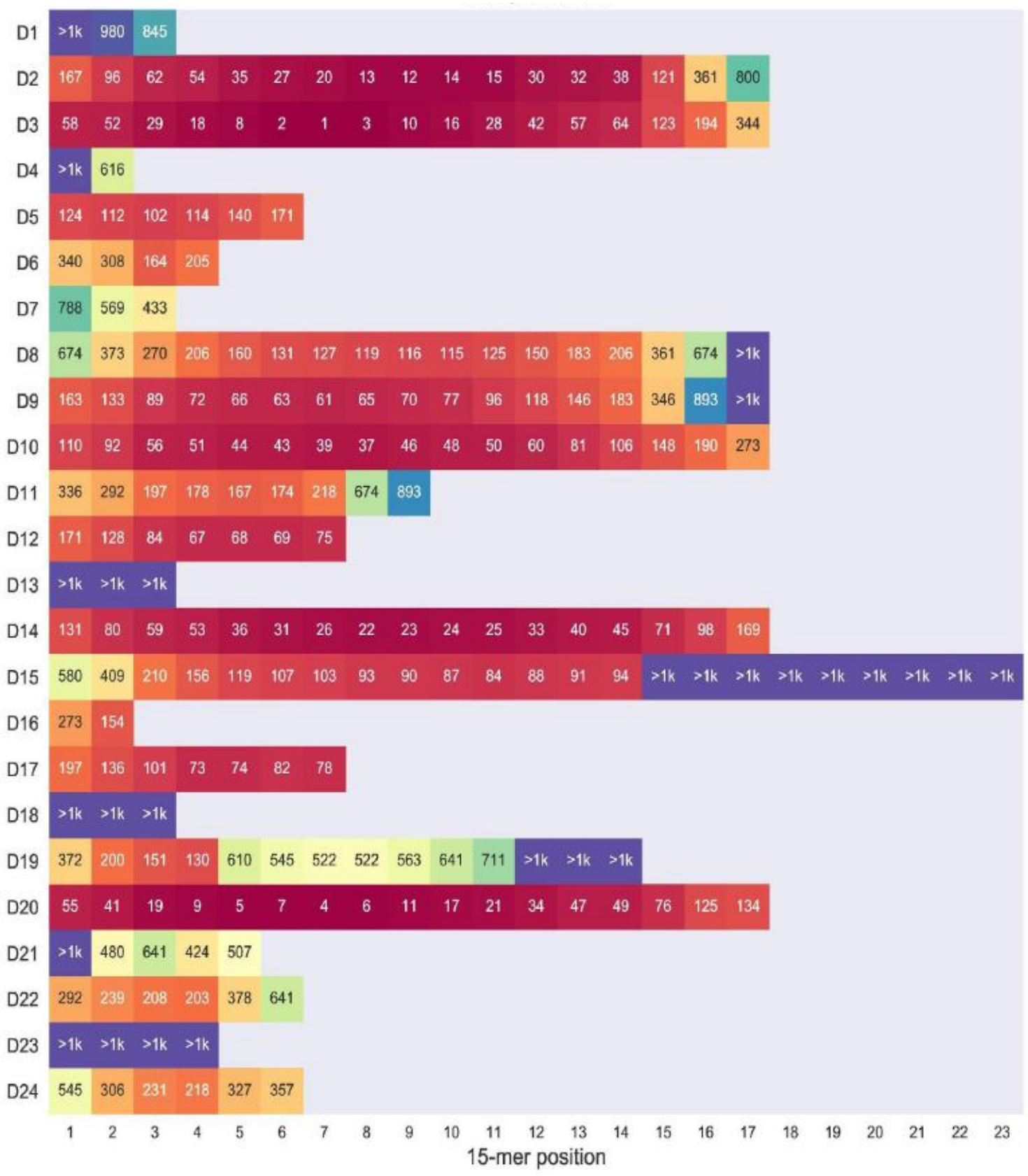

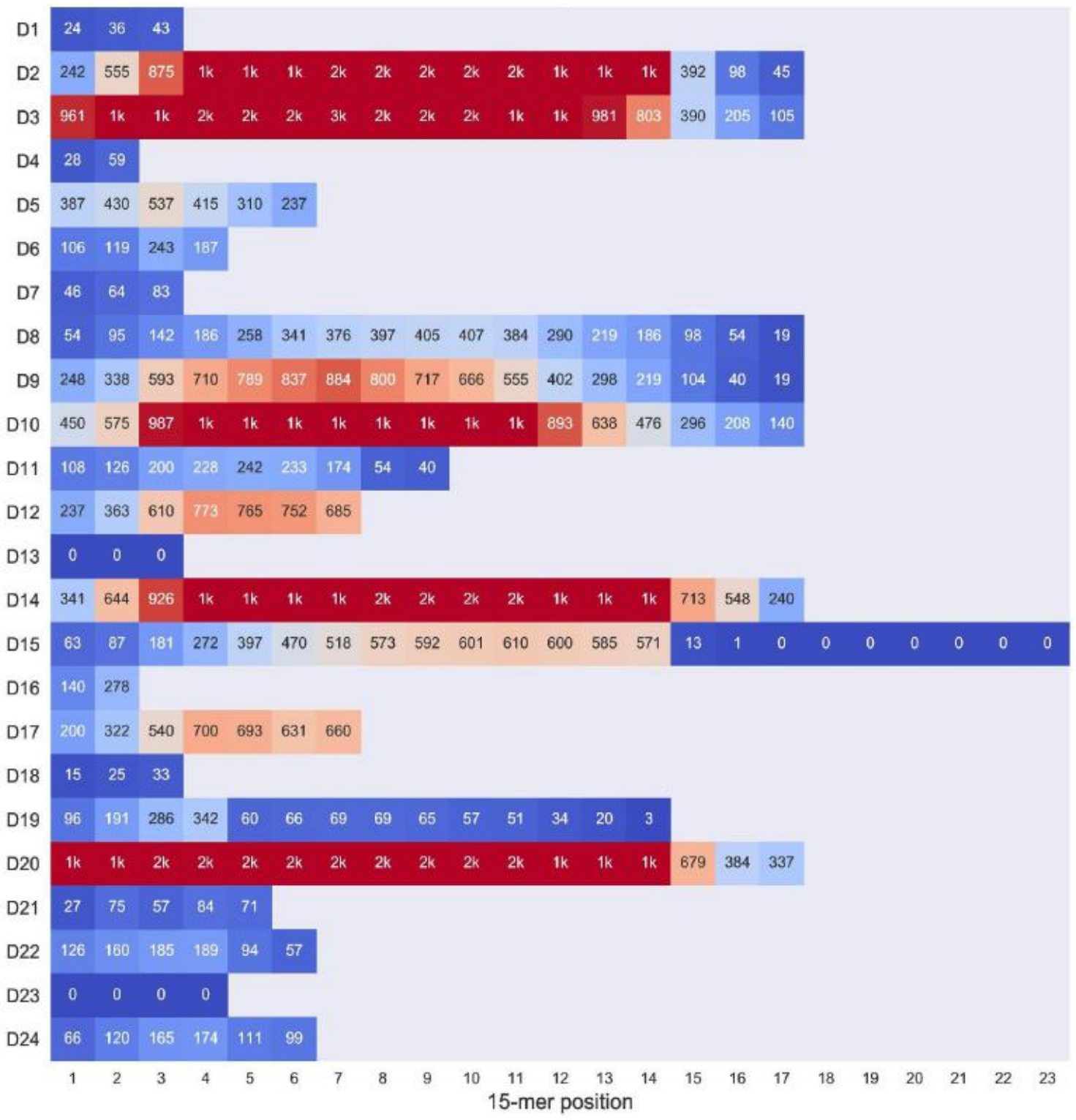

**Figure A3. Ranks (top) and abundances (bottom) of 15-mers from human D genes computed for the *CDR3$_{trimmed}$* dataset.** Each D gene of length *t* is shown as a sequence of *t*-14 cells representing its 15-mers. For example, the gene D1 of length 17 is shown as a sequence of 3 cells. A number within a cell represents the rank (top) or abundance (bottom) of the corresponding 15-mer. For example, D3 contains the most abundant 15-mer with rank 1 and abundance 314. Red (blue) cells correspond to low (high) values of ranks and high (low) values of abundances. 11-nucleotide long gene D25 is not shown. Genes D13, D23 and D25 do not contribute 15-mers to the *CDR3$_{trimmed}$* dataset.

## Supplemental Note: IgScout pseudocode

IgScout takes (i) a set *Strings* representing trimmed CDR3s, (ii) the *k*-mer size, (iii) the information content threshold *IC*, and (iv) the minimum multiplicity *minMultiplicity* of *k*-mers threshold as the input parameters (Figure A4). It uses the **PrefixExtension** and **SuffixExtension** subroutines for extending the selected *k*-mers to the left and to the right and generating the putative D genes. See Supplemental Note “IgScout parameters.”

```
IgScout(Strings, k, IC, minMultiplicity)
RemainingStrings ← Strings
D-genes ← empty set
while forever
  D ← a most frequent k-mer in RemainingStrings
  if frequency of D in Strings exceeds minMultiplicity
    D ← PrefixExtension(RemainingStrings, D, k, IC)
    D ← SuffixExtension(RemainingStrings, D, k, IC)
    add string D to the set D-genes
    Strings(D) ←all strings in RemainingStrings containing k-mers from the string D
    remove all strings from Strings(D) from the set RemainingStrings
  else
    return D-genes
```

```
PrefixExtension(Strings, D, k, IC)
  prefix ← first k-mer in string D
  Strings(prefix) ← all strings in Strings containing prefix
  Alignment ← prefix-anchored alignment of all strings in Strings(prefix)
```

```
  previousColumn ← the column in Alignment preceding the first position of D
    if information content of previousColumn exceeds IC
      consensus ← a most frequent nucleotide in previousColumn
      D ← the prefix-extension of the string D by the nucleotide consensus
      PrefixExtension(Strings, D, k, IC)
    else
      return D
```

```
SuffixExtension(Strings, D, k, IC)
  suffix ← last k-mer in string D
  Strings(suffix) ← all strings in Strings containing suffix
  Alignment ← suffix-anchored alignment of all strings in Strings(suffix)
  nextColumn ← the column in Alignment following the last position of D
    if information content of nextColumn exceeds IC
      consensus ← a most frequent nucleotide in nextColumn
      D ← the suffix-extension of the string D by the nucleotide consensus
      SuffixExtension(Strings, D, k, IC)
    else
      return D
```

**Figure A4. IgScout pseudocode.**

## Supplemental Note: IgScout parameters

IgScout stops when the most frequent *k*-mer in the remaining CDR3s has abundance below *minMultiplicity=fraction**$|CDR3_{trimmed}|$. We applied IgScout with various values of the parameters *fraction* = {0.0005, 0.001, 0.0015, 0.002, 0.0025, 0.003, 0.0035, 0.004} and *IC* = {30%, 40%, 50%, 60%, 70%, 80%} to 14 HEALTHY immunosequencing datasets. For each launch of IgScout, we computed the following metrics:

- the number of reconstructed D genes (we classify a segment as a reconstructed D gene if it is a substring of this D gene),
- the average number of nucleotides missing at the start/end of the reconstructed D genes,
- the number of novel genes (we classify a segment as novel if it does not represent a substring of a known D gene). Note that novel D genes may represent both allelic variants and false positive inferences.

Figure A5 shows distributions of values of these metrics (averaged over 14 HEALTHY datasets) for each pair of *fraction* and *IC* values and illustrates that fraction = 0.001 and IC = 0.5 represent suitable parameters.

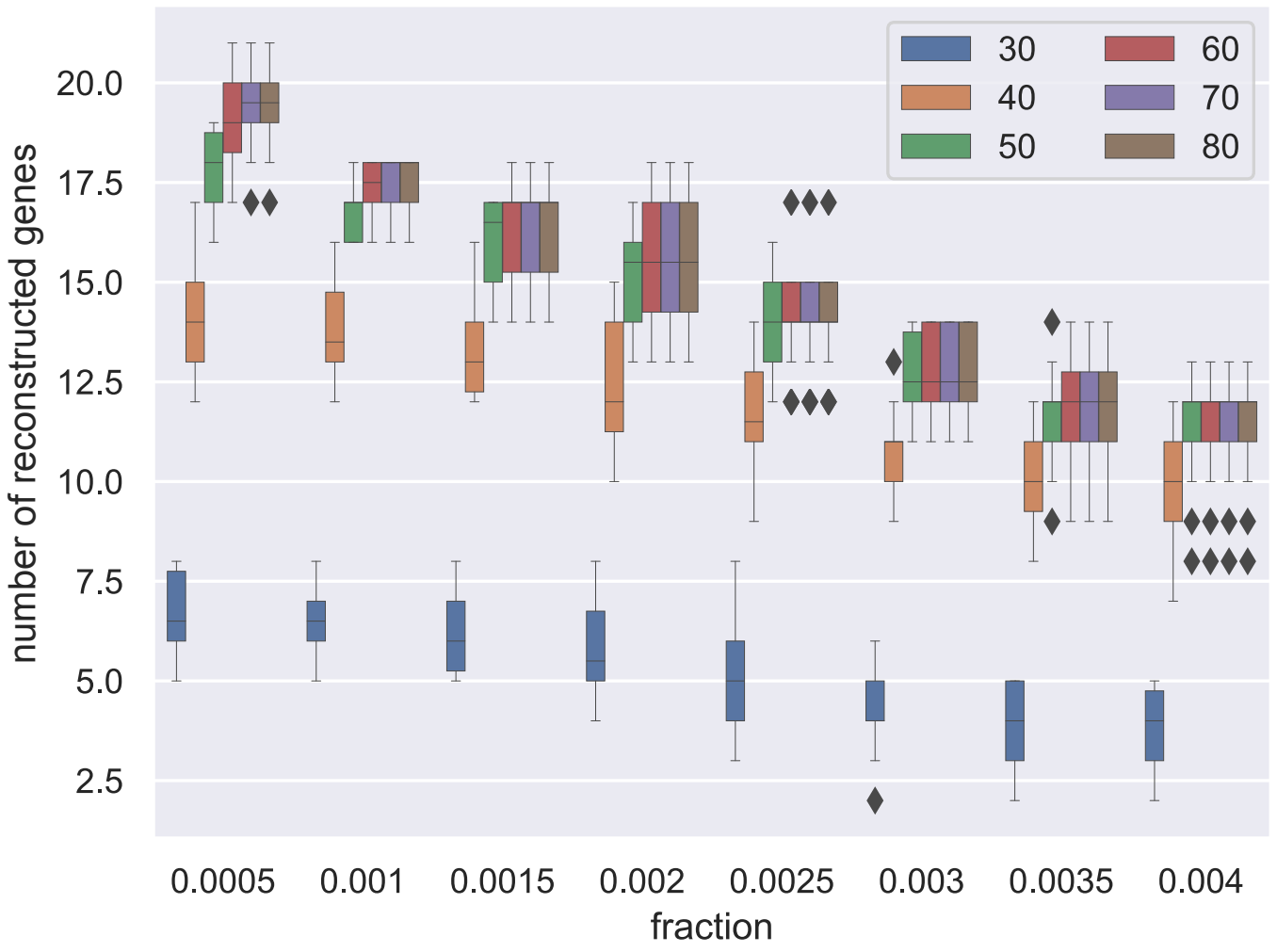

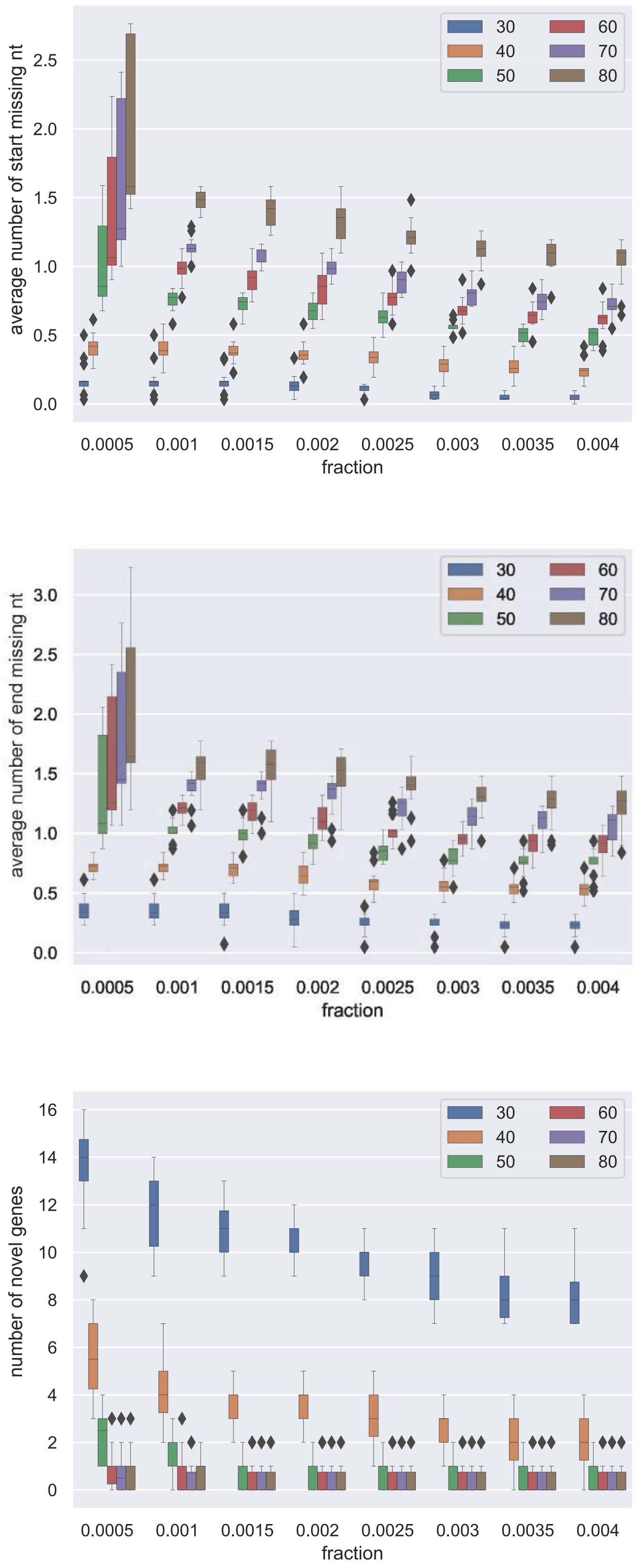

**Figure A5. IgScout results for various values of parameters *fraction* and *IC* (the HEALTHY datasets). Colors correspond to various values of *IC* (in percentages).** The distribution of the metric values for each pair of *fraction* and *IC* is illustrated as an error bar. The bar shows the quartiles of the distribution, the whiskers demonstrate the rest of the distribution, except for points that are determined as outliers.

Figure A6 shows multiplicities and ranks of known 15-mers after 17 iterations of IgScout (the Set 1 dataset). IgScout stops before reconstructing the D7 gene because its most abundant 15-mer occurs in less than *fraction*=0.1% of strings in the $CDR3_{trimmed}$ dataset. Although decreasing the *fraction* threshold would lead to reconstructing additional D genes, it may also add false positive reconstructions. Figure A7 shows abundances of common 15-mers in the set of CDR3s that remain after IgScout completed its work.

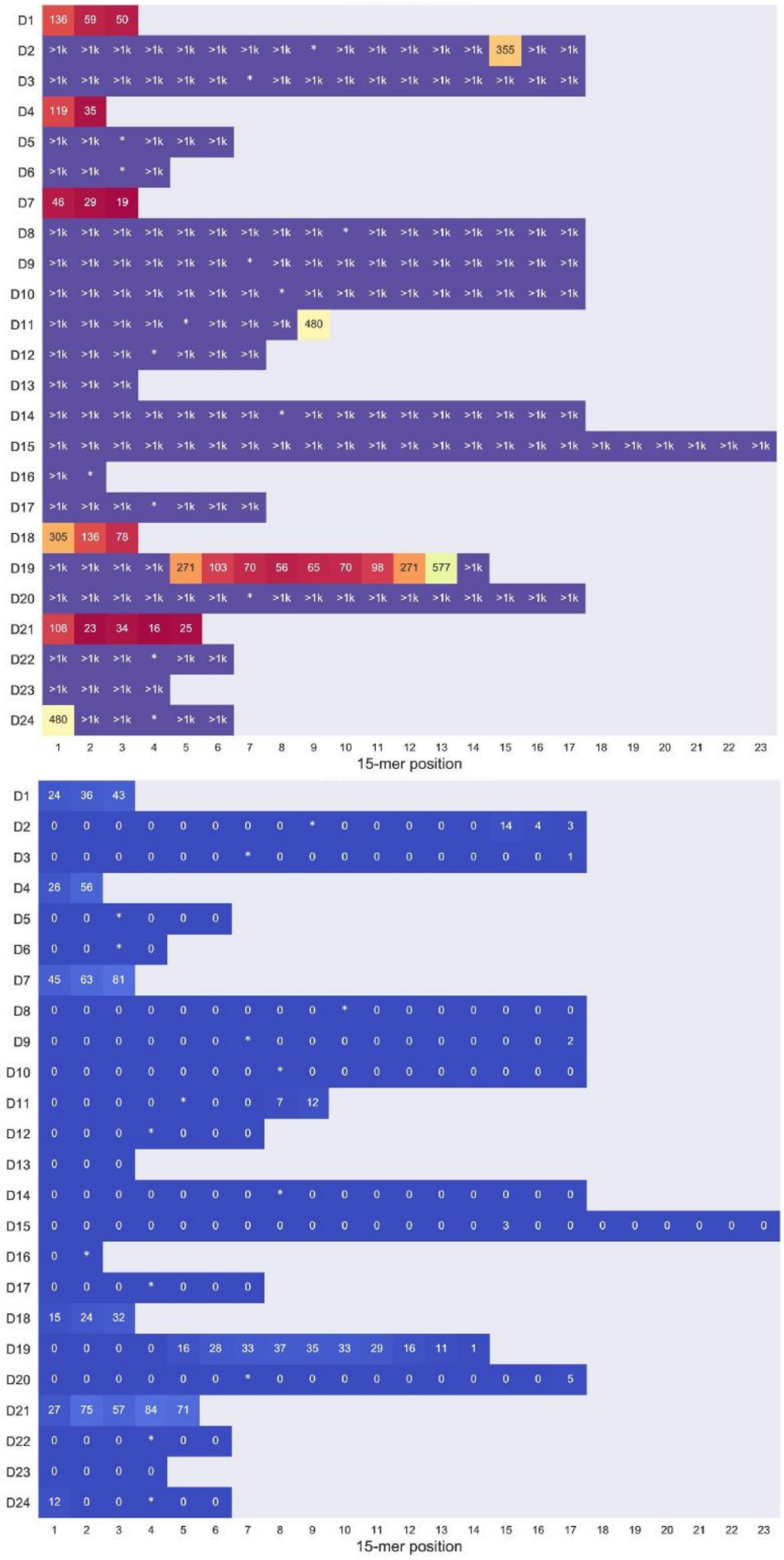

**Figure A6. Ranks (top) and abundances (bottom) of 15-mers from known D genes constructed for the CDR3s remaining after IgScout completed its work (for the Set 1 dataset).** 15-mers used for inference of D genes are marked with "*".

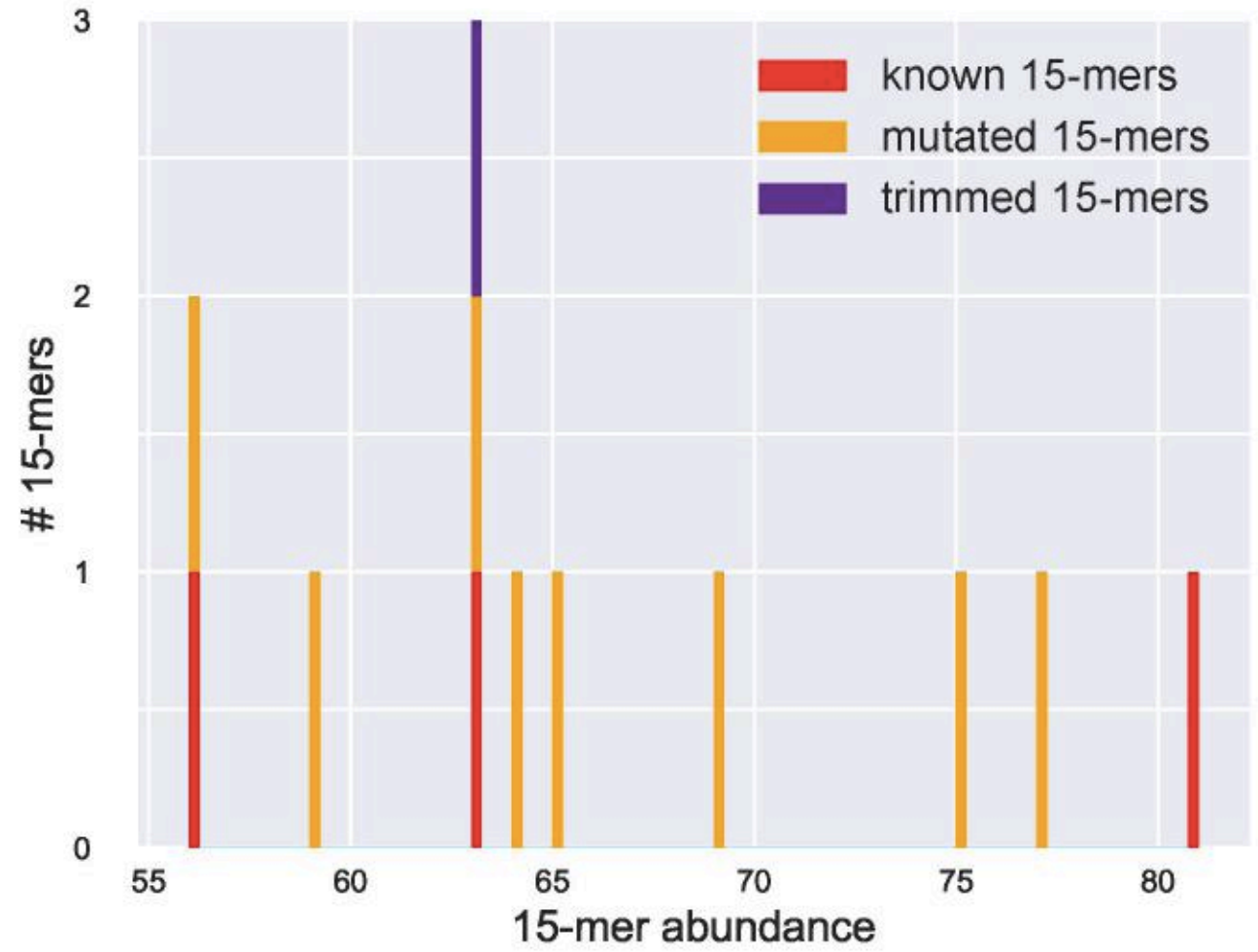

**Figure A7. Abundances of all 12 common 15-mers in the *CDR3*$_{trimmed}$ set after 17 iterations of IgScout.** We removed CDR3s participating in the inference of novel D segments during the first 17 steps of the IgScout algorithm and analyzed all common 15-mers in the remaining 55,514 CDR3s. These 12 common 15-mers have abundances varying from 56 to 81. The *y*–axis represents the number of common 15-mers with the given abundance. Red, orange, and violet bars represent the number of common 15-mers with given abundance among common 15-mers. There exist 3 known (red bars), 8 mutated (orange bars) and 1 trimmed (violet bars) 15-mers. There are no foreign 15-mers among common 15-mers after the IgScout run. The three known 15-mers belong to genes D7 and D14.

## Supplemental Note: Simulating CDR3 datasets

To generate simulated immunosequencing datasets, we used the IgSimulator tool (Safonova et al., 2015) with the set of 25 human D genes (D1-D25). For the sake of simplicity, we first assumed that all 25 D genes participate in VDJ recombination with the same probability 0.04 (uniform distribution of abundances) and later analyzed a non-uniform distribution of abundances. As a variable parameter of the simulation, we used the maximal length of the exonuclease removal (*ERmax*). To simulate a substring of a D gene in the VDJ recombination, we randomly selected integers *ERstart* and *ERend* (uniformly distributed between 0 to *ERmax*), cropped the sequence of a D gene by *ERstart* nucleotides from the start and *ERend* nucleotides from the end, and added random insertions (with the length uniformly distributed from 0 to 10 nucleotides) on both ends. We varied *ERmax* from 1 to 10 (according to Ralph and Matsen, 2016, *ERmax* typically does not exceed 10 nucleotides for all D genes).

IgSimulator simulates SHMs as low-frequency random mutations. Since such SHMs are unlikely to change the IgScout results (most mutated D segments in CDR3 will be simply excluded from analysis since they do not preserve *k*-mers), we decided not to simulate clonal lineages with abundant SHMs. Indeed, a new Supplemental Note “How IgScout results are affected by the number of consensus CDR3s and cell types?” demonstrates that IgScout shows better results on datasets with high diversity of VDJ recombination (datasets from PBMC / naïve / memory B cells) compared to highly mutated datasets with low diversity of VDJ recombination (e.g., datasets from specific plasma B cells).

We simulated 10 datasets with 100,000 CDR3s each (*ERmax* varies from 1 to 10) for and applied IgScout with default parameters ($k$ = 15, *fraction* = 0.001, *IC* = 0.5) to each of them. We assume that IgScout reconstructs a D gene if it reports its unique substring (i.e., a substring that does not appear in other D genes). Below we discuss how the length of the reconstructed D genes influences downstream applications of IgScout.

To analyze how our ability to reconstruct low-usage D genes is affected by non-uniform distribution of usage of D gene, we have simulated a repertoire with a high usage of a single gene (we selected the longest D gene D15) and low usages of all other genes (Table A7).

| D gene | Usage | D gene | Usage |
|---|---|---|---|
| D1 | 0.001 | D14 | 0.0205 |
| D2 | 0.0025 | D15 | 0.5620 |
| D3 | 0.0040 | D16 | 0.0220 |
| D4 | 0.0055 | D17 | 0.0235 |
| D5 | 0.0070 | D18 | 0.0250 |
| D6 | 0.0085 | D19 | 0.0265 |
| D7 | 0.0100 | D20 | 0.0280 |
| D8 | 0.0115 | D21 | 0.0295 |
| D9 | 0.0130 | D22 | 0.0310 |
| D10 | 0.0145 | D23 | 0.0325 |
| D11 | 0.0160 | D24 | 0.0340 |
| D12 | 0.0175 | D25 | 0.0355 |
| D13 | 0.0190 | | |

**Table A7. Simulating CDR3s with non-uniform usage of D genes.** Abundance of a D gene shows the fraction of CDR3s in the simulated repertoire formed by this D gene. We arbitrarily assigned abundances varying from 0.001 to 0.025 (with step 0.0015) to all genes except for D15. The sum of these abundances is 0.438 and abundance of D15 was set to 1 – 0.438 = 0.562.

## Supplemental Note: Benchmarking IgScout on simulated immunosequencing datasets

We applied IgScout to simulated CDR3 datasets with uniform and non-uniform usage of D genes (see Supplemental Note "Simulating CDR3 datasets").

**IgScout results in the case of the uniform distribution of usage of D genes.** IgScout reconstructed 24 out of 25 D genes for all values of *ERmax* (short 11-nucleotide long gene D25 cannot be detected with $k = 15$). On average, IgScout misses one nucleotide at the start of D gene and one nucleotide at the end D gene for all values of *ERmax*. In all simulations, IgScout returned erroneous D genes only for unrealistically small values *ERmax* = 1 and 2. Our simulation suggests that IgScout would likely reconstruct all D genes (except for a short D25) if their abundances were uniformly distributed.

**IgScout results in the case of a non-uniform distribution of usage of D genes.** Figure A8 shows that IgScout reconstructs long D genes (> 20 nt) even if they are presented in less than 1% of CDR3s. IgScout missed short D genes (< 20 nt) when their abundance falls below 2.5% (D1, D4, D5, D6, D7, D13, D16, D18). IgScout also missed D25 because it is shorter than the default value of $k$=15.

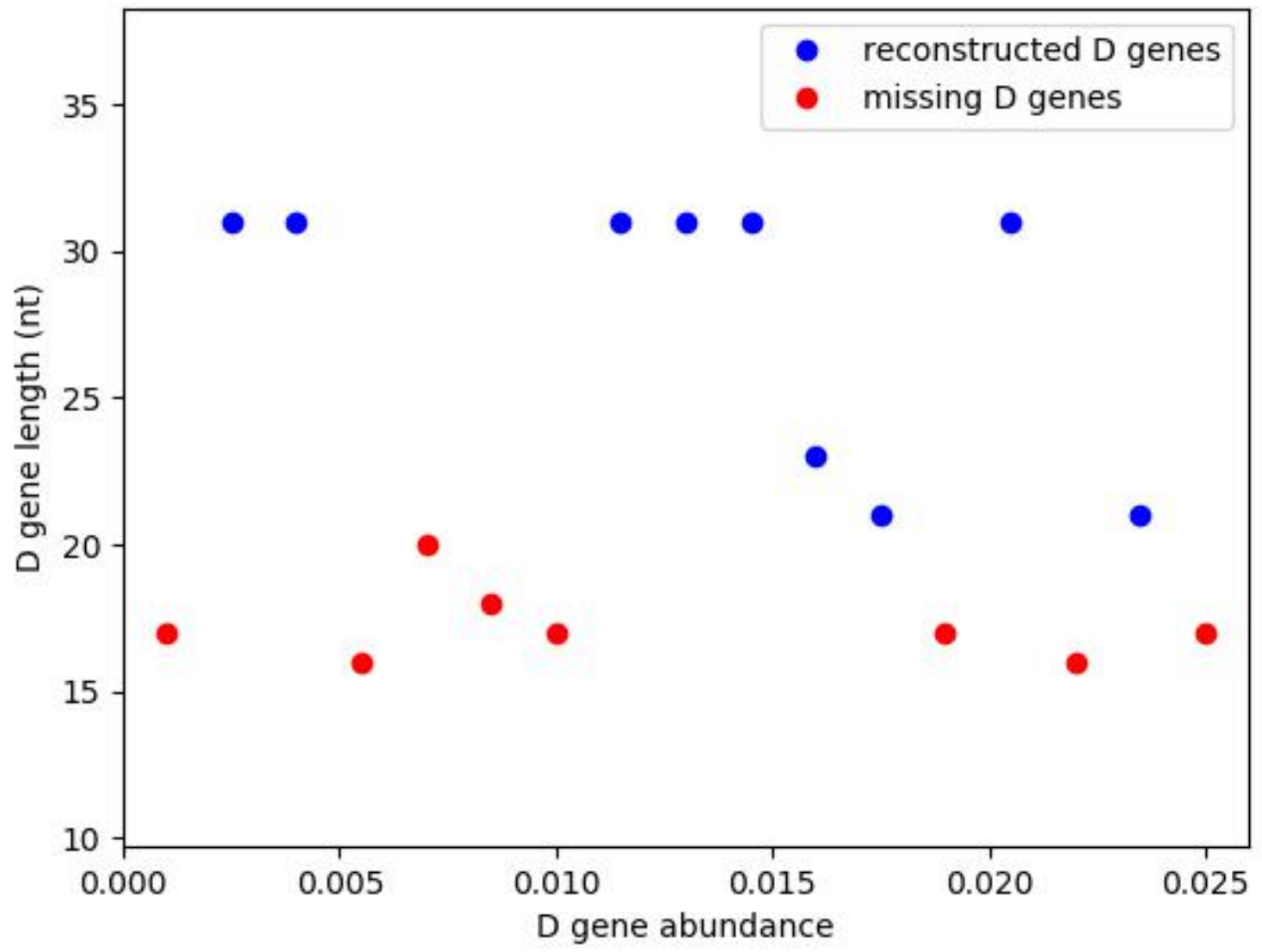

**Figure A8. Performance of IgScout on simulated CDR3s with non-uniform usage of D genes.** D genes reconstructed (missing) by IgScout are shown by blue (red) dots. Usages of D genes are shown in Table A7.

## Supplemental Note: How trimmed (rather than complete) D genes affect the downstream analysis of immunosequencing datasets

**Two modes of IgScout applications.** IgScout can be applied for:

- inference of novel variants of known D genes (for species with known germline genes) and further population analysis of antibody repertoires,
- inference of D genes (for species with unknown germline genes) and further VDJ classification (i.e., finding V, D, and J genes explaining the observed VDJ recombination).

The first *reference-based* mode does not require inference of full-length D genes because they can be reconstructed from the trimmed D genes by aligning against the known variants of D genes (IgScout has a reference-based mode "*--d-genes*" that compares the reconstructed D genes with known ones). However, the negative impact of the reduced lengths of the inferred D genes on the quality of the D gene classification is unclear. Below we show that this impact is very small.

**Defining a match between a CDR3 and a D gene.** Existing VDJ classification tools search for a match between a CDR3 and a D gene with the score exceeding a threshold *L*. To estimate the accuracy of the D gene classification, we used the datasets simulated with *ERmax* = 10 (that we refer to as the SIMULATED dataset) with uniform abundances of D genes (see Supplemental Notes "Simulating CDR3 datasets").

We analyzed a simple scoring based a longest match between each CDR3 from the SIMULATED dataset and each D gene from a database. We say that a CDR3 is *generated* by a specific D gene if this D gene results in a longest match with this CDR (over all D genes from the database). If several D genes provide the longest matches, we say that all of them generated a given CDR3. Since we did not simulate SHMs, we compute only the exact matches and thus produce more accurate results for the SIMULATED dataset compared to IgBlast that allows mismatches and indels. Note that an algorithm for D gene classification that takes into account mismatches and indels might generate less accurate results than an algorithm based on exact matches since it may be confused by highly similar D genes (e.g., it can extend an exact match by mismatches and report an incorrect D gene).

**False positive and false negative CDR3 classifications.** Given a simulated CDR3 (referred to as *CDR3*), we refer to the D gene it originated from as *D(CDR3)* and to D genes with a longest match against this CDR3 as *D*=D*(CDR3).* We refer to this match as *match(CDR3,D*).* We further check if the found longest match exceeds the length threshold *L* and classify *CDR3* as follows:

1. If $|match(CDR3,D^*)| \geq L$ and *match(CDR3,D*)* represents a unique substring of *D(CDR3),* we classify *CDR3* as a true positive (TP) and as false negative (FN) otherwise.
2. If $|match(CDR3,D^*)| < L$, we classify *CDR3* as a true negative (TN), and a false positive (FP) otherwise.

Using the classification of all CDRs, we compute the *sensitivity* as #TP / (#TP + #FN) and the *specificity* as #TN / (#TN + # FP) for values of the length threshold *L* varying from 5 to 15 nucleotides.

We applied this procedure to the following three datasets of D genes to study the negative effects of trimmed D genes as compared to the full-length D genes:

- 25 complete human D genes (referred to as the COMPLETE database)
- 24 *trimmed* human D genes inferred by IgScout for the SIMULATED dataset (referred to as the TRIMMED database)
- 17 *highly trimmed* D genes represented by the most abundant 15-mers selected by IgScout for inference of these D genes in the HEALTHY dataset (referred to as $TRIMMED^+$ database)

**Sensitivity and specificity of the D gene classification.** Figure A9 shows the sensitivity and specificity of the classification of an arbitrarily selected single D gene (D12) in all CDR3s generated by this D gene from the SIMULATED dataset. As Figure A9 illustrates, the sensitivity and specificity of the complete and trimmed D genes are very similar. On the other hand, the low sensitivity of the D gene classification using 15-mers in the TRIMMED* dataset demonstrates that the extension of abundant 15-mers by IgScout is an important step that significantly improves the D gene classification.

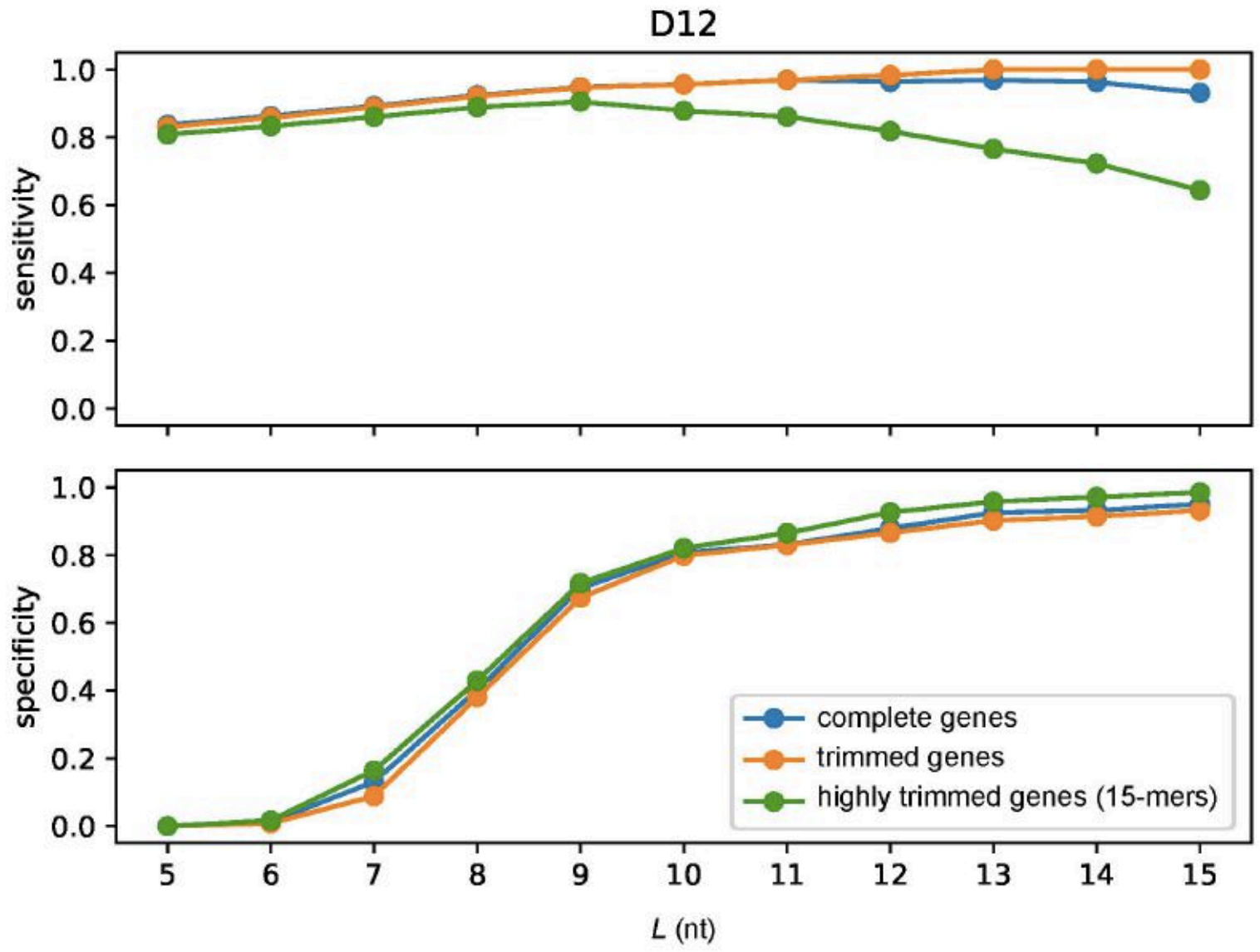

**Figure A9. Sensitivity and specificity of classifying the D12 gene in the CDR3s from the SIMULATED dataset.** We used the COMPLETE (blue), TRIMMED (orange), and TRIMMED+ (green) datasets of D genes for classifying CDR3s. The length threshold *L* varied from 5 to 15 nucleotides.

Figure A10 illustrates that the sensitivity/specificity of the D gene classification of complete D genes and trimmed D genes are nearly identical (for all values of *L*). Thus, trimming 1-2 nucleotides at the start/end of D genes hardly affects the accuracy of the D gene classification.

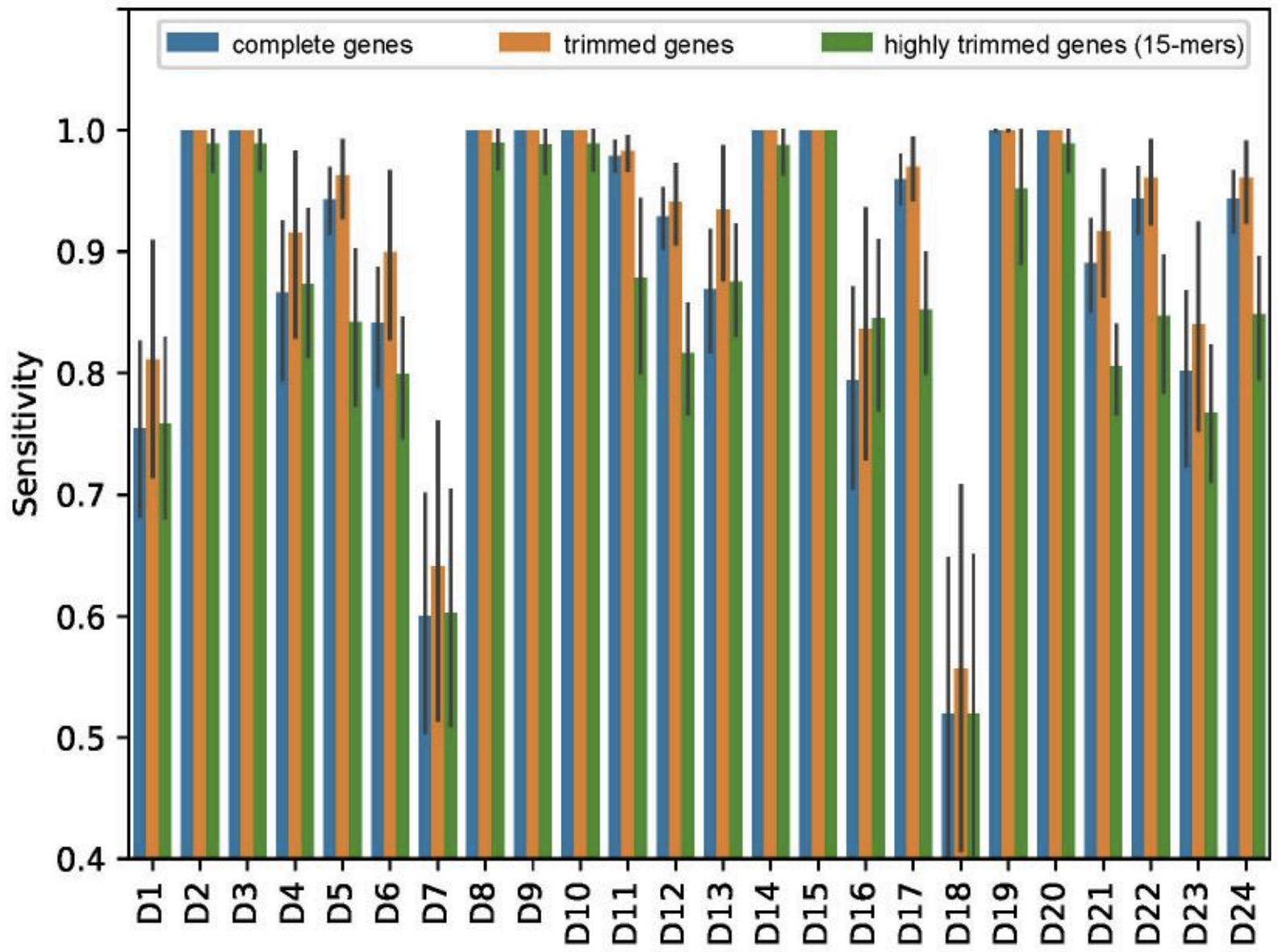

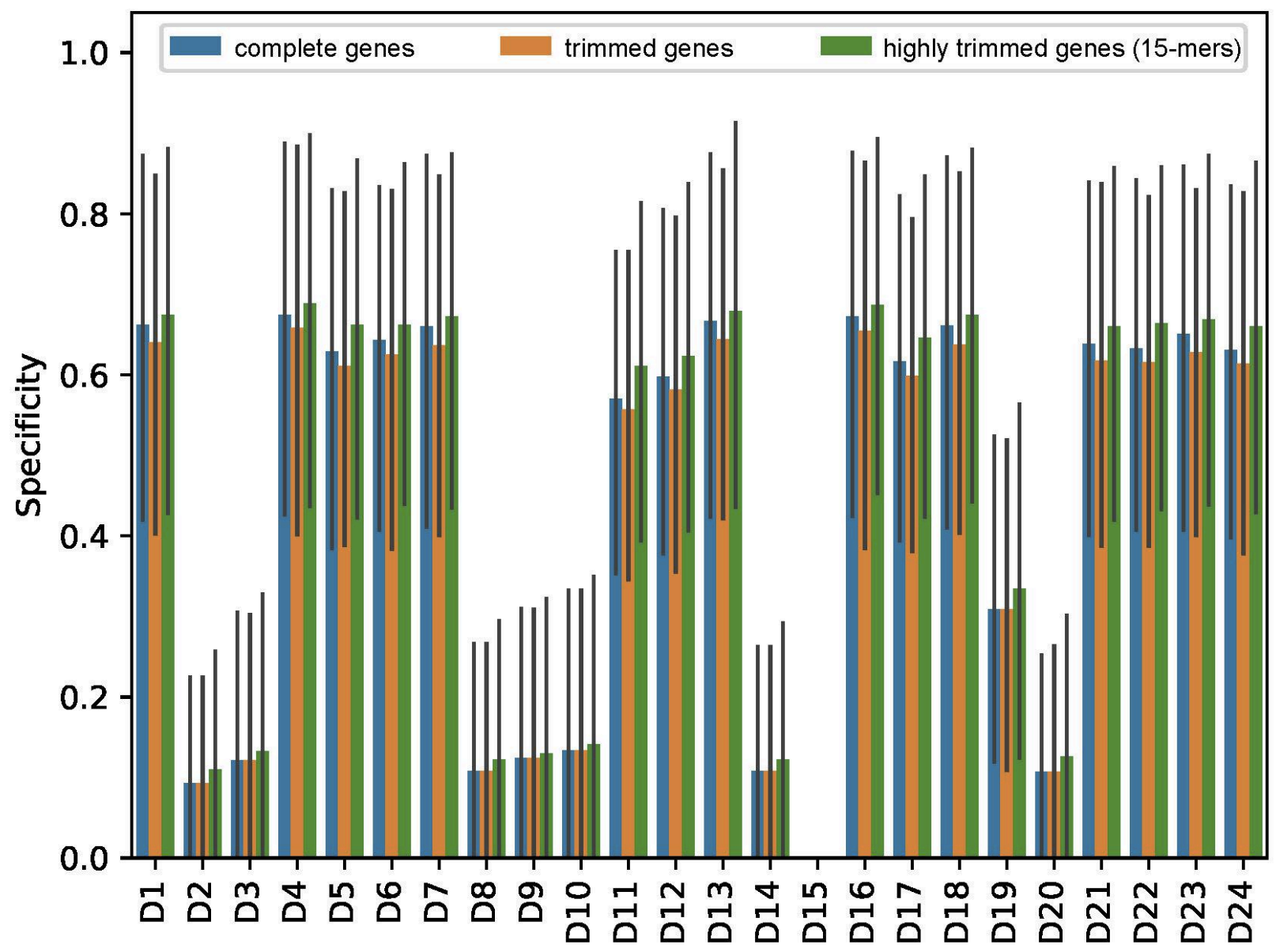

**Figure A10. Sensitivity (top) and specificity (bottom) of classifying 24 inferred D genes in the SIMULATED dataset.** Each bar represents the sensitivity (specificity) for various values of the length threshold *L*. Height of a bar shows the average value of sensitivity (specificity). Error bars shown by black lines correspond to the minimal and maximal values of sensitivity (specificity).

## Supplemental Note: Reconstructing variants of human D genes

Figure A11 and Figure A12 present information about reconstructed D genes across the HEALTHY, ALLERGY, and HIV datasets.

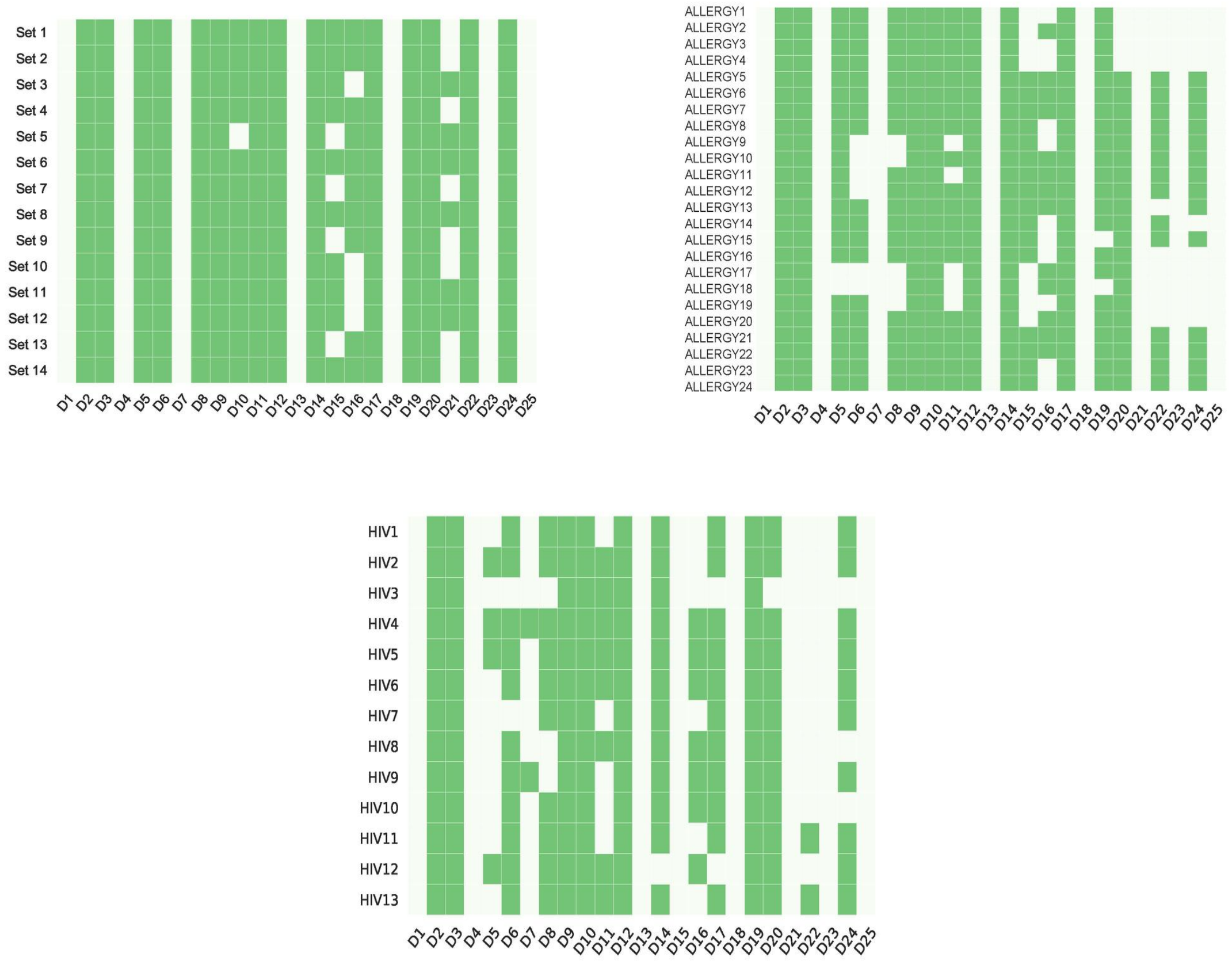

**Figure A11. Human D genes that were reconstructed (green cells) and missed (white cells) by IgScout in all HEALTHY (top left), ALLERGY (top right), and HIV (bottom) dataset.**

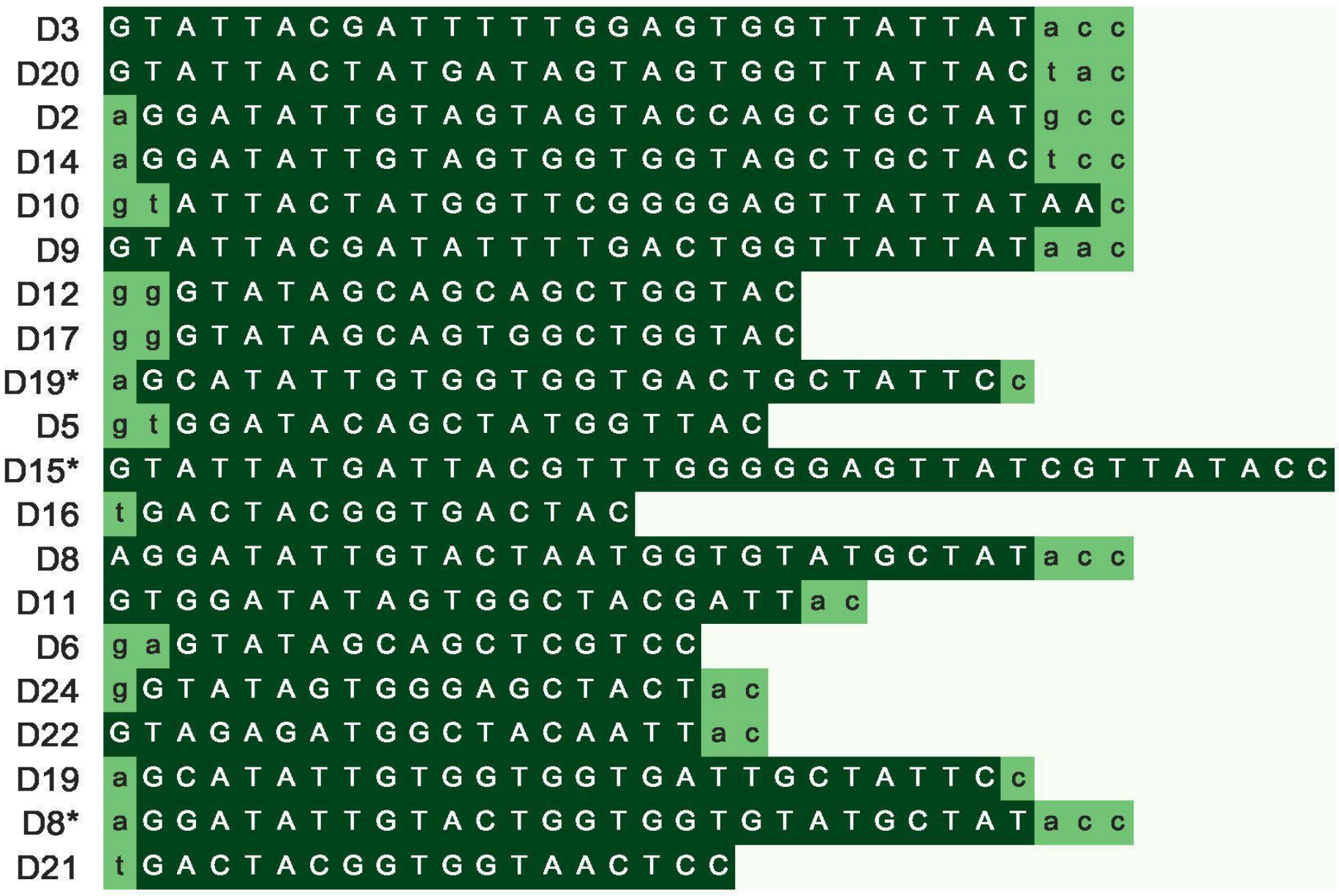

**Figure A12. Information about D genes reconstructed by IgScout across all HEALTHY datasets.** Position in a D gene is colored in dark green if it was reconstructed in at least one of the HEALTHY datasets. Positions in D genes that were not reconstructed in all datasets are shown in light green. Ordering of rows reflects the order in which IgScout discovers various D genes, e.g., D3 appears in the first row because it was reconstructed at the first step of IgScout in 7 out of 14 datasets, D20 appears in the second row because it was reconstructed at the first step of IgScout in 6 out of 14 datasets. If IgScout took $n$ steps to analyze the $i$-th dataset and reconstructed a gene $D$ at the $j$-step we assign $index(D,i)=j$ and assign $index(D,i)=n+1$ if IgScout failed to reconstruct the gene $D$ in the $i$-th dataset. All D genes are arranged from top to bottom in the increasing order of the average values of their indices across all fourteen datasets. Genes D1, D4, D7, D13, D18, D23, and D25 are not shown since they were not discovered in any of the HEALTHY datasets.

Table A8 illustrates that IgScout finds novel variants D10$^+$ (D15$^+$) in 50 (46) datasets from 600 PROJECTS10 immunosequencing datasets. It also found novel D10$^{++}$ and D15$^{++}$ in two datasets from the PRJNA308566 project and one dataset from the PRJNA308641 project, respectively (Figure A13). All variants of the D15 gene (D15, D15*, D15$^+$, and D15$^{++}$) differ from each other in two positions that can be described as 0-0, 0-1, 1-1, and 1-0 haplotypes for D15, D15*, D15$^+$, and D15$^{++}$, respectively.

| NCBI project | Reference | # datasets | # datasets supporting D10+ | # datasets supporting D15+ |
|---|---|---|---|---|
| PRJEB18926 | *(21)* | 24 | 8 | 8 |
| PRJNA396773 | *(28)* | 13 | 13 | 13 |
| PRJNA308641 | *(37)* | 107 | 9 | 8 |
| PRJNA324093 | *(39)* | 95 | – | 3 |
| PRJNA248475 | *(40)* | 32 | – | – |
| PRJNA308566 | *(41)* | 142 | 6 | 3 |
| PRJNA355402 | *(42)* | 93 | 1 | 1 |
| PRJNA393446 | *(43)* | 42 | 8 | 7 |
| PRJNA349143 | *(44)* | 24 | – | – |
| PRJNA430091 | *(45)* | 28 | 5 | 3 |

**Table A8. Information about immunosequencing datasets in the PROJECTS10 collection supporting the novel D10$^+$ and D15$^+$ variants.** The “# datasets” column shows the total number of datasets in each project.

```
D10+   GTATTACTATGGTTCAGGGAGTTATTATAAC   D15+   GTATTATGATTACATTTGGGGGAGTTATCGTTATACC
D10++  GTATTACTATGGGTCGGGGAGTTATTATAAC   D15++  GTATTATGATTACATTTGGGGGAGTTATGCTTATACC
D10    GTATTACTATGGTTCGGGGAGTTATTATAAC   D15    GTATTATGATTACGTTTGGGGGAGTTATGCTTATACC
D10*   GTATTACTATGTT-CGGGGAGTTATTATAAC   D15*   GTATTATGATTACGTTTGGGGGAGTTATCGTTATACC
```

**Figure A13. Novel variants of D10 and D15 genes.** The $D10^{++}$ variant was inferred from the datasets SRR3099127 and SRR3099139 (project PRJNA308566) corresponding to the same individual. The $D15^{++}$ variant was inferred from the SRR3099414 dataset (project PRJNA308641).

## Supplemental Note: Summary of IgScout results across diverse immunosequencing datasets

We applied IgScout to 361 Rep-seq datasets from ten independent immunosequencing projects corresponding to diverse immunogenomics studies (Table A8). Figure A14 shows the sets of reconstructed D genes for each dataset and illustrates that 20 D genes were reconstructed across all datasets. In addition to 18 D genes inferred from the HEALTHY datasets, IgScout reconstructed D4 (in 3 datasets) and D7 (in 5 datasets). Five D genes (D1, D13, D18, D23, and D25) are missing in all analyzed datasets. These five genes are also reported as missing in multiple studies on analyzing the usage of D genes: Souto-Carneiro et al., 2005, Briney et al., 2012, Elhanati et al., 2015, and Kidd et al., 2016. For example, Briney et al., 2012 reported three D genes (D13, D18, D25) as not contributing to VDJ recombination in all their datasets, while Elhanati et al., 2015 reported the same three genes as well as six other D genes as (D4, D5, D7, D11, D22, D23) as missing in their datasets.

As Figure A14 illustrates, even the most abundant D genes are missing in some datasets, e.g., D20 was identified in all HEALTHY datasets but was not identified in 10% of the 361 datasets. These datasets likely represent repertoires where a single D gene with a very high usage overpowers all others D genes because of a clonal selection.

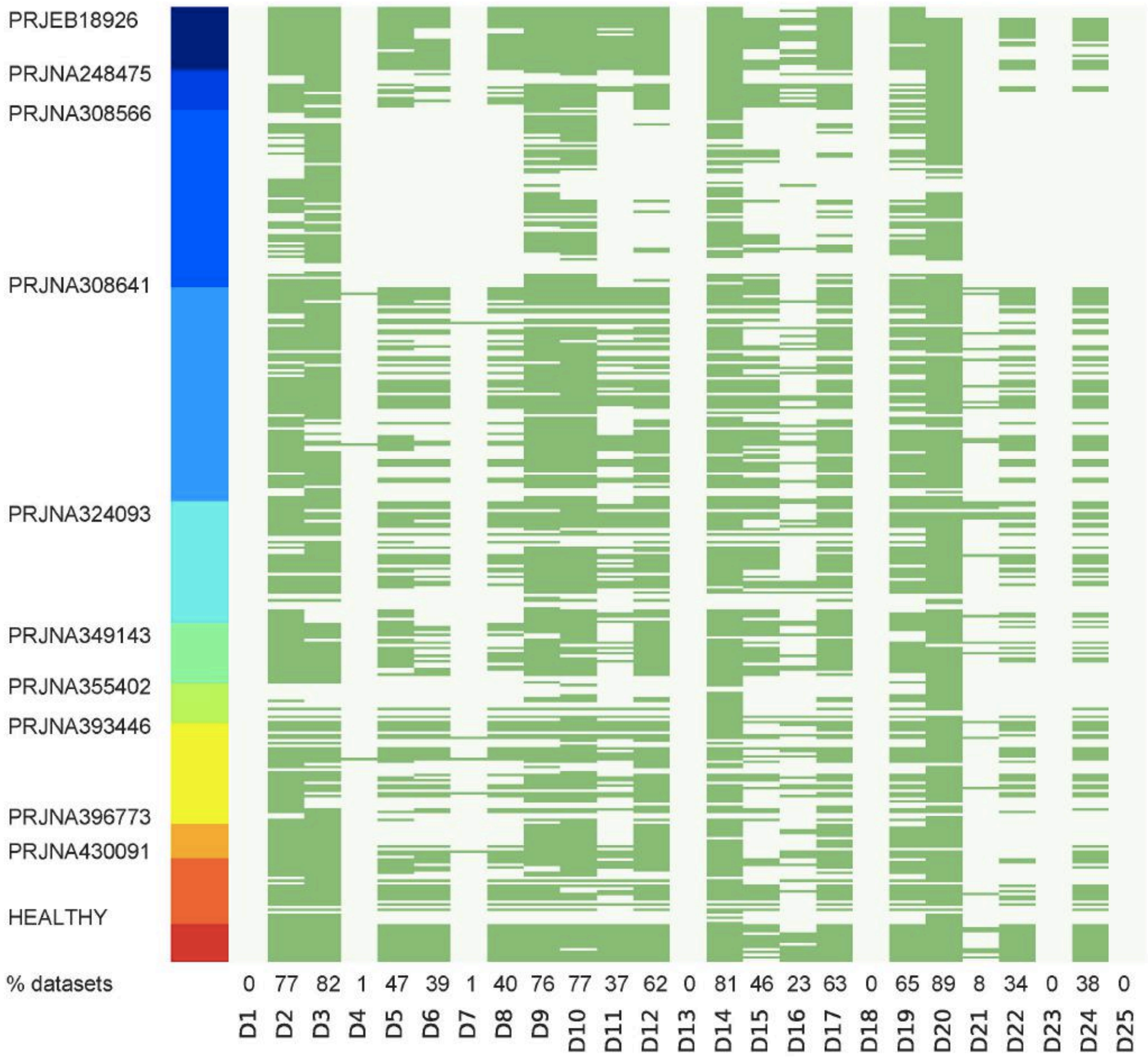

**Figure A14. Human D genes that were reconstructed (green cells) and missed (white cells) by IgScout in 361 immunosequencing datasets from ten NCBI projects (Table A8) and the HEALTHY datasets.** Datasets corresponding to the same NCBI project are grouped together and shown by a colored bar on the left. 14 HEALTHY datasets are shown at the bottom of the table. The percentage of the datasets supporting inference of each of 25 human D genes is shown in the row "% datasets".

## Supplemental Note: How IgScout results are affected by the number of consensus CDR3s and cell types?

To evaluate how IgScout results are affected by the number of consensus CDR3s and cell types, we analyzed two NCBI immunosequencing projects containing 242 datasets with B cells sorted by their type and antigen specificity (Table A9).

| NCBI project | # datasets | description | analyzed B cells | analyzed tissues |
|---|---|---|---|---|
| PRJNA308566 | 142 | Hepatitis study | PBMC, HBsAg+ cells, HLA-DR+ plasma cells | blood |
| PRJNA324093 | 100 | Flu vaccination study (healthy donors) | PBMC, memory, naïve, HA+ ASCs, HA– ASCs | blood |

**Table A9. Information about the PRJNA308566 and PRJNA324093 projects with 242 immunosequencing datasets.** HBsAg+ / HLA-DR+ / HA+ refer to cells with positive response to HBsAg, HLA-DR, and hemagglutinin, respectively. HA– refers to cells with negative response to hemagglutinin. ASC refers to antibody secreting cells.

Some of the datasets from PRJNA308566 and PRJNA308566 projects are characterized by a low number of consensus CDR3s (< 5000). Such low-diversity datasets likely correspond to situations when one clonal lineage (or a few clonal lineages) has an extremely high abundance as compared to all other clonal lineages. Since IgScout was not designed to analyze such datasets, it did not reconstruct any D genes in 75 datasets from PRJNA308566 and 49 datasets from PRJNA308566. Figure A15 presents the summary of IgScout results for all other datasets.

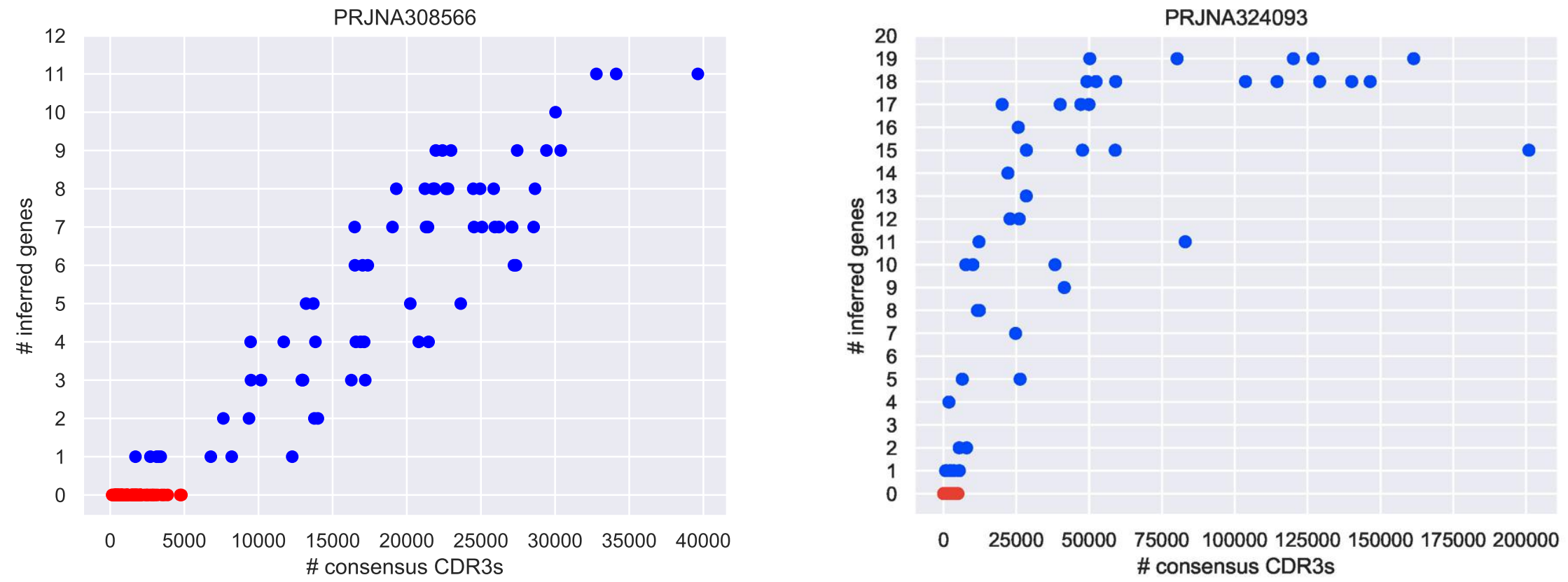

**Figure A15. The number of inferred D genes vs the number of consensus CDR3s for datasets from the PRJNA308566 (left) and PRJNA324093 (right) projects.** Each dataset corresponds to a single dot. Datasets without inferred D genes are shown as red dots.

We also analyzed how the type of cells in a dataset affects the IgScout results. Figure A16 shows that IgScout results depend mainly on the number of consensus CDR3s in a dataset rather than the type of B cells in this dataset. As Figure A16 illustrates, IgScout reconstructed D genes even from datasets corresponding to highly specific B cells (e.g., HBsAg+ or ASC-HA+). However, it is important to take into account that the number of consensus CDR3s is correlated with the number of different VDJ recombinations in a dataset. Thus, a small number of VDJ recombinations, occurring in datasets with highly specific B cells, may lead to a small number of inferred D genes.

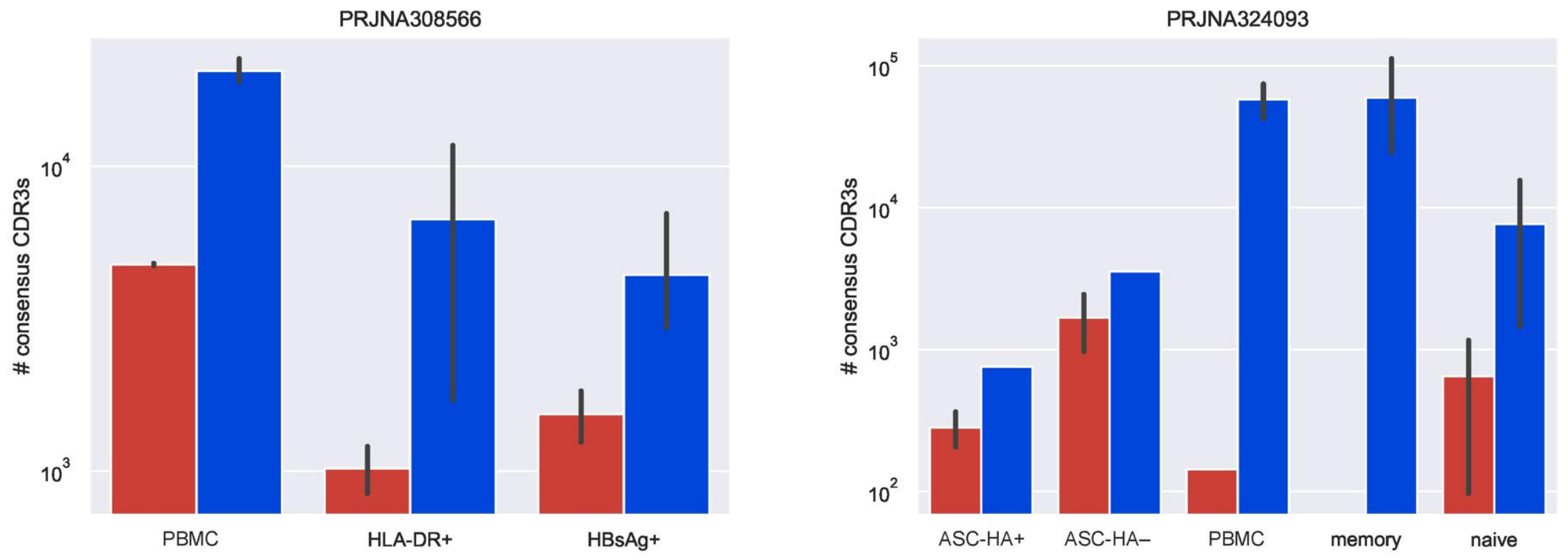

**Figure A16. IgScout results on datasets corresponding to various types of B cells in the PRJNA308566 (left) and PRJNA324093 (right) projects.** Each bar represents datasets corresponding to the same type of B cells. The height of a bar shows the average number of consensus CDR3s in these datasets (in the logarithmic scale), the error bars show the distribution of the numbers of consensus CDR3s. Red bars correspond to datasets where IgScout did not infer any D genes and blue bars correspond to datasets where IgScout inferred some D genes.

## Supplemental Note: Reconstructing camel D genes

IgScout reconstructed four D genes in the case of the Camel 1VH dataset that we refer to as D1, D2, D3, and D4 (Figure A17). IgScout reconstructed the same or very similar putative D genes in all camel datasets (Table A10) but missed D4 in datasets 2VHH, 3VH, and 3VHH (all camel D genes are shared between the VH and VHH datasets). Table A10 shows abundances of common 15-mers in the Camel 1VH dataset before and after the IgScout run.

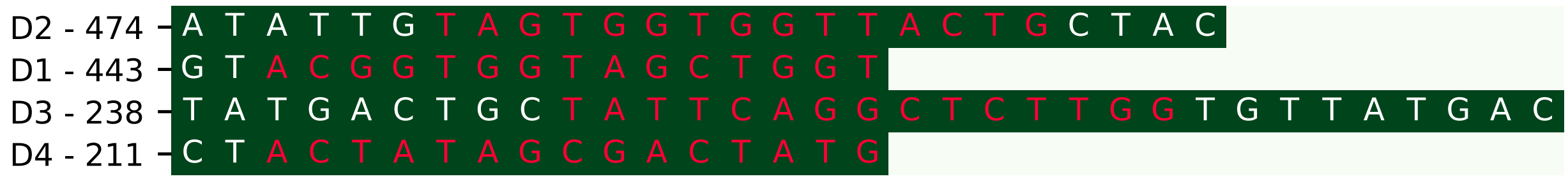

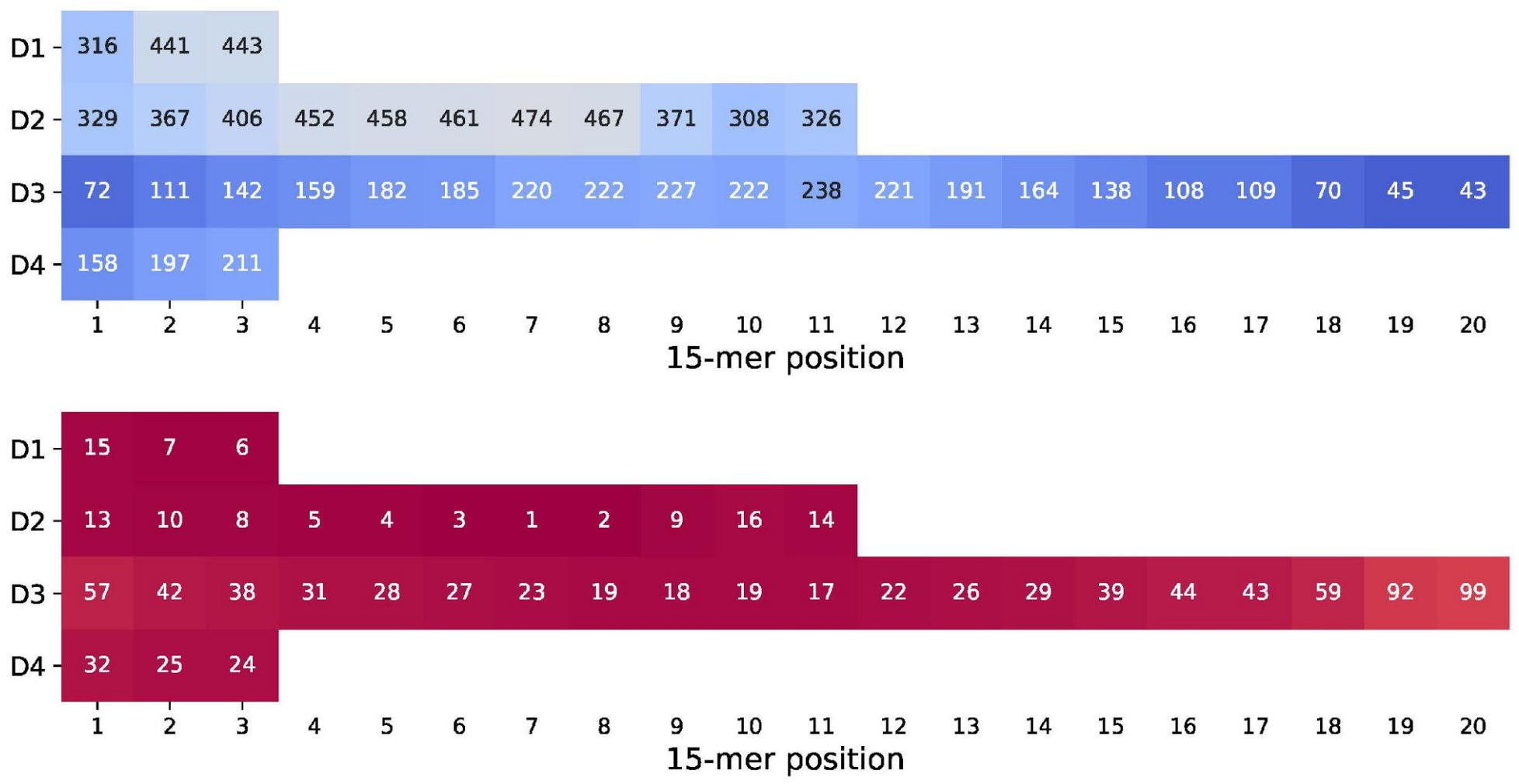

**Figure A17. Results of the IgScout algorithm on the Camel 1VH dataset.** Four inferred D genes in the Camel 1VH dataset (top), abundances (middle) and ranks (bottom) of 15-mers from the inferred camel D genes. Details of this visualization are described in the legends for Figure 2 and Figure A3.

| D1 | | D2 | |
|---|---|---|---|
| **Camel 1VH** | **GTACGGTGGTAGCTGGT** | **Camel 1VH** | **ATATTGTAGTGGTGGTTACTGCTAC** |
| Camel 1VHH | GTACGGTGGTAGCTGGT | Camel 1VHH | ATATTGTAGTGGTGGTTACTGCTAC |
| Camel 2VH | GTACGGTGGTAGCTGGT | Camel 2VH | CGCATACTATAGTGGTGGTTACTACTAC |
| Camel 2VHH | GTACGGTGGTAGCTGGT | Camel 2VHH | ATATTGTAGTGGTGGTTACTGC--- |
| Camel 3VH | GTACGGTGGTAGCTGGT | Camel 3VH | ATATTGTAGTGGTGGTTACTGCTAC |
| Camel 3VHH | GTACGGTGGTAGCTGGT | Camel 3VHH | CATATTGTAGTGGTGGTTACTGC--- |
| **D3** | | **D4** | |
| **Camel 1VH** | **TATGACTGCTATTCAGGCTCTTGGTGTTATGAC** | **Camel 1VH** | **CTACTATAGCGACTATG** |
| Camel 1VHH | GTATGACTGCTATTCAGGCTCTTGGTGTTATGAC | Camel 1VHH | CTACTATAGCGACTATG |
| Camel 2VH | GTATGACTACTGTTCAGGCTCTTGGTGTTATG-- | Camel 2VH | CTACTATAACGAATATGAC |
| Camel 2VHH | GTATGACTGCTATTCAGGCTCTTGGT-------- | | |
| Camel 3VH | TATGACTGCTATTCAGGCTCTTGGTGTTATG-- | | |
| Camel 3VHH | -ATGACTGCTATTCAGGCTCTTGGTG------- | | |

**Table A10. Constructing four putative camel D genes.** Strings inferred for the Camel 1VH are shown in bold. Differences between the strings inferred for the Camel 1VH dataset and strings inferred from other datasets are shown in red.

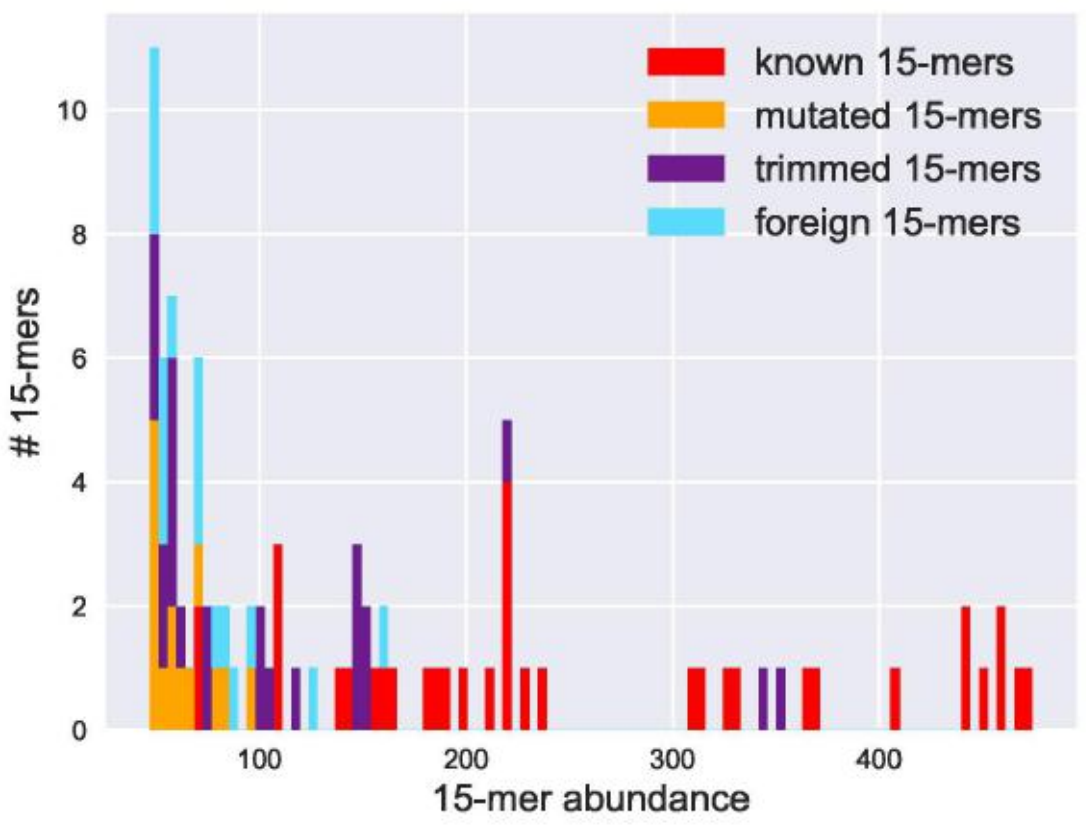

**Figure A18. Abundances of all common 15-mers in the Camel 1VH dataset.** 89 common 15-mers in the Camel 1VH dataset have abundances varying from 48 to 474. The *y*–axis represents the number of common 15-mers with the given abundance. Red, yellow, and violet bars represent the number of common 15-mers with given abundance among known, mutated, and trimmed 15-mers. There exist 35 known (red bars), 14 mutated (orange bars), 24 trimmed (violet bars), and 16 foreign (blue bars) common 15-mers.

Table A11 provides information about the fraction of traceable and tandem CDR3s in various camel datasets. Figure A19 provides information about the usage of four inferred camel D genes. Although all four D genes occur in both VH and VHH antibodies, their usage varies depending on the antibody type. In *(35)*, authors hypothesized that the same D and J genes are used for forming both the VH and VHH camel antibodies. If this hypothesis is correct, then the variations in the usage of D genes in the VH and VHH antibodies are most likely caused by differences between the RSSs in the V genes in the VH and VHH loci.

| **dataset** | **traceable CDR3s** | | **tandem CDR3s** | | **non-traceable CDR3s** | |
|---|---|---|---|---|---|---|
| | **# (%)** | **avg. length** | **# (%)** | **avg. length** | **# (%)** | **avg. length** |
| Camel 1VH | 10,224 (22%) | 55 | 176 (0.4%) | 69 | 35,926 (77.6%) | 54 |
| Camel 1VHH | 8443 (21.2%) | 60 | 222 (0.6%) | 73 | 31,036 (78.2%) | 60 |
| Camel 2VH | 12,158 (21%) | 54 | 183 (0.3%) | 70 | 45,777 (78.7%) | 53 |
| Camel 2VHH | 17,356 (23.3%) | 56 | 1292 (1.7%) | 61 | 56,198 (76%) | 54 |
| Camel 3VH | 17,289 (21.8%) | 51 | 1124 (1.4%) | 59 | 60,969 (76.8%) | 46 |
| Camel 3VHH | 13546 (23%) | 56 | 1068 (1.8%) | 62 | 44708 (75.2%) | 53 |

**Table A11. Classification of CDR3s across six camel immunosequencing datasets.** Since only a small percentage of CDR3s in camel immunosequencing datasets contains 15-mers from the inferred camel D genes, we defined a traceable CDR3 as a CDR3 that include 15-mers with up to two mutations from 15-mers from the inferred camel D genes.

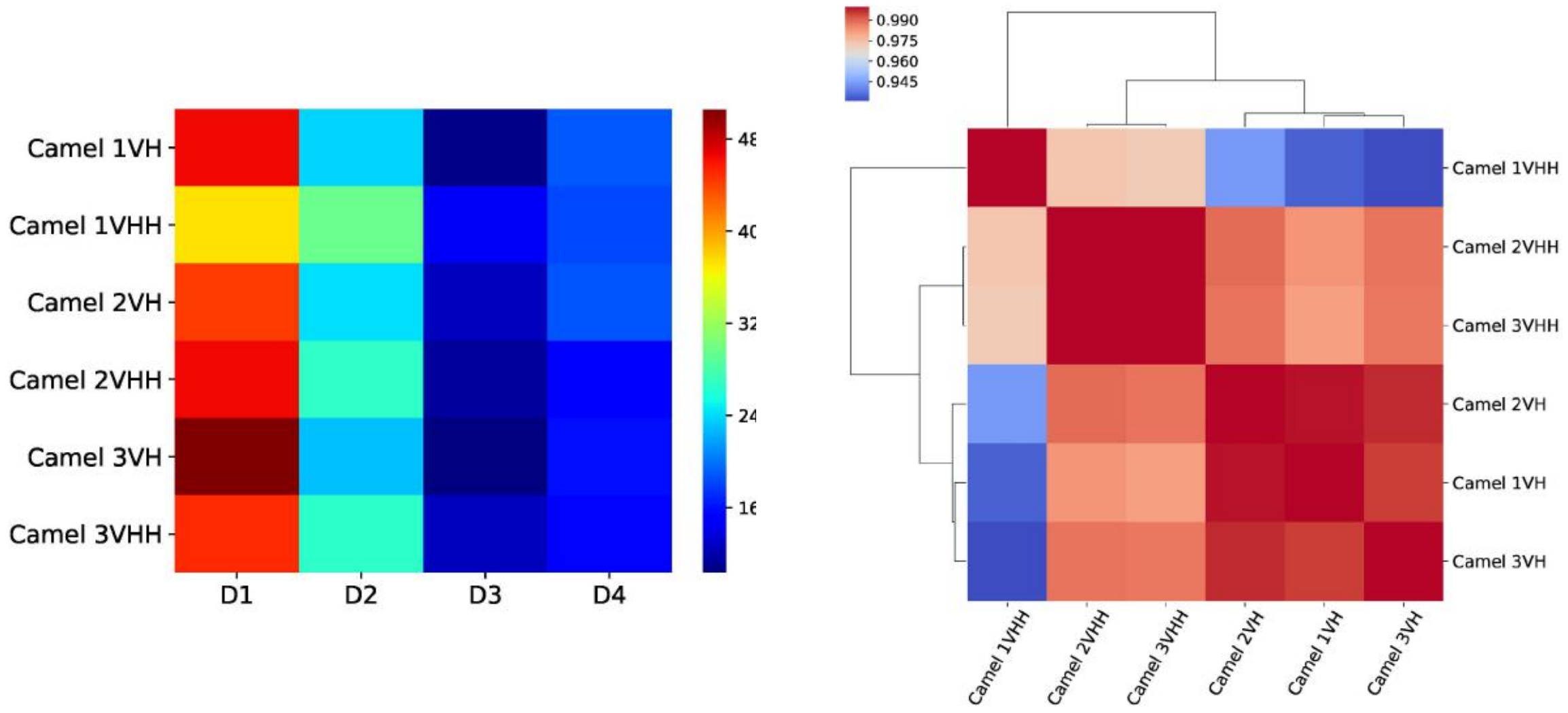

**Figure A19. Usage of four camel D genes across six camel datasets (left) and the similarity matrix of camel datasets constructed based on usages of their D genes (right).**

Figure A20 presents comparison between the four camel D genes and eight alpaca D genes listed in the IMGT database (camel and alpaca split ≈16 million years ago). For each camel D genes, there exists a similar alpaca D gene with percent identity varying from 82% to 94%.

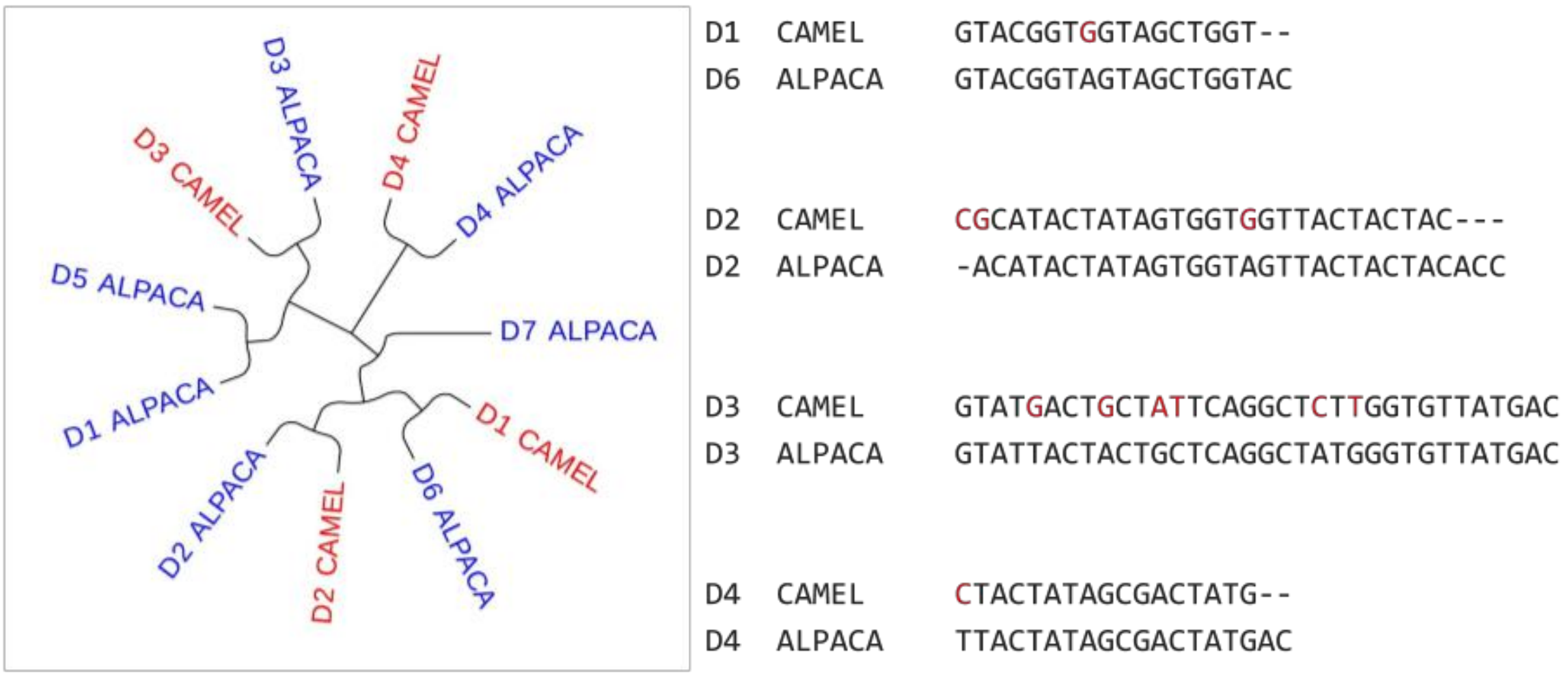

**Figure A20**. **Comparison of four camel D genes with eight alpaca D genes.** (Left) Phylogenetic tree for combined camel (blue) and alpaca (red) D genes. (Right) Alignment of four pairs of similar camel and alpaca genes. Differences between camel and alpaca D genes are shown in red.

## Supplemental Note: Traceable CDR3s

Figure A21 illustrates the usage of all human D genes across all HEALTHY datasets. Table A12 illustrates that the percentage of traceable (tandem) CDR3s varies from 43% to 55% (0.1 – 0.2%) across all HEALTHY datasets. The average length of traceable, tandem, and non-traceable CDR3s is 53, 71, and 40 nucleotides, respectively.

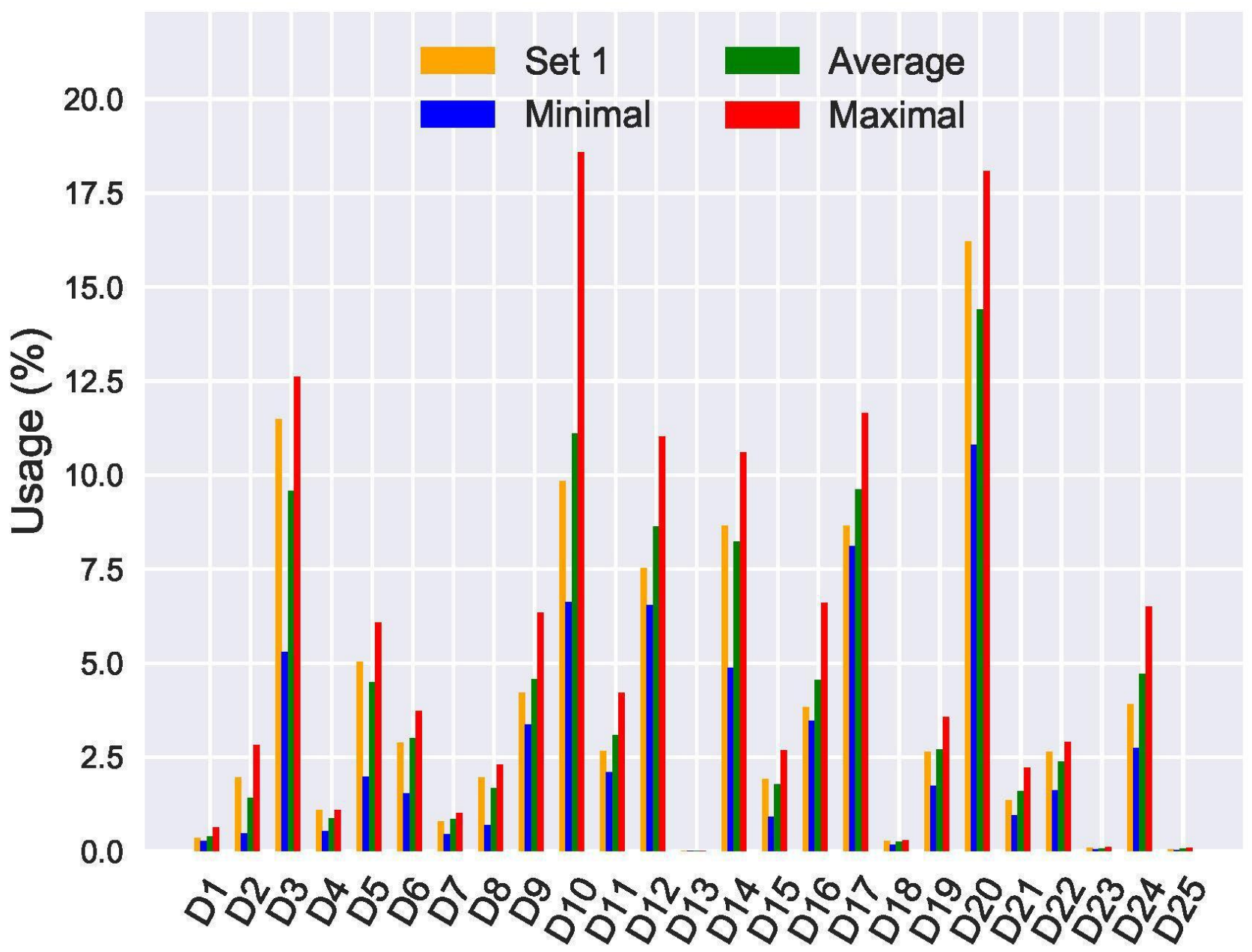

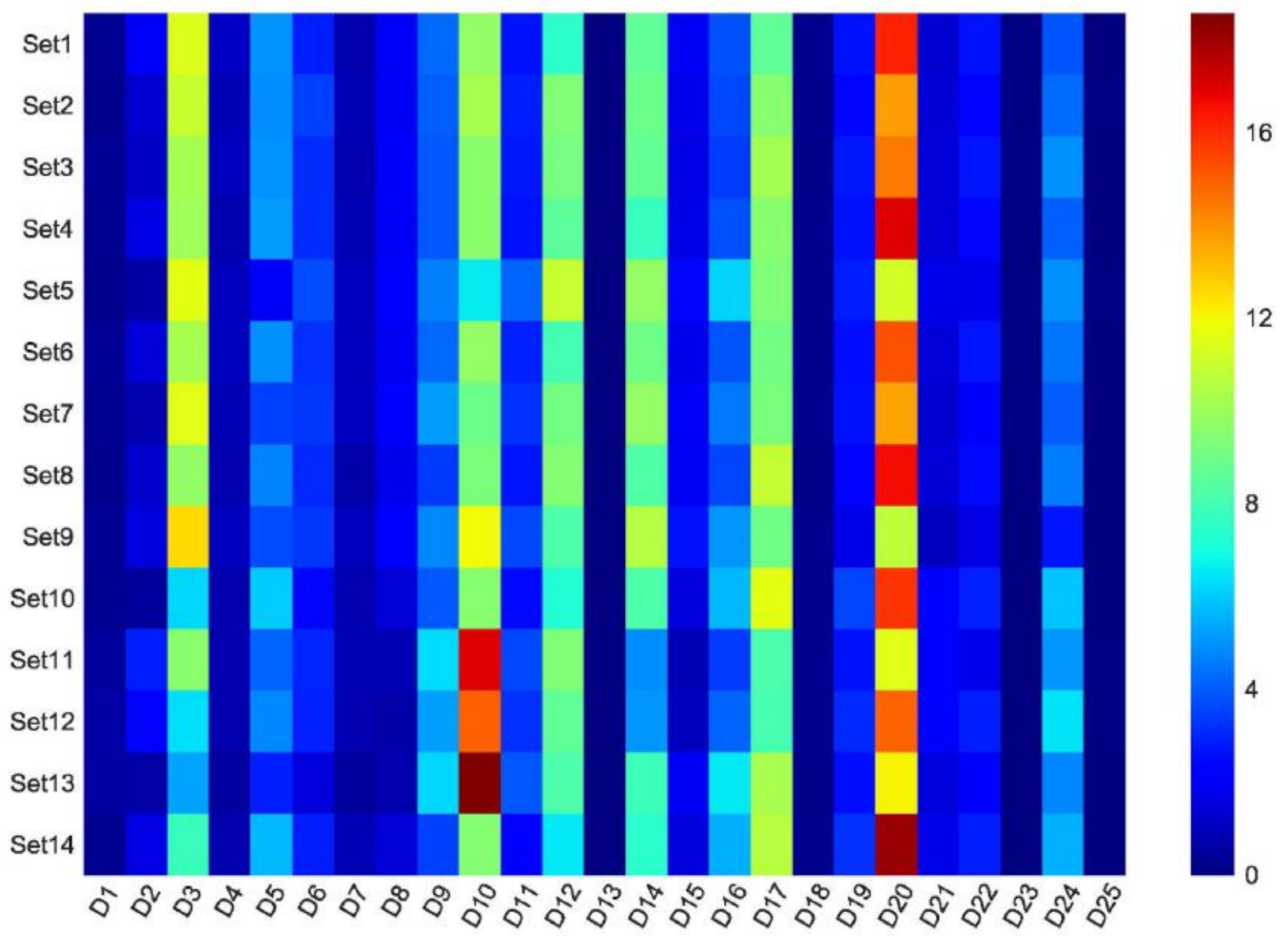

**Figure A21. Usage of human D genes across all HEALTHY datasets.** (Top) Usage of human D genes in the Set 1 dataset (yellow bars) compared to the minimal (blue bars), average (green bars), and maximal (red bars) usage of D genes across all HEALTHY datasets. (Bottom) Each cell shows the percentage of CDR3s formed by the corresponding D gene (*x*-axis) in the corresponding dataset (*y*-axis).

| dataset | traceable CDR3s | | tandem CDR3s | | non-traceable CDR3s | |
|---|---|---|---|---|---|---|
| | **# (%)** | **avg. length** | **# (%)** | **avg. length** | **# (%)** | **avg. length** |
| Set 1 | 37938 (46%) | 54 | 114 (0.1%) | 72 | 44528 (54%) | 49 |
| Set 2 | 34768 (46%) | 54 | 161 (0.2%) | 71 | 40470 (54%) | 48 |
| Set 3 | 14492 (44%) | 53 | 45 (0.1%) | 70 | 18552 (56%) | 48 |
| Set 4 | 47764 (50%) | 53 | 159 (0.1%) | 73 | 47296 (60%) | 48 |
| Set 5 | 54997 (46%) | 53 | 145 (0.1%) | 68 | 63600 (54%) | 47 |
| Set 6 | 34900 (46%) | 54 | 122 (0.1%) | 71 | 40886 (54%) | 48 |
| Set 7 | 54180 (43%) | 53 | 123 (0.1%) | 70 | 70435 (56%) | 47 |
| Set 8 | 31072 (46%) | 53 | 94 (0.1%) | 70 | 36157 (54%) | 48 |
| Set 9 | 68664 (52%) | 54 | 263 (0.2%) | 72 | 63563 (48%) | 49 |
| Set 10 | 56873 (55%) | 52 | 127 (0.1%) | 69 | 45700 (44%) | 47 |
| Set 11 | 38381 (45%) | 53 | 87 (0.1%) | 70 | 45916 (54%) | 48 |
| Set 12 | 54674 (48%) | 51 | 110 (0.1%) | 70 | 60195 (52%) | 46 |
| Set 13 | 64696 (50%) | 52 | 109 (0.1%) | 70 | 65721 (50%) | 46 |
| Set 14 | 63610 (53%) | 52 | 151 (0.1%) | 72 | 55933 (47%) | 47 |

**Table A12. Information about traceable, tandem, and non-traceable CDR3s across all HEALTHY datasets.**

## Supplemental Note: D gene classification by IgScout and IgBlast

We compared the D gene classification results generated by IgScout and IgBlast. Since IgBlast computes alignments for full-length immunoglobulin sequences, we analyzed raw reads for the Set1 dataset. 1,414,503 out of 1,611,497 reads (87%) were classified as CDR3-containing reads by both IgBlast and DiversityAnalyzer.

We classify a CDR3 as *non-traceable* if IgBlast reports several best D hits with the same alignment score. 550,514 out of 1,414,503 CDR3s (39%) are non-traceable. We also discarded 287,881 CDR3s (20%) because the D hits found by IgBlast are short and thus unreliable (shorter than 11 nt). For the remaining 576,108 CDR3s with putative D hits, we compared hits reported by IgBlast with hits reported by IgScout. For 504,028 out of 576,108 CDR3s (87%), IgBlast and IgScout report identical D hits. For 4613 CDR3s, IgScout reported tandem D genes (1%). The vast majority of the remaining 12% of CDR3s (where IgBlast and IgScout disagreed) correspond to similar D genes (e.g., the 31-nucleotide long IGHD3-22 and IGHD3-9 that share a 7-mer and a 9-mer). In this case, different scoring schemes produce slightly different results and it is not clear how to select the best one.

## Supplemental Note: Analysis of tandem CDR3s

IgScout identified 1900 tandem CDR3s in fourteen immunosequencing datasets corresponding to 225 distinct pairs of D genes (*D-pairs*). For each D-pair, we define the *D-pair abundance* as the number of tandem CDR3s formed by D genes in the pair and classify *abundant D-pairs* as the D-pairs with abundances exceeding 1% of the number of all tandem CDR3s. 27 abundant D-pairs include 15 D genes and form 916 out of 1900 tandem CDR3. Figure A22 presents a graph with 16 vertices corresponding to D genes participating in abundant D-pairs (gene D5 corresponds to two vertices since it occurs twice in the IGH locus) and 27 edges (corresponding to abundant D-pairs). This graph turned out to be an acyclic directed graph and its topological order is the same as the order of D genes in the IGH locus. Thus, our analysis agrees with conclusion in *(26)* that the order of D genes forming tandem CDR3s follows their order in the IGH locus.

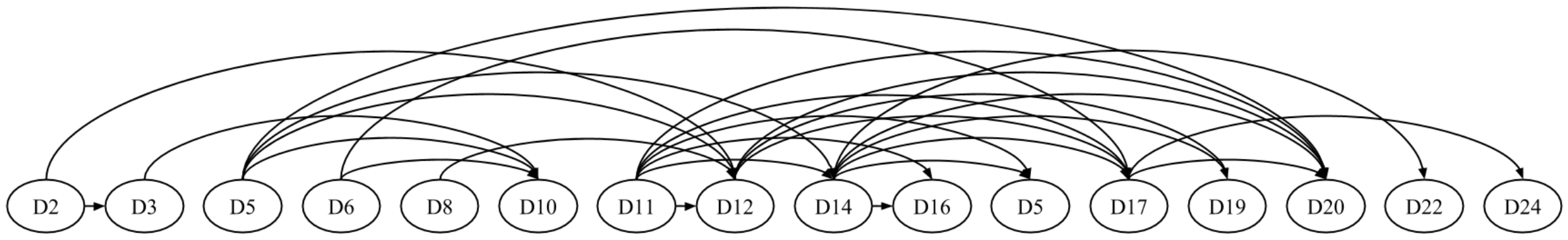

**Figure A22. Graph on 16 vertices and 27 edges corresponding to abundant D-pairs.** Each abundant D-pair is represented by an edge from its first D gene to its second D gene.

We compared usage of D genes in traceable CDR3s with usage of the first and second D genes in D-pairs. Six D genes with high usage (>5%) in traceable CDR3s (D3, D10, D12, D14, D17, and D20) also have high usage (>5%) in tandem CDR3s (Figure A23). However, eight abundant D genes in tandem CDR3s (D2, D5, D6, D8, D11, D16, D19, and D24) are not abundant in traceable CDR3s.

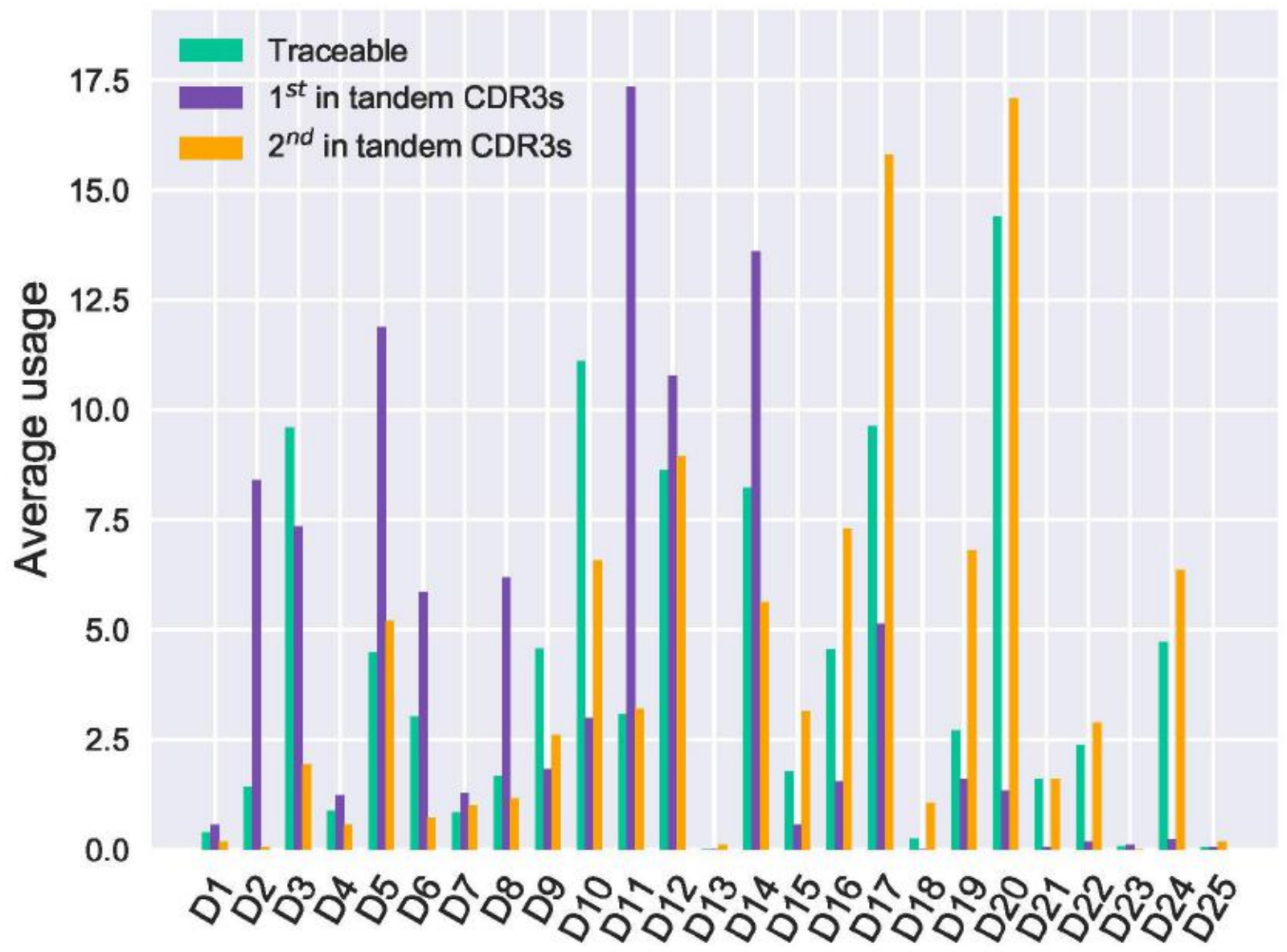

**Figure A23. Usage of known D genes in traceable CDR3s (green) and tandem CDR3s as the 1st (purple) and 2nd (orange) D gene forming a tandem CDR3.** The average usage was computed across usages in all HEALTHY datasets.

Table A13 presents the tandem bias and fraction of tandem CDR3s for all HEALTHY datasets. We classify a pair (*D*, *D'*) of D genes as *direct* (*reverse*) if all occurrences of *D* precede (follow) all occurrences of *D'* in the IGH locus. A pair (*D*, *D'*) is classified as *ambiguous* if it is neither direct, not reverse. Note that only pairs including D4 or D5 gene (that have two copies in the IGH locus) can be classified as ambiguous. We classify a tandem CDR3 as direct/reverse/ambiguous if it is formed by direct/reverse/ambiguous pair of D genes. The average percentages of direct, reverse and ambiguous CDR3s across all HEALTHY datasets are 82%, 6%, and 12%.

| dataset | tandem bias | % of tandem CDR3s | dataset | tandem bias | % of tandem CDR3s |
|---|---|---|---|---|---|
| Set 1 | 0.10 | 0.20 | Set 8 | 0.10 | 0.18 |
| Set 2 | 0.20 | 0.30 | Set 9 | 0.04 | 0.25 |
| Set 3 | 0.07 | 0.18 | Set 10 | 0.06 | 0.16 |
| Set 4 | 0.10 | 0.21 | Set 11 | 0.16 | 0.14 |
| Set 5 | 0.05 | 0.16 | Set 12 | 0.09 | 0.13 |
| Set 6 | 0.10 | 0.21 | Set 13 | 0.06 | 0.11 |
| Set 7 | 0.08 | 0.14 | Set 14 | 0.09 | 0.16 |

**Table A13. The tandem bias and the percentage of tandem CDR3s for all HEALTHY datasets.** The column "% of tandem CDR3" shows the percentage of tandem CDR3 among all traceable CDR3s for each immunosequencing dataset.

Figure A24 demonstrates that the percentage and length of tandem CDR3s in the ALLERGY BM datasets is higher than in the ALLERGY PBMC and HIV PBMC datasets.

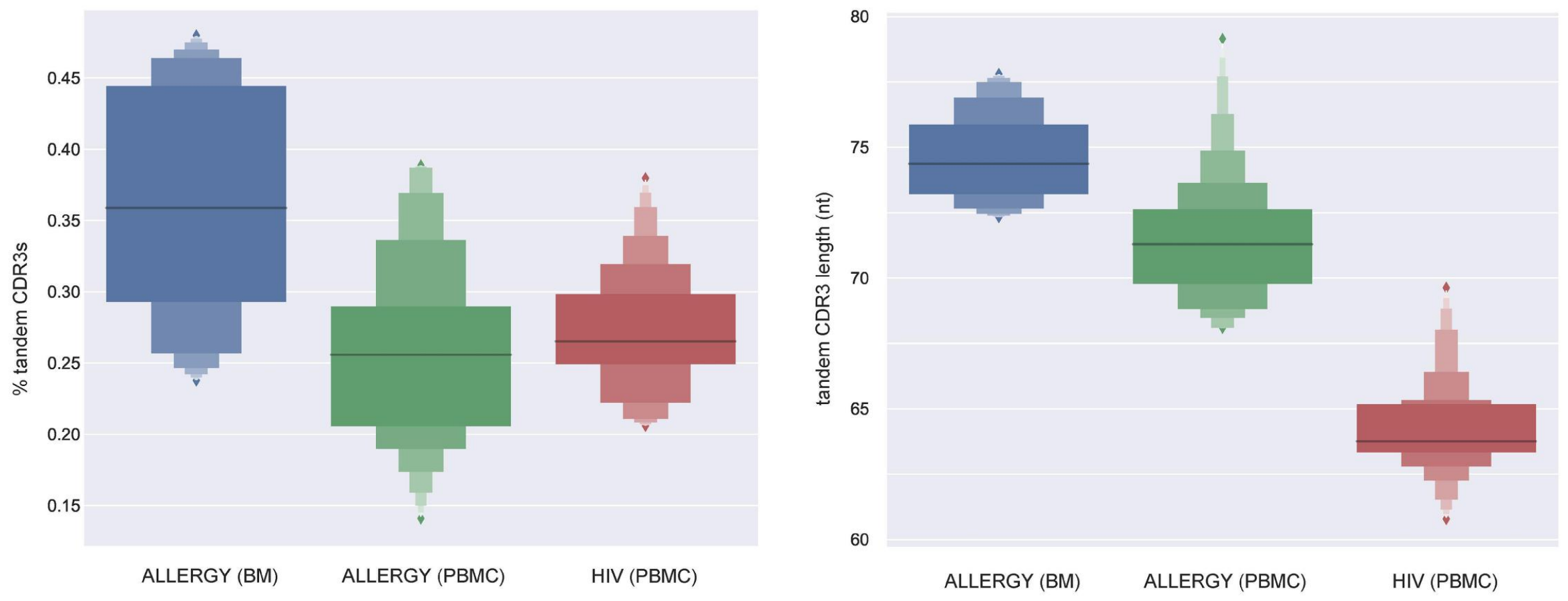

**Figure A24. The percentages (left) and lengths (right) of tandem CDR3s in ALLERGY (bone marrow), ALLERGY (PBMC), and HIV (PBMC) datasets.**

Figure A25 shows the tandem matrix constructed based on pairs of D genes forming tandem CDR3s in 15 datasets corresponding to the hepatitis patient 1776 analyzed in *(37)*. The large number of entries in the D20 row in the lower part of this matrix suggests that the D20 gene is duplicated in this patient.

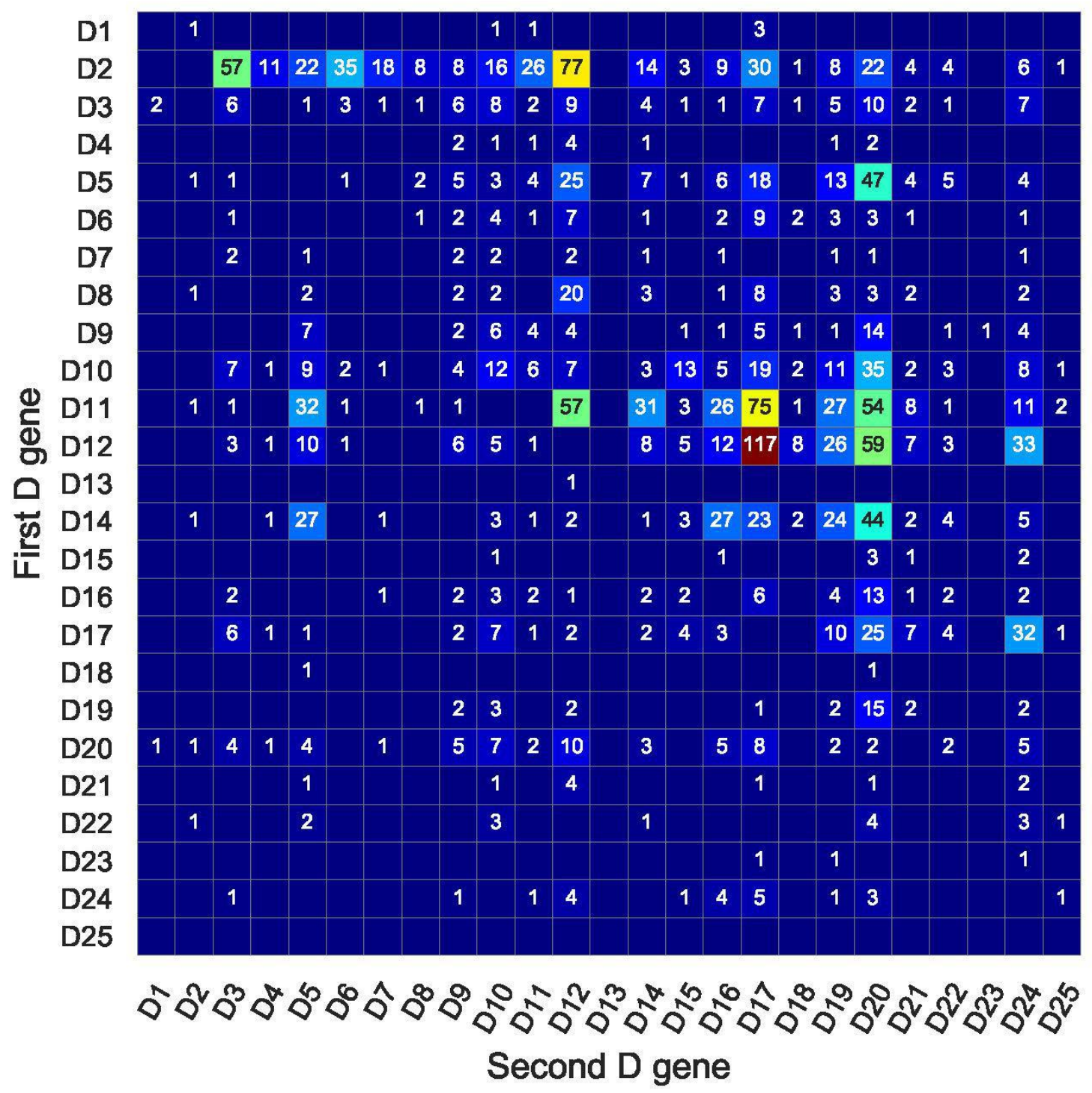

**Figure A25. The tandem matrix for D genes forming tandem CDR3s in the datasets corresponding to the hepatitis patient 1776 *(37)*.**

## Supplemental Note: Ultra-long tandem CDR3s

We analyzed inter-D insertions in all 1900 tandem CDR3s across all HEALTHY datasets. These tandem CDR3s contain 1081 distinct inter-D insertions, varying in length from 0 to 153 nucleotides (Figure A26). 384 of them have length at least 10 nucleotides. Since most of them do not share significant similarities, they likely correspond to randomly generated sequences.

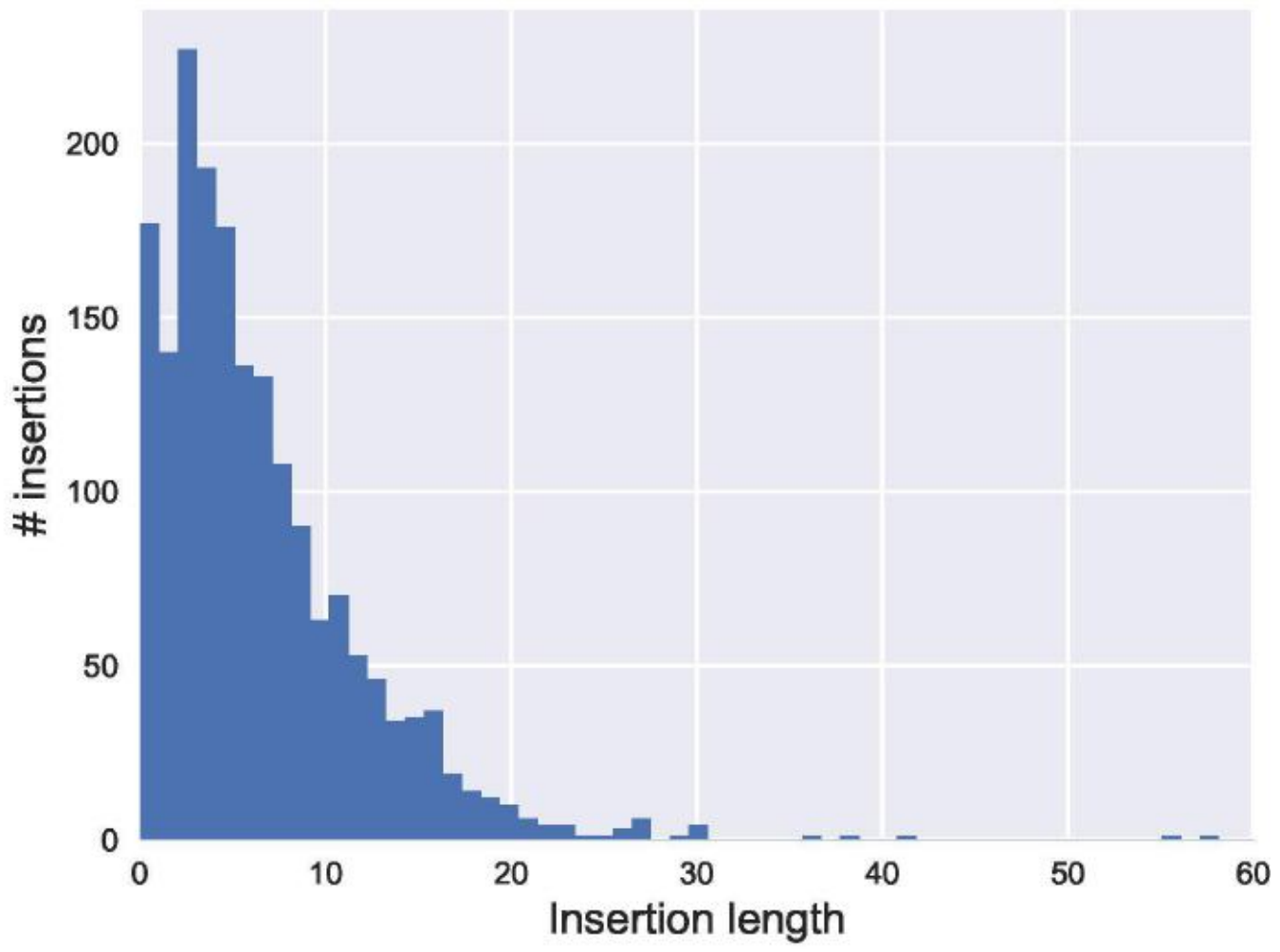

**Figure A26. Distribution of lengths of inter-D insertions in tandem CDR3s across all HEALTHY datasets.** Two ultra-long inter-D insertions (of length 153 nucleotides) are not shown.

Two longest inter-D insertions (denoted $I_1$ and $I_2$) appear in the Set 1 and have length 153 nucleotides. They are formed by genes D9 and D10, differ by a single nucleotide, and appear in CDR3s differing by six nucleotides (Figure A27, top). The inter-D insertion $I_2$ starts with the right RSS of D9 and ends with the left RSS of D10 (Figure A27, bottom).

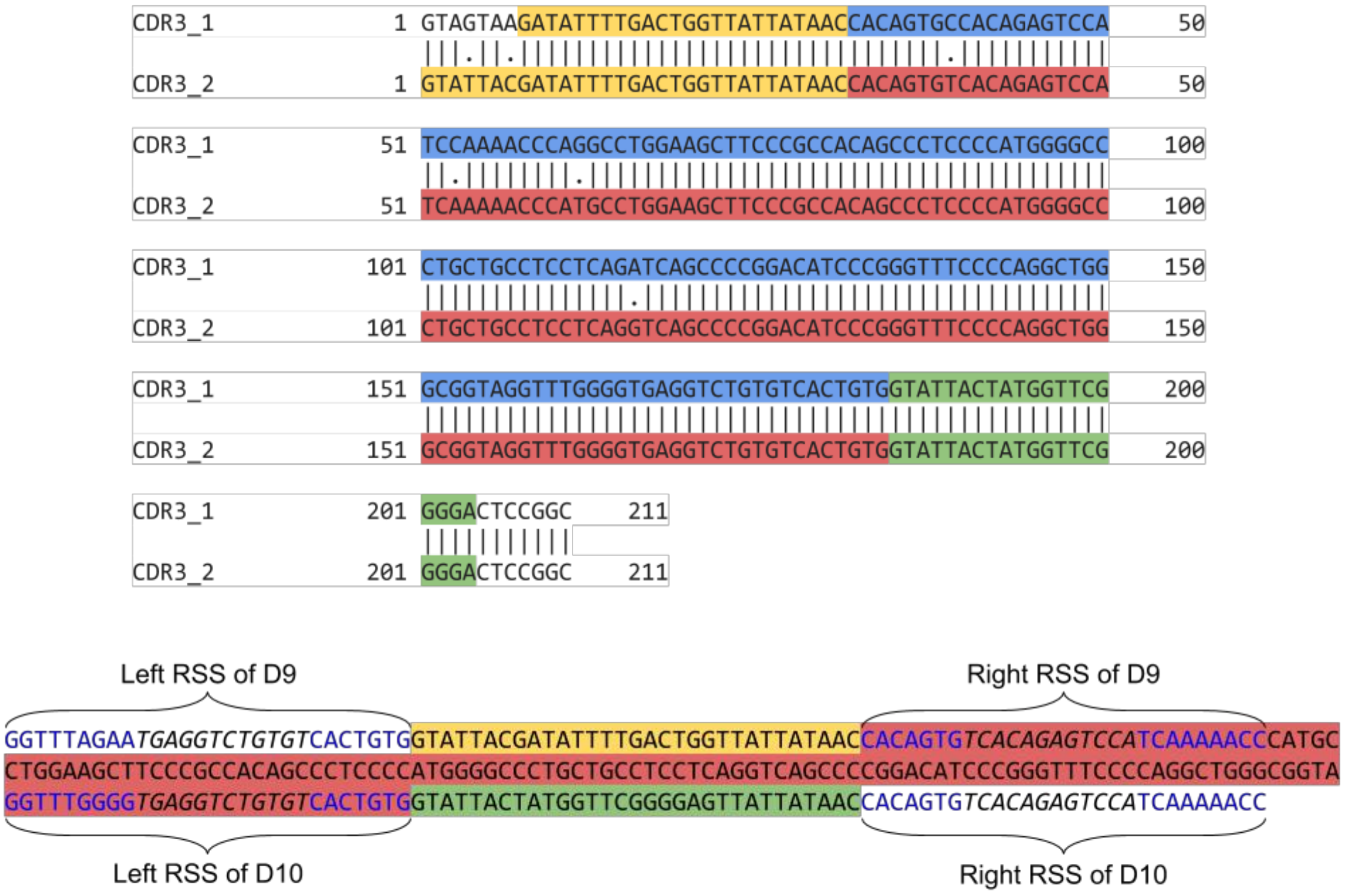

**Figure A27. Ultra-long tandem CDR3s.** (Top) Alignment of tandem CDR3s formed by genes D9 (yellow) and D10 (green) and containing ultra-long inter-D insertions $I_1$ (blue) and $I_2$ (red). (Bottom) Fragment of the IGH locus starting with the left RSS sequence of the D9 gene and ending with the right RSS sequence of the D10 gene. Left RSS sequences are shown as concatenates of a nonamer (shown in blue), a 12-nucleotide long spacer (shown in italic), and a heptamer (shown in blue). Right RSS sequences are shown as concatenates of a heptamer (shown in blue), a 12-nucleotide long spacer (shown in italic), and a nanomer (shown in blue).

## Supplemental Note: *De novo* reconstruction of human J genes

All human J genes are located in a 2 kb long region in the human IGH locus (Table A14). Table A15 and Figure A28 show allelic variants of human J genes listed in the IMGT database.

| Name | IMGT name | Position (bp) | Length (nt) |
|---|---|---|---|
| J1 | IGHJ1 | 1,178,312 | 52 |
| J2 | IGHJ2 | 1,178,520 | 53 |
| J3 | IGHJ3 | 1,179,132 | 50 |
| J4 | IGHJ4 | 1,179,504 | 48 |
| J5 | IGHJ5 | 1,179,905 | 51 |
| J6 | IGHJ6 | 1,180,521 | 63 |

**Table A14. Positions and lengths of J genes on the 14th chromosome in the human genome.** Since the IGH locus starts at the end of the 14th chromosome, positions are given with respect to its reverse complementary sequence.

| J gene | IMGT allele | ID |
|---|---|---|
| J1 | IGHJ1*01 | J1 |
| J2 | IGHJ2*01 | J2 |
| J3 | IGHJ3*01 | J3* |
| | IGHJ3*02 | J3 |
| J4 | IGHJ4*01 | J4* |
| | IGHJ4*02 | J4 |
| | IGHJ4*03 | J4** |
| J5 | IGHJ5*01 | J5 |
| | IGHJ5*02 | J5* |
| J6 | IGHJ6*01 | J6* |
| | IGHJ6*02 | J6 |
| | IGHJ6*03 | J6** |
| | IGHJ6*04 | J6*** |

**Table A15. Information about variants of J genes and their correspondence with alleles listed in the IMGT database.**

```
J3     TGATGCTTTTGATATCTGGGGCCAAGGGACAATGGTCACCGTCTCTTCAG
J3*    TGATGCTTTTGATGTCTGGGGCCAAGGGACAATGGTCACCGTCTCTTCAG

J4     ACTACTTTGACTACTGGGGCCAGGGAACCCTGGTCACCGTCTCCTCAG
J4*    ACTACTTTGACTACTGGGGCCAAGGAACCCTGGTCACCGTCTCCTCAG
J4**   GCTACTTTGACTACTGGGGCCAAGGGACCCTGGTCACCGTCTCCTCAG

J5     ACAACTGGTTCGACTCCTGGGGCCAAGGAACCCTGGTCACCGTCTCCTCAG
J5*    ACAACTGGTTCGACCCCTGGGGCCAGGGAACCCTGGTCACCGTCTCCTCAG

J6     ATTACTACTACTACTACGGTATGGACGTCTGGGGCCAAGGGACCACGGTCACCGTCTCCTCAG
J6*    ATTACTACTACTACTACGGTATGGACGTCTGGGGGCAAGGGACCACGGTCACCGTCTCCTCAG
J6**   ATTACTACTACTACTACTACATGGACGTCTGGGGCAAAGGGACCACGGTCACCGTCTCCTCAG
J6***  ATTACTACTACTACTACGGTATGGACGTCTGGGGCAAAGGGACCACGGTCACCGTCTCCTCAG
```

**Figure A28. Allelic variants of human J genes.** Differences between various variants are highlighted in red.

*De novo* reconstruction of J genes requires immunosequencing reads that cover the entire J genes. This is not the case for many immunosequencing datasets, including all HEALTHY datasets. We benchmarked how IgScout reconstructs human J genes using the ALLERGY (rather than HEALTHY) datasets since reads in these datasets cover the entire J segment.

We combined four datasets corresponding to the first allergic donor and identified 5,940,059 fragments of J genes in reads from this dataset using IgReC, resulting in the set of strings $J_{trimmed}$. Afterwards, we applied IgScout to infer J genes from these strings. Since J genes are longer than D genes, we increased the parameter *k-mer size* from 15 to 40 (all 40-mers in J genes are unique, i.e., they appear in a single J gene). The human J genes (from J1 to J6) contain 83 40-mers (192 40-mers including their alleles listed in the IMGT database). The $J_{trimmed}$ dataset contains all 40-mers appearing in six human J genes.

We classify a *k*-mer as *known* if it occurs in a human J gene (from J1 to J6), *mutated* if it differs from a known *k*-mer by a single nucleotide, and *trimmed* if it contains a known (*k*-2)-mer. All other *k*-mers are classified as *foreign.* 41% of strings in the $J_{trimmed}$ dataset contain a known 40-mer. 43% strings in the $J_{trimmed}$ dataset contain either a known, or a mutated, or a trimmed 40-mer.

Since the number of J genes is smaller than the number of D genes (6 vs 25), we increased the *fraction* parameter to 0.02 for the case of the J gene finding, i.e., a *k*-mer is classified as *common* if its abundance exceeds 2% of the number of sequences in the $J_{trimmed}$ set. Figure A29 presents distribution of abundances of all 47 common 40-mers in the $J_{trimmed}$ set. Figure A30 shows that the usages of human J genes are similar for various ALLERGY datasets.

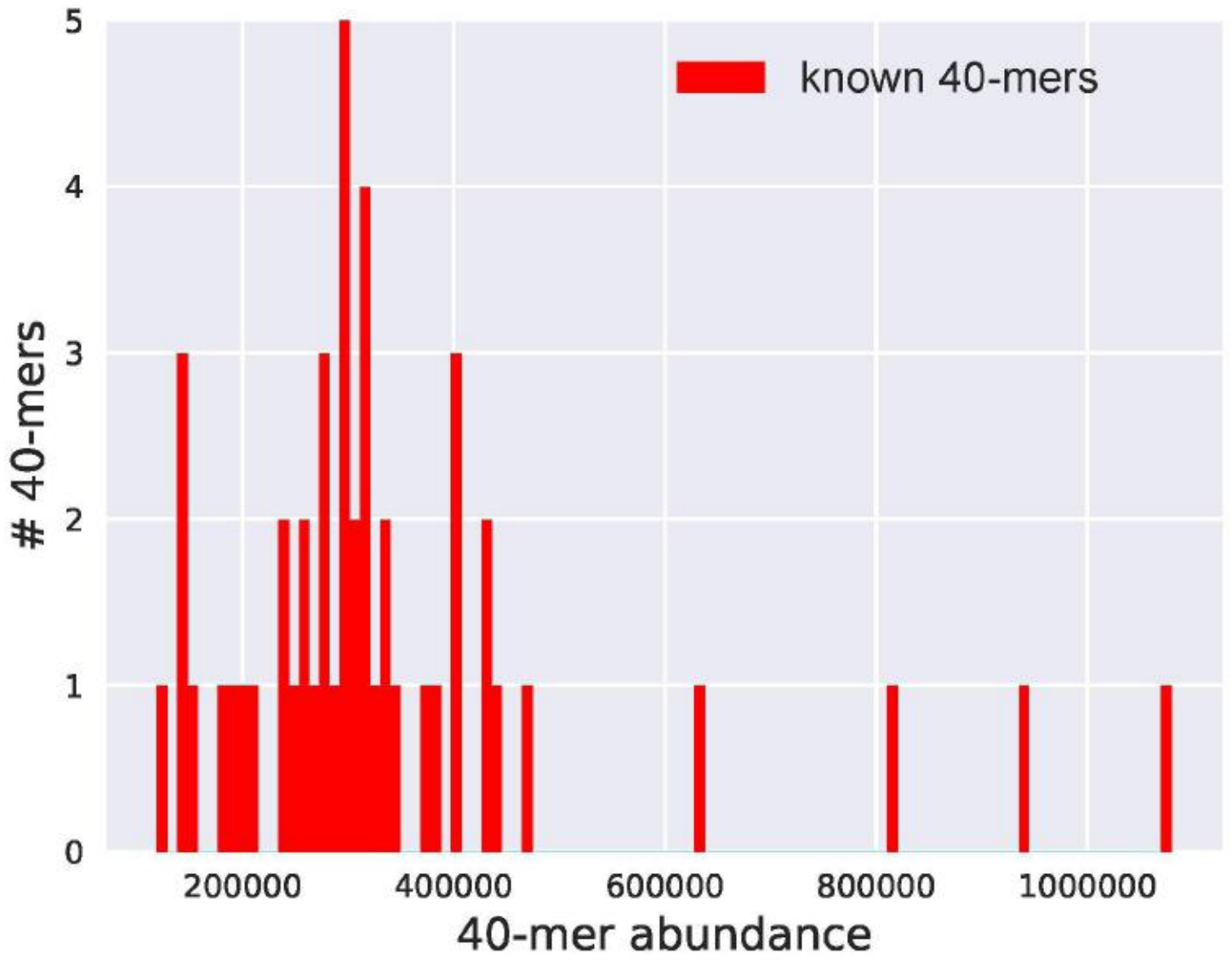

**Figure A29. Abundances of all 47 common 40-mers in the $J_{trimmed}$ set.** 47 common 40-mers in the $J_{trimmed}$ set have abundances varying from 119,082 to 1,079,233. The *y*–axis represents the number of common 40-mers with given abundance. All 47 common 40-mers are known (shown as red bars). Each bar represents the number of common 40-mers with given abundance. The histogram represents 29 bins of width 10,000 each.

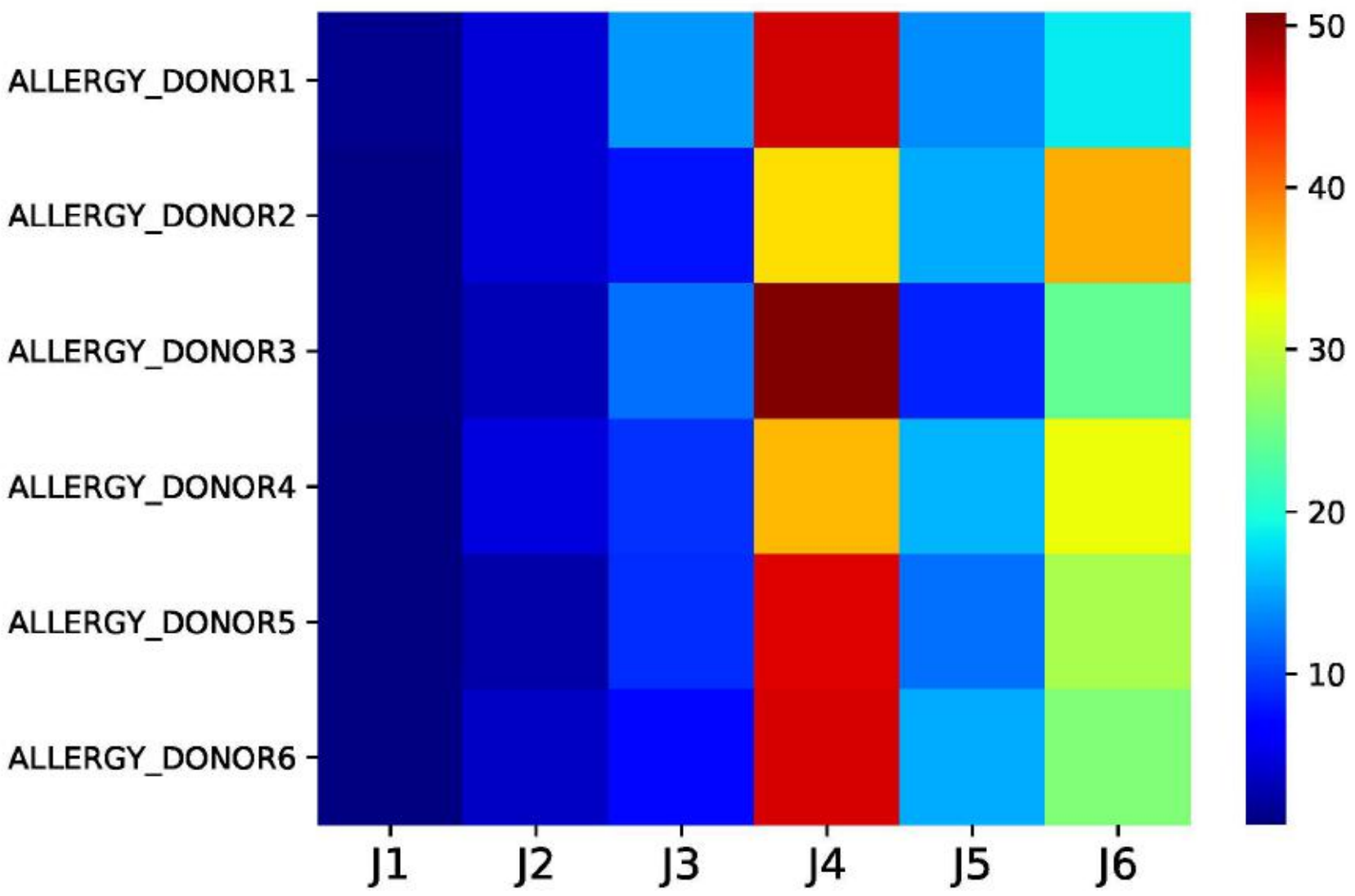

**Figure A30. Usage of human J genes in the ALLERGY datasets.** We merged four datasets corresponding to each of six ALLERGY donors and computed the J gene usage for each donor.

We applied IgScout to the $J_{trimmed}$ dataset with $k$ = 40. Ranks and abundances of known 40-mers are shown in Figure A31. IgScout reconstructed four strings representing the complete sequences of the J3, J4, J5, and J6 genes (Figure A32). The J1 and J2 genes were not reconstructed by IgScout since their most abundant 40-mers do not pass the *fraction* threshold (the most abundant 40-mer from the J1 and J2 genes are supported by 30,000 and 99,000 CDR3s, respectively.

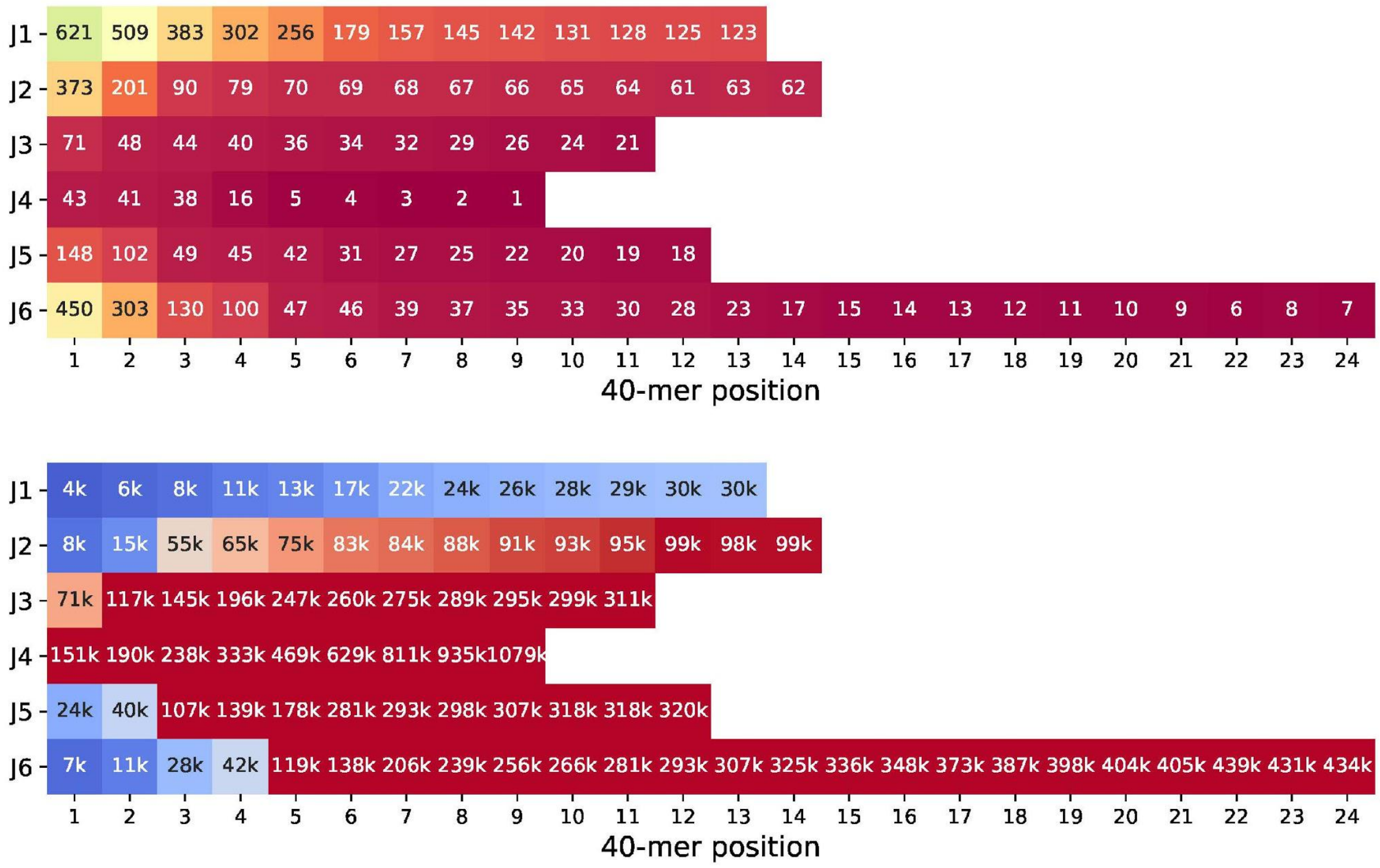

**Figure A31. Ranks (top) and abundances (bottom) of 40-mers from human J genes (*J<sub>trimmed</sub>* set).** Details of this visualization are described in the legend for Figure A3. Abundances exceeding 100,000 are shown in red.

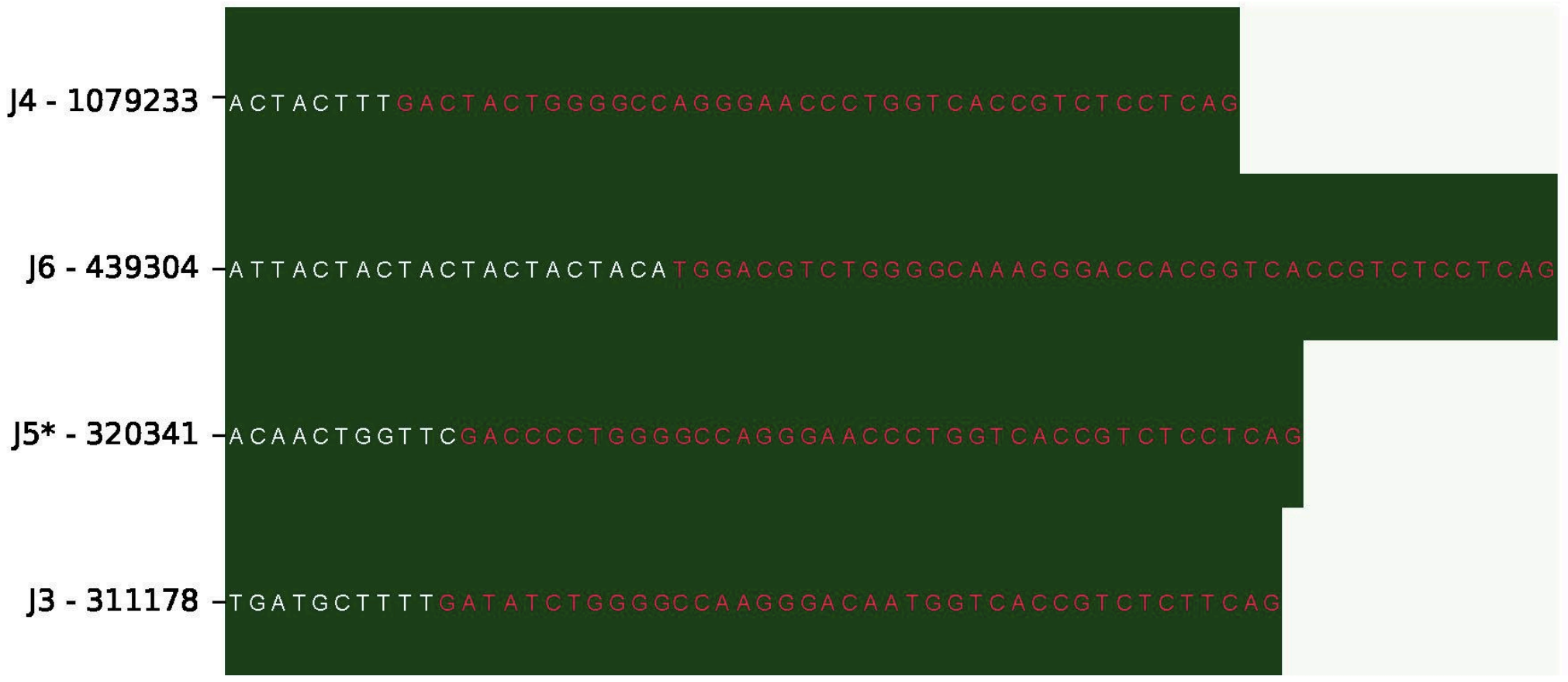

**Figure A32. Results of the IgScout algorithm on the $J_{trimmed}$ dataset.** Details of this visualization are described in the legend for Figure 2.

Since IgDiscover *(18)* is limited to *de novo* reconstruction of V genes, IMPre *(19)* is the only available tool for *de novo* reconstruction of J genes. Since IMpre demonstrates the best results on sequences trimmed by the ends of J genes, we trimmed suffixes of reads corresponding to constant regions. We applied IgScout and IMPre to 6 donors from the ALLERGY dataset and compared the J genes inferred by IgScout and IMPre (Figure A33).

We classify an inferred segment as *erroneous* if it was formed by an addition of incorrect nucleotides to the start of a J gene. IgScout reconstructed four out of six human J genes over their entire lengths (including the known variant J6**) and made no errors. IMPre reconstructed all six J genes (including the known variants J3* and J6**) but made seven errors. While IgScout reconstructs complete D genes, IMPre misses 2.3 nucleotides on average at the start of the reconstructed J genes.

Both IgScout and IMPre reported 3 novel variants of J genes each. However, since all these six variants are different, they are likely caused by frequent SHMs in J genes and thus require additional tuning of both tools for *de novo* reconstruction of J genes.

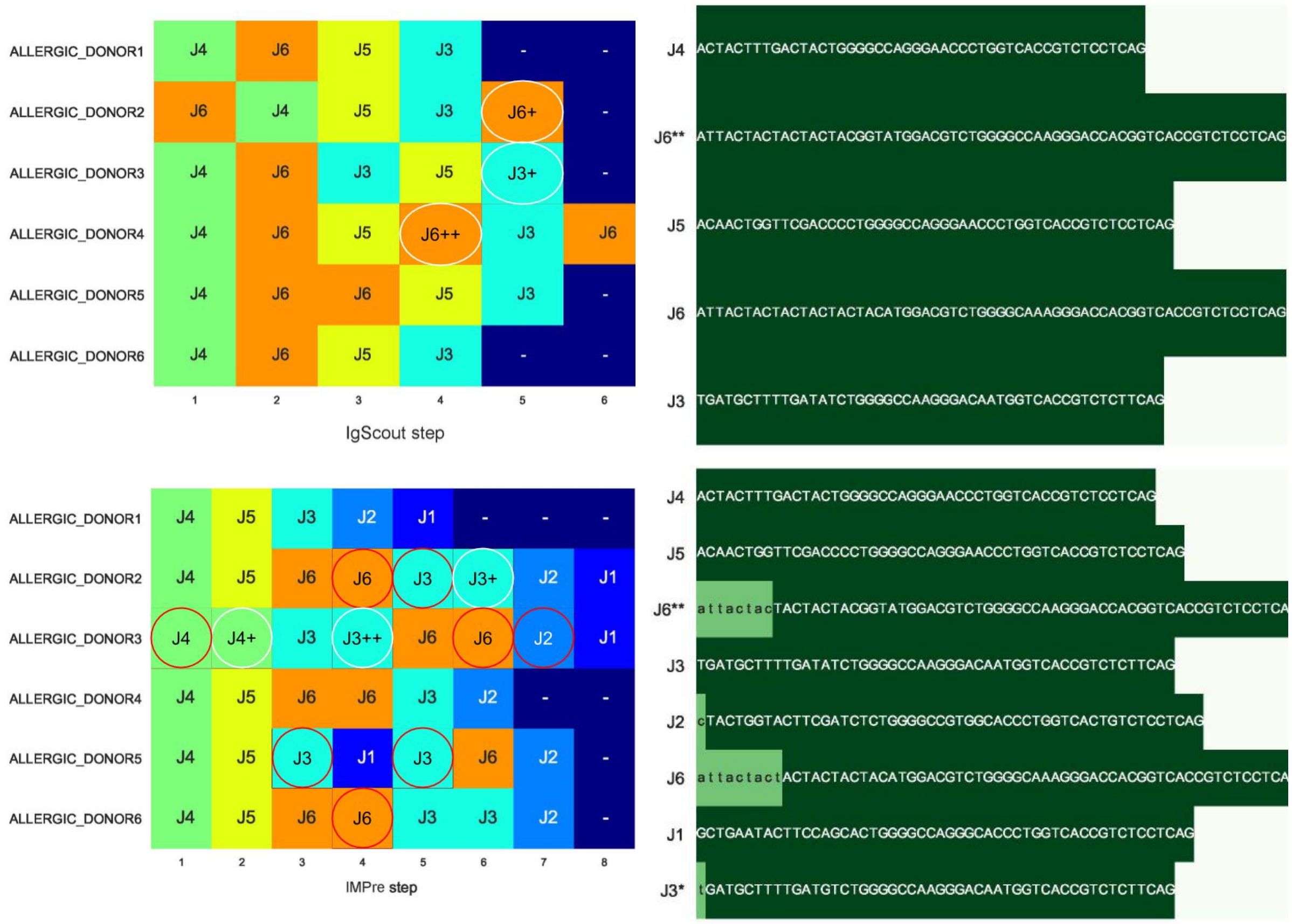

**Figure A33.** ***De novo*** **reconstructions of J genes for six ALLERGY patients using using IgScout (top) and IMPre (bottom).** Details of this visualization are described in the legends for Figure 3 and Figure A3. Some reported sequences represent inaccurately reconstructed J genes (e.g., a J gene with several added nucleotides at the start) that we represented using red circles.

## Supplemental Note: List of tandem CDR3s

Figure A34 lists all 114 tandem CDR3s in the Set 1 dataset.

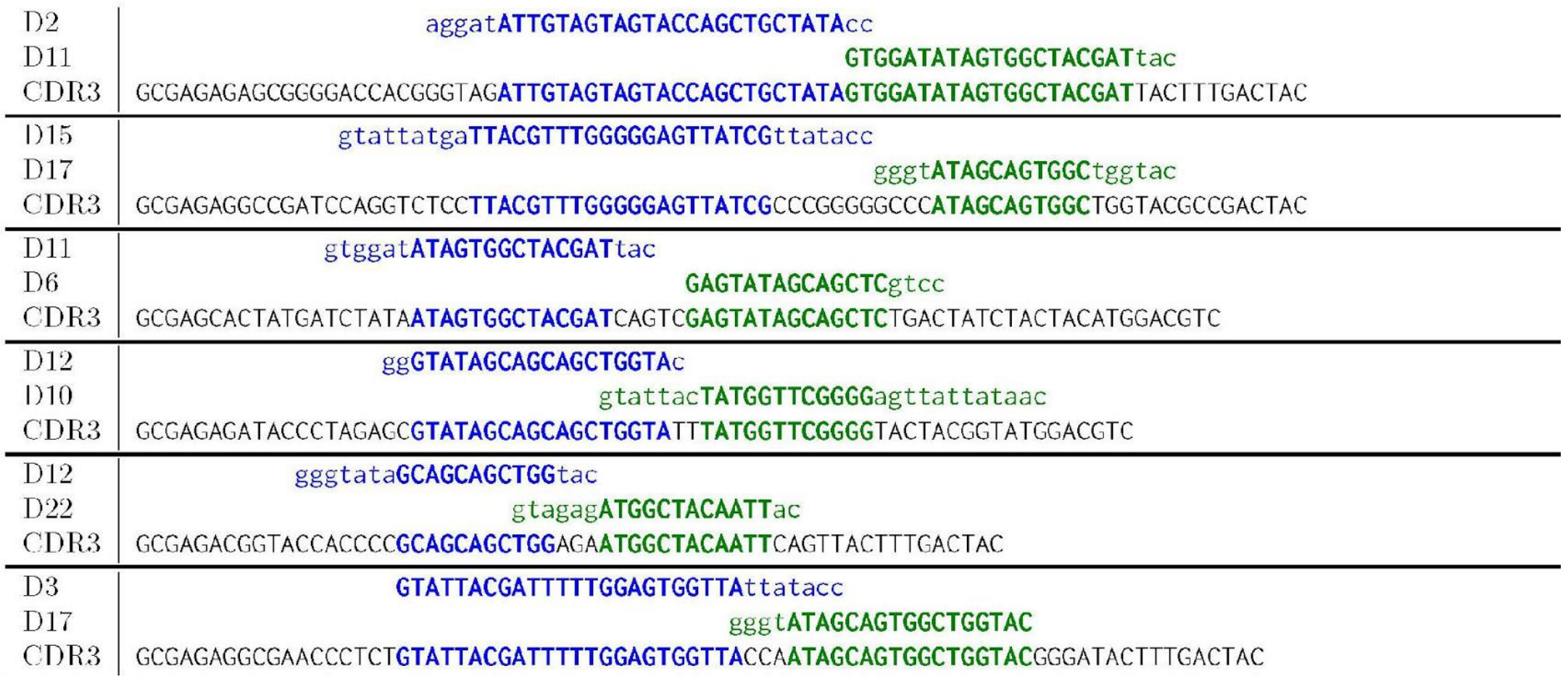

**Figure A34. List of 114 tandem CDR3s in the Set 1 dataset.** All 114 tandem CDR3s can be found at https://drive.google.com/open?id=141FtIZBH3RMNyVBmNbfJIucn4lYBeHgB.